\RequirePackage{ifpdf}
\ifpdf 
\documentclass[pdftex]{sigma}
\else
\documentclass{sigma}
\fi

\numberwithin{equation}{section}

\def\Z{\mathbb Z}

\def\p{\partial}
\def\r{\rangle}
\def\l{\langle}
\def\p{\partial}

\def\diag{\operatorname{diag}}

\begin{document}

\allowdisplaybreaks

\renewcommand{\PaperNumber}{023}

\FirstPageHeading

\ShortArticleName{Antisymmetric Orbit Functions}

\ArticleName{Antisymmetric Orbit Functions}

\Author{Anatoliy KLIMYK~$^\dag$ and Jiri PATERA~$^\ddag$}
\AuthorNameForHeading{A. Klimyk and J. Patera}

\Address{$^\dag$~Bogolyubov Institute for Theoretical Physics,
       14-b  Metrologichna Str., Kyiv 03143, Ukraine}
\EmailD{\href{mailto:aklimyk@bitp.kiev.ua}{aklimyk@bitp.kiev.ua}}

\Address{$^\ddag$~Centre de Recherches Math\'ematiques,
         Universit\'e de Montr\'eal,\\
$\phantom{^\ddag}$~C.P.6128-Centre ville,         Montr\'eal,
H3C\,3J7, Qu\'ebec, Canada}
\EmailD{\href{mailto:patera@crm.umontreal.ca}{patera@crm.umontreal.ca}}

\ArticleDates{Received December 25, 2006; Published online February 12, 2007}

\Abstract{In the paper, properties of antisymmetric orbit
functions are reviewed and further developed. Antisymmetric orbit
functions on the Euclidean space $E_n$ are antisymmetrized
exponential functions. Antisymmetrization is fulf\/illed by a Weyl
group, corresponding to a Coxeter--Dynkin diagram. Properties of
such functions are described. These functions are closely related
to irreducible characters of a compact semisimple Lie group~$G$ of
rank $n$. Up to a sign, values of antisymmetric orbit functions
are repeated on copies of the fundamental domain~$F$ of the
af\/f\/ine Weyl group (determined by the initial Weyl group) in
the entire Euclidean space~$E_n$. Antisymmetric orbit functions
are solutions of the corresponding Laplace equation in~$E_n$,
vanishing on the boundary of the fundamental domain~$F$.
Antisymmetric orbit functions determine a so-called
antisymmetrized Fourier transform which is closely related to
expansions of central functions in characters of irreducible
representations of the group $G$. They also determine a transform
on a f\/inite set of points of $F$ (the discrete antisymmetric
orbit function transform). Symmetric and antisymmetric
multivariate exponential, sine and cosine discrete transforms are
given.}

\Keywords{antisymmetric orbit functions; signed orbits; products
of orbits; orbit function transform; f\/inite orbit function
transform; f\/inite Fourier transforms; f\/inite cosine
transforms; f\/inite sine transforms; symmetric functions}

\Classification{33-02; 33E99; 42B99; 42C15; 58C40}

\section{Introduction}
In \cite{KP06} and \cite{P04} we considered properties and
applications of symmetric orbit functions (they are called there
orbit functions; without the word ``symmetric'').  They are
closely related to f\/inite groups $W$ of geometric symmetries
generated by ref\/lection transformations $r_i$ (that is, such
that $r_i^2=1$), $i=1,2,\dots ,n$, of the $n$-dimensional
Euclidean space $E_n$ with respect to $(n{-}1)$-dimensional
subspaces containing the origin. In order to obtain a symmetric
orbit function we take a point $\lambda \in E_n$ and act upon
$\lambda$ by all elements of the group $W$. If $O(\lambda)$ is the
$W$-orbit of the point $\lambda$, that is the set of all
dif\/ferent points of the form $w\lambda$, $w\in W$, then the
symmetric orbit function, determined by $\lambda$, coincides with
\[
\phi_\lambda(x)=\sum_{\mu\in O(\lambda)} e^{2\pi{\rm i}\langle
\mu,x\rangle},
\]
where $\langle \mu,x\rangle$ is the scalar product on $E_n$. These
functions are invariant with respect to the action by elements of
the group $W$: $\phi_\lambda(wx)=\phi_\lambda(x)$, $w\in W$. If
$\lambda$ is an integral point of $E_n$, then $\phi_\lambda(x)$ is
invariant with respect to the af\/f\/ine Weyl group $W^{\rm aff}$
corresponding to $W$. Since in the simplest case symmetric orbit
functions coincide with the cosine function, sometimes they are
called $C$-functions.

Symmetric orbit functions are multivariate functions having many
beautiful and useful pro\-per\-ties and applicable both in
mathematics and engineering. For this reason, they can be treated
as special functions \cite{P04}.

Symmetricity is the main property of symmetric orbit functions,
considered in \cite{KP06} and \cite{P04}, which make them useful
in applications. Being a modif\/ication of monomial symmetric
functions, they are directly related to the theory of symmetric
(Laurent) polynomials \cite{Mac1, Mac2, Mac3, KVl} (see Section~11
in \cite{KP06}).

Symmetric orbit functions $\phi_\lambda(x)$ for integral $\lambda$
are closely related to the representation theory of compact groups
$G$. In particular, they were ef\/fectively used for dif\/ferent
calculations in representation theory \cite{MP84, MMP85, MMP86,
PS, GP}. They are constituents of traces (characters) of
irreducible unitary representations of $G$. Although characters
contain all (or almost all) information about the corresponding
irreducible representations, they are seldom used as special
functions. The reason is that a construction of characters is
rather complicated, whereas orbit functions have much more simple
structure.

The symmetric orbit function $\phi_\lambda(x)$ is a symmetrized
(by means of the group $W$) exponential function $e^{2\pi{\rm
i}\langle \lambda,x\rangle}$ on $E_n$. For each transformation
group $W$, the symmetric orbit functions form a complete
orthogonal basis in the space of symmetric (with respect to $W$)
polynomials in~$e^{2\pi{\rm i}x_j}$, $j=1,2,\ldots,n$, or in the
Hilbert space obtained by closing this space with respect to an
appropriate scalar product. Orbit functions $\phi_\lambda(x)$,
when $\lambda$ runs over integral weights, determine so-called
(symmetric) orbit function transform, which is a symmerization of
the usual Fourier series expansion on $E_n$. If $\lambda$ runs
over the space $E_n$, then $\phi_\lambda(x)$ determines a
symmetric orbit function transform, which is a symmetrization of
the usual continuous Fourier expansion in $E_n$ (that is, of the
Fourier integral).

In the same way as the Fourier transform leads to discrete Fourier
transform, the symmetric orbit function transform leads to a
discrete analogue of this transform (which is a generalization of
the discrete cosine transform \cite{R}). This discrete transform
is useful in many things related to discretization (see \cite{AP,
Pat-Z-1, Pat-Z-2, Appl-1, Appl-2, Appl-3, Appl-4, Appl-5}).
Construction of the discrete orbit function transform is
fulf\/illed by means of the results of paper \cite{MP87}.

In this paper we consider antisymmetric orbit functions (since in
the simplest case they coincide with the sine function, sometimes
they are called $S$-functions). They are given by
\[
\varphi_\lambda(x)=\sum_{w\in W} (\det w) e^{2\pi{\rm i}\langle
w\lambda,x\rangle},\qquad x\in E_n,
\]
where $\lambda$ is a strictly dominant weight and $\det w$ is a
determinant of the transformation $w$ (it is equal to 1 or $-1$,
depending on either $w$ is a product of even or odd number of
ref\/lections). The orbit functions $\varphi_\lambda$ have many
properties that the symmetric orbit functions $\phi_\lambda$ do.
But antisymmetricity leads to some new properties which are useful
for applications \cite{PZ-06}. For integral $\lambda$,
antisymmetric orbit functions are closely related to characters of
irreducible representations of the corresponding compact Lie group
$G$. Namely, the character $\chi_\lambda$ of the irreducible
representation $T_\lambda$, $\lambda\in P_+$, coincides with
$\varphi_{\lambda+\rho}/\varphi_{\rho}$, where $\rho$ is a half of
a sum of positive roots related to the Weyl group $W$.

Symmetric orbit functions are a generalization of the cosine
function, whereas antisymmetric orbit functions are a generalization
of the sine function. A generalization of the exponential functions
are called $E$-orbit functions. A detailed description of these
functions for rank two see in \cite{Pat-Kash}.

Our goal in this paper is to bring together in full generality the
diverse facts about antisymmetric orbit functions, many of which
are not found in the literature, although they often are
straightforward consequences of known facts. In general, for a
given transformation group $W$ of the Euclidean space $E_n$, most
of the properties of antisymmetric orbit functions, which are
described in this paper, are implications of properties either of
the orbits of the group $W$ or of the usual exponential function
on $E_n$.

For dominant elements $\lambda$, antisymmetric orbit functions
$\varphi_\lambda(x)$ are antisymmetric with respect to elements of
the corresponding Weyl group, that is, $\varphi_\lambda(wx)=(\det
w)\varphi_\lambda(x)$ for any $w\in W$. For this reason,
antisymmetric orbit functions are def\/ined only for strictly
dominant elements $\lambda$ (a dominant element $\lambda$ is
strictly dominant if $w\lambda =\lambda$ means that $w=1$). If
$\lambda$ is {\it integral} strictly dominant element, then the
corresponding antisymmetric orbit function $\varphi_\lambda(x)$ is
antisymmetric also with respect to elements of the af\/f\/ine Weyl
group $W^{\rm aff}$, corresponding to the Weyl group $W$.
Antisymmetricity is a main property of antisymmetric orbit
functions. Because of antisymmetricity, it is enough to determine
$\varphi_\lambda(x)$ only on a fundamental domain of the
af\/f\/ine Weyl group $W^{\rm aff}$ (if $\lambda$ is integral).

 In the case when the group $W$ is a direct product of its
subgroups, say $W=W_1\times W_2$, the fundamental domain is the
Cartesian product of fundamental domains for $W_1$ and~$W_2$.
Similarly, antisymmetric orbit functions of $W$ are products of
antisymmetric orbit functions of $W_1$ and~$W_2$. Hence it
suf\/f\/ices to carry out our considerations for groups $W$ which
cannot be represented as
 a~product of its subgroups (that is, for such $W$ for which a
corresponding Coxeter--Dynkin diagram is connected).

In the article many examples of dimensions 2 and 3 are shown
because they are likely to be used more often.

We shall need a general information on Weyl groups, af\/f\/ine
Weyl groups, root systems and their properties. We have given this
information in \cite{KP06}. In order to make this paper
self-contained we repeat a part of that information in
Section~\ref{section2}.

In Section~\ref{section3} we describe signed Weyl group orbits.
They dif\/fer from the Weyl group orbits by a~sign (equal to $+1$
or $-1$) assigned to each point of an orbit. To each signed orbit
there corresponds an antisymmetric orbit function if a
dominant element of the orbit is strictly dominant.

Section~\ref{section4} is devoted to description of antisymmetric
orbit functions. Antisymmetric orbit functions, corresponding to
Coxeter--Dynkin diagrams, containing only two nodes, are given in
an explicit form. In this section we also give explicit formulas
for antisymmetric orbit functions, corresponding to
Coxeter--Dynkin diagrams of $A_n$, $B_n$, $C_n$ and $D_n$, in the
corresponding ortho\-gonal coordinate systems. In
Section~\ref{section5} properties of antisymmetric orbit functions
are described. If $\lambda$ is integral, then a main property of
the antisymmetric orbit function $\varphi_\lambda$ is an
invariance with respect to the af\/f\/ine Weyl group $W^{\rm
aff}$. We also give here the symmetric and antisymmetric orbit
functions $\phi_\rho(x)$ and $\varphi_\rho(x)$, corresponding to
the half-sum $\rho$ of positive roots, in a form of products of
the cosine and sine functions of certain angles depending on $x$.
Specif\/ic properties of antisymmetric orbit functions of the
Coxeter--Dynkin diagram $A_n$ are given in Section~\ref{section6}.

In Section~\ref{section7} we consider expansions of products of
symmetric (antisymmetric) orbit functions into a sum of symmetric
or antisymmetric orbit functions. These expansions are closely
related to properties of (signed) $W$-orbits. Many examples for
expansions in the case of Coxeter--Dynkin diagrams $A_2$ and $C_2$
are considered. Section~\ref{section8} is devoted to expansion of
antisymmetric $W$-orbit functions into a sum of antisymmetric
$W'$-orbit functions, where $W'$ is a subgroup of the Weyl group
$W$. Many particular cases are studied in detail.

Connection between antisymmetric orbit functions
$\varphi_\lambda(x)$ with integral $\lambda$ and characters of
f\/inite dimensional irreducible representations of the
corresponding simple compact Lie groups is studied in
Section~\ref{section9}. In particular, the well-known Weyl formula
for characters of such representations contains antisymmetric
orbit functions.

In Section~\ref{section10} we expose antisymmetric orbit function
transforms. There are two types of such transforms. The f\/irst
one is an analogue of the expansion into Fourier series and the
second one is an analogue of the Fourier integral transform. In
Section~\ref{section11} a description of an antisymmetric
generalization of the multi-dimensional f\/inite Fourier transform
is given. This analogue is connected with grids on the
corresponding fundamental domains for the af\/f\/ine Weyl groups
$W^{\rm aff}$. Symmetric and antisymmetric multivariate
exponential discrete transforms, as well as symmetric and
antisymmetric multivariate sine and cosine discrete transforms are
given in this section.

In Section~\ref{section12} we show that antisymmetric orbit
functions are solutions of the Laplace equation on the
corresponding $n$-dimensional simplex vanishing on a boundary of
the simplex. It is shown that antisymmetric orbit functions are
eigenfunctions of other dif\/ferential operators.

Section~\ref{section13} is devoted to exposition of symmetric and
antisymmetric functions, which are symmetric and antisymmetric
analogues of special functions of mathematical physics or
ortho\-gonal polynomials. In particular, we f\/ind eigenfunctions
of antisymmetric and symmetric orbit function transforms. These
eigenfunctions are connected with classical Hermite polynomials.

\section[Weyl groups and affine Weyl groups]{Weyl groups and af\/f\/ine Weyl groups}
\label{section2}

\subsection{Coxeter--Dynkin diagrams and Cartan matrices}\label{section2.1}
The sets of symmetric or antisymmetric orbit functions on the
$n$-dimensional Euclidean space~$E_n$ are determined by f\/inite
transformation groups $W$, generated by ref\/lections $r_i$,
$i=1,2,\dots ,n$ (a characteristic property of ref\/lections is
the equality $r^2_i=1$); the theory of such groups see, for
example, in \cite{Kane} and \cite{Hem-1}. We are interested in
those groups $W$ which are Weyl groups of semisimple Lie groups
(semisimple Lie algebras). It is well-known that such Weyl groups
together with the corresponding systems of ref\/lections $r_i$,
$i=1,2,\dots ,n$, are determined by Coxeter--Dynkin diagrams.
There are 4 series and~5 separate simple Lie algebras, which
uniquely determine their Weyl groups $W$. They are denoted as
\[
A_n\ (n\geq1),\ \ B_n\ (n\geq3),\ \ C_n\ (n\geq2),\ \ D_n\
(n\geq4), \ \ E_6,\ \ E_7,\ \ E_8,\ \ F_4,\ \ G_2 .
\]
To these Lie algebras there correspond connected Coxeter--Dynkin
diagrams. To semisimple Lie algebras (they are direct sums of
simple Lie subalgebras) there correspond Coxeter--Dynkin diagrams,
which consist of connected parts, corresponding to simple Lie
subalgebras; these parts are not connected with each other (a
description of the correspondence between simple Lie algebras and
Coxeter--Dynkin diagrams see, for example, in~\cite{Hem-2}). Thus,
we describe only Coxeter--Dynkin diagrams, corresponding to simple
Lie algebras. They are of the form

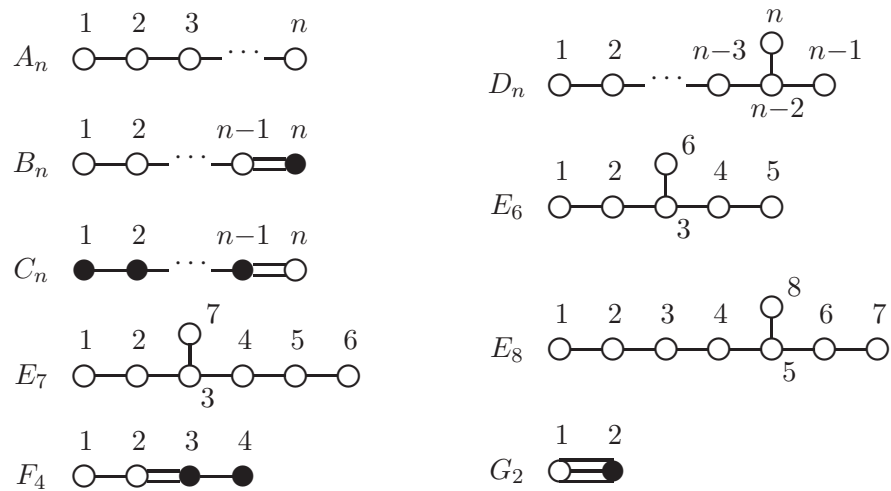
\begin{center}
\parbox{.6\linewidth}{\setlength{\unitlength}{2pt}
\def\kr{\circle{4}}
\def\cr{\circle*{4}}
\thicklines
\begin{picture}(180,90)

\put(0,80){\makebox(0,0){${A_n}$}} \put(10,80){\kr} \put(9,85){1}
\put(20,80){\kr} \put(19,85){2} \put(30,80){\kr} \put(29,85){3}
\put(50,80){\kr} \put(49,85){$n$} \put(12,80){\line(1,0){6}}
\put(22,80){\line(1,0){6}} \put(32,80){\line(1,0){4}}
\put(37,80){$\cdots$} \put(44,80){\line(1,0){4}}

\put(0,60){\makebox(0,0){${B_n}$}}   
\put(10,60){\kr} \put(9,65){1}   \put(20,60){\kr} \put(19,65){2}
\put(40,60){\kr}  \put(35,65){$n{-}1$}  \put(50,60){\cr*}
\put(49,65){$n$} \put(12,60){\line(1,0){6}}
\put(22,60){\line(1,0){4}} \put(27,60){$\cdots$}
\put(34,60){\line(1,0){4}} \put(42,61){\line(1,0){6}}
\put(42,59){\line(1,0){6}}

\put(0,40){\makebox(0,0){${C_n}$}}   
\put(10,40){\cr*}  \put(9,45){1}   \put(20,40){\cr*}
\put(19,45){2} \put(40,40){\cr*}  \put(35,45){$n{-}1$}
\put(50,40){\kr}  \put(49,45){$n$} \put(12,40){\line(1,0){6}}
\put(22,40){\line(1,0){4}}  \put(27,40){$\cdots$}
\put(34,40){\line(1,0){4}} \put(42,41){\line(1,0){6}}
\put(42,39){\line(1,0){6}}

\put(90,75){\makebox(0,0){${D_n}$}}    
\put(100,75){\kr}  \put(99,80){1}  \put(110,75){\kr}
\put(109,80){2}
  \put(130,75){\kr} \put(125,80){$n{-}3$}
\put(140,75){\kr}  \put(136,69){$n{-}2$}  \put(150,75){\kr}
\put(147,80){$n{-}1$} \put(140,83){\kr}  \put(139,87){$n$}
\put(102,75){\line(1,0){6}} \put(112,75){\line(1,0){4}}
\put(117,75){$\cdots$} \put(124,75){\line(1,0){4}}
\put(132,75){\line(1,0){6}} \put(142,75){\line(1,0){6}}
\put(140,77){\line(0,1){4}}

\put(90,52){\makebox(0,0){${E_6}$}}     
\put(100,52){\kr} \put(99,57){1}  \put(110,52){\kr}
\put(109,57){2} \put(120,52){\kr} \put(122,46){3}
\put(130,52){\kr}  \put(129,57){4} \put(140,52){\kr}
\put(139,57){5}  \put(120,60){\kr} \put(123,62){6}
\put(102,52){\line(1,0){6}} \put(112,52){\line(1,0){6}}
\put(122,52){\line(1,0){6}} \put(132,52){\line(1,0){6}}
\put(120,54){\line(0,1){4}}

\put(0,20){\makebox(0,0){${E_7}$}}   
\put(10,20){\kr} \put(9,25){1}  \put(20,20){\kr}  \put(19,25){2}
\put(30,20){\kr}  \put(32,14){3}  \put(40,20){\kr}  \put(39,25){4}
\put(50,20){\kr}  \put(49,25){5}  \put(60,20){\kr}  \put(59,25){6}
\put(30,28){\kr}   \put(33,30){7} \put(12,20){\line(1,0){6}}
\put(22,20){\line(1,0){6}} \put(32,20){\line(1,0){6}}
\put(42,20){\line(1,0){6}} \put(52,20){\line(1,0){6}}
\put(30,22){\line(0,1){4}}

\put(90,25){\makebox(0,0){${E_8}$}} \put(100,25){\kr}
\put(99,30){1}  \put(110,25){\kr}  \put(109,30){2}
\put(120,25){\kr}  \put(119,30){3}  \put(130,25){\kr}
\put(129,30){4} \put(140,25){\kr}  \put(142,19){5}
\put(150,25){\kr}  \put(149,30){6} \put(160,25){\kr}
\put(159,30){7}  \put(140,33){\kr}  \put(143,35){8}
\put(102,25){\line(1,0){6}} \put(112,25){\line(1,0){6}}
\put(122,25){\line(1,0){6}} \put(132,25){\line(1,0){6}}
\put(142,25){\line(1,0){6}} \put(152,25){\line(1,0){6}}
\put(140,27){\line(0,1){4}}

\put(0,1){\makebox(0,0){${F_4}$}}   
\put(10,1){\kr}   \put(9,6){1}   \put(20,1){\kr}  \put(19,6){2}
\put(30,1){\cr*}  \put(29,6){3}   \put(40,1){\cr*} \put(39,6){4}
\put(12,1){\line(1,0){6}} \put(22,2){\line(1,0){6}}
\put(22,0){\line(1,0){6}} \put(32,1){\line(1,0){6}}
\put(90,2){\makebox(0,0){${G_2}$}} \put(100,2){\kr} \put(99,7){1}
\put(110,2){\cr*}  \put(109,7){2} \put(102,2){\line(1,0){6}}
\put(100,4){\line(1,0){10}} \put(100,0){\line(1,0){10}}
\end{picture}}
\end{center}

A diagram determines a certain non-orthogonal basis
$\{\alpha_1,\alpha_2,\dots,\alpha_n\}$ in the Euclidean
spa\-ce~$E_n$. Each node is associated with a basis vector
$\alpha_k$, called a {\it simple root}. A direct link between two
nodes indicates that the corresponding basis vectors are not
orthogonal. Conversely, an absence of a direct link between nodes
implies orthogonality of the corresponding vectors. Single,
double, and triple links indicate that the relative angles between
the two simple roots are $2\pi/3$, $3\pi/4$, $5\pi/6$,
respectively. There can be only two cases: all simple roots are of
the same length or there are only two dif\/ferent lengths of
simple roots. In the f\/irst case all simple roots are denoted by
white nodes. In the case of two lengths, shorter roots are denoted
by black nodes and longer ones by white nodes. Lengths of roots
are determined uniquely up to a common constant. For the cases
$B_n$, $C_n,$ and $F_4$, the squared longer root length is double
the squared shorter root length. For $G_2$, the squared longer
root length is triple the squared shorter root length.

If two nodes are connected by a single line, then the angle
between the corresponding simple roots is $2\pi/3$. If nodes are
connected by a double line, then the angle is $3\pi/4$. A triple
line means that the angle is $5\pi/6$. Simple roots of the same
length are orthogonal to each other or an angle between them is
$2\pi/3$.

To each Coxeter--Dynkin diagram there corresponds a Cartan
matrix $M$, consisting of the entries
\begin{gather}\label{Mmatrix}
M_{jk}=\frac{2\l\alpha_j ,\alpha_k\r}
            {\l\alpha_k ,\alpha_k\r},\qquad
            j,k\in\{1,2,\dots,n\},
\end{gather}
where $\l x, y\r$ denotes the scalar product of $x,y\in E_n$.
Cartan matrices of simple Lie algebras are given in many places
(see, for example, \cite{BMP}). We recall them here for ranks $2$
and $3$ because of their usage later on:
\begin{gather*}
 A_2 :\ \left(
 \begin{array}{cc}
 2&-1\\ -1&2\
 \end{array} \right) ,\qquad
C_2 :  \left(
 \begin{array}{cc}
 2&-1\\ -2&2\
 \end{array} \right) ,\qquad
 G_2 :
 \left(
 \begin{array}{cc}
 2&-3\\ -1&2\
 \end{array} \right) ,
\\
 A_3 :
\left(
 \begin{array}{ccc}
 2&-1&0\\ -1&2&-1\\ 0&-1&2
 \end{array} \right) ,\qquad
 B_3 :
\left(
 \begin{array}{ccc}
 2&-1&0\\ -1&2&-2\\ 0&-1&2
 \end{array} \right) ,\qquad
 C_3 :
\left(
 \begin{array}{ccc}
 2&-1&0\\ -1&2&-1\\ 0&-2&2
 \end{array} \right) .
\end{gather*}

Lengths of the basis vectors $\alpha_i$ are f\/ixed by the
corresponding Coxeter--Dynkin diagram up to a constant. We adopt
the standard choice in the Lie theory, namely
\[
\l\alpha ,\alpha\r=2
\]
for all simple roots of $A_n$, $D_n$, $E_6$, $E_7$, $E_8$ and for
the longer simple roots of $B_n$, $C_n$, $F_4$, $G_2$.

\subsection{Weyl group}\label{section2.2}
A Coxeter--Dynkin diagram determines uniquely the corresponding
transformation group of $E_n$, generated by ref\/lections $r_i$,
$i=1,2,\dots ,n$. These ref\/lections correspond to simple roots
$\alpha_i$, $i=1,2,\ldots,n$. Namely, the transformation $r_i$
corresponds to the simple root $\alpha_i$ and is the ref\/lection
with respect to $(n-1)$-dimensional linear subspace (hyperplane)
of $E_n$ (containing the origin), orthogonal to $\alpha_i$. It is
well-known that such ref\/lections are given by the formula
\begin{gather}\label{reflection}
r_ix=x-\frac{2\l x, \alpha_i\r}{\l\alpha_i, \alpha_i\r}\alpha_i,
\qquad i= 1,2,\ldots,n,\quad x\in E_n .
\end{gather}
Each ref\/lection $r_i$ can be thought as attached to the $i$-th
node of the corresponding diagram.

A f\/inite group $W$, generated by the ref\/lections $r_i$,
$i=1,2,\dots ,n$, is called a {\it Weyl group}, corresponding to a
given Coxeter--Dynkin diagram. If a Weyl group $W$ corresponds to
a Coxeter--Dynkin diagram of a simple Lie algebra $L$, then this
Weyl group is often denoted by $W(L)$. Properties of Weyl groups
are well known (see \cite{Kane} and \cite{Hem-1}). The orders
(numbers of elements) of Weyl groups are given by the formulas
\begin{alignat}{3}
&|W(A_n)|=(n+1)!,\quad &&|W(B_n)|=|W(C_n)|=2^nn!,\quad
                              &&|W(D_n)|=2^{n-1}n!,\notag\\
&|W(E_6)|=51\,840 ,\quad &&|W(E_7)|=2\, 903\,040,\quad
                                    &&|W(E_8)|=696\,729\ 600,\\
&|W(F_4)|=1\,152,\quad &&|W(G_2)|=12. &&\notag
\end{alignat}
In particular,
\[
|W(A_2)|=6,\qquad  |W(C_2)|=8,\qquad |W(A_3)|=24,\qquad
|W(C_3)|=48.
\]

\subsection{Roots and weights}\label{section2.3}
A Coxeter--Dynkin diagram determines a system of simple roots in
the Euclidean space $E_n$. Acting by elements of the Weyl group
$W$ upon simple roots we obtain a f\/inite system of vectors,
which is invariant with respect to $W$. A set of all these vectors
is called a {\it system of roots} associated with a given
Coxeter--Dynkin diagram. It is denoted by $R$. As we see, a system
of roots $R$ is calculated from simple roots by a straightforward
algorithm.

It is proved (see, for example, \cite{Hem-2}) that roots of $R$
are linear combinations of simple roots with integral
coef\/f\/icients. Moreover, there exist no roots, which are linear
combinations of $\alpha_i$, $i=1,2,\dots ,n$, both with positive
and negative coef\/f\/icients. Therefore, the set of roots $R$ can
be represented as a union $R=R_+\cup R_-$, where $R_+$
(respectively $R_-$) is the set of roots which are linear
combination of simple roots with positive (negative)
coef\/f\/icients. The set $R_+$ (the set~$R_-$) is called a {\it
set of positive (negative) roots}.

As mentioned above, a set $R$ of roots is invariant under the
action of elements of the Weyl group $W(R)$. However, $wR_+\ne
R_+$ if $w$ is not a trivial element of $W$. The following
proposition holds:

\begin{proposition}\label{prop1} A reflection $r_i\in W$, corresponding to a simple root $\alpha_i$,
maps $\alpha_i$ into $-\alpha_i$ and reflects the set
$R_+\backslash \{ \alpha_i\}$ of all other roots of $R_+$ onto
itself.
\end{proposition}

Let $X_\alpha$ be the $(n-1)$-dimensional linear subspace
(hyperplane) of $E_n$ which contains the origin and is orthogonal
to the root $\alpha$. Clearly, $X_\alpha=X_{-\alpha}$. The set of
ref\/lections with respect to $X_\alpha$, $\alpha\in R_+$,
coincides with the set of all ref\/lections of the corresponding
Weyl group $W$. The hyperplane $X_\alpha$ consists of all points
$x\in E_n$ such that $\langle x,\alpha \rangle=0$.

The subspaces $X_\alpha$, $\alpha\in R_+$, split the Euclidean
space $E_n$ into connected parts which are called {\it Weyl
chambers}. A number of Weyl chambers coincides with the number of
elements of the Weyl group $W$. Elements of the Weyl group permute
Weyl chambers. A part of a Weyl chamber, which belongs to some
hyperplane $X_\alpha$ is called a {\it wall} of this Weyl chamber.
If for some element $x$ of a Weyl chamber we have $\langle
x,\alpha \rangle=0$ for some root $\alpha$, then this point
belongs to a wall. The Weyl chamber consisting of points $x$ such
that
\[
\langle x,\alpha_i \rangle\ge 0, \qquad i=1,2,\dots,n,
\]
is called the {\it dominant Weyl chamber}. It is denoted by $D_+$.
Elements of $D_+$ are called {\it dominant}. If $\langle
x,\alpha_i \rangle > 0$, $i=1,2,\dots,n$, then $x$ is called {\it
strictly dominant element}.

The set $Q$ of all linear combinations
\[
Q=\left\{ \sum_{i=1}^n a_i\alpha_i \ | \ a_i\in {\mathbb
Z}\right\}\equiv \bigoplus_i {\mathbb Z}\alpha_i
\]
is called a {\it root lattice} corresponding to a given
Coxeter--Dynkin diagram. Its subset
\[
Q_+=\left\{ \sum_{i=1}^n a_i\alpha_i \ | \ a_i=0,1,2,\dots\right\}
\]
is called a {\it positive root lattice}.

To each root $\alpha\in R$
there corresponds the coroot $\alpha^\vee$ def\/ined by the
formula
\[
\alpha^\vee =\frac{2\alpha}{\l \alpha,\alpha\r} .
\]
It is easy to see that $\alpha^{\vee\vee} =\alpha$. The set
$Q^\vee$ of all linear combinations
\[
Q^\vee=\left\{ \sum_{i=1}^n a_i\alpha^\vee_i \ | \ a_i\in {\mathbb
Z}\right\}\equiv \bigoplus_i {\mathbb Z}\alpha^\vee_i
\]
is called a {\it coroot lattice} corresponding to a given
Coxeter--Dynkin diagram. The subset
\[
Q^\vee_+=\left\{ \sum_{i=1}^n a_i\alpha^\vee_i \ | \
a_i=0,1,2,\dots\right\}
\]
is called a {\it positive coroot lattice}.

As noted above, the set of simple roots $\alpha_i$, $i=1,2,\dots
,n$, form a basis of the space $E_n$. In addition to the
$\alpha$-basis, it is convenient to introduce the so-called
$\omega$-basis, $\omega_1,\omega_2,\dots ,\omega_n$ (also called
the {\it basis of fundamental weights}). The two bases are dual to
each other in the following sense:
\begin{gather}\label{kronecker}
\frac{2\l\alpha_j ,\omega_k\r} {\l\alpha_j,\alpha_j\r}\equiv
\l\alpha^\vee_j ,\omega_k\r =\delta_{jk},\qquad
j,k\in\{1,2,\dots,n\}.
\end{gather}
The $\omega$-basis (as well as the $\alpha$-basis) is not
orthogonal.

Note that the factor $2/\l\alpha_j,\alpha_j\r$ can take only three
values. Indeed, with the standard normalization of root lengths,
we have
\begin{alignat*}{3}
& \frac2{\l\alpha_k,\alpha_k\r}
 =1 \quad&& \text{for roots of}\quad
                A_n,\ D_n,\ E_6,\ E_7,\ E_8,&
\\
& \frac2{\l\alpha_k,\alpha_k\r}=1  \quad&&
 \text{for long roots of}\quad B_n, \ C_n,\ F_4,\ G_2,&
\\
& \frac2{\l\alpha_k,\alpha_k\r}=2
 \quad&& \text{for short roots of}\quad B_n,\ C_n,\ F_4,&
\\
& \frac2{\l\alpha_k,\alpha_k\r}= 3 \quad&& \text{for short roots
of}\quad G_2 .&
\end{alignat*}
For this reason, we get
\begin{alignat*}{3}
& \alpha^\vee_k=\alpha_k  \quad&& \text{for roots of}\quad
                A_n,\ D_n, \ E_6,\ E_7,\ E_8,&
\\
& \alpha^\vee_k=\alpha_k  \quad&&
 \text{for long roots of}\quad B_n,\ C_n,\ F_4,\ G_2,&
\\
& \alpha^\vee_k=2\alpha_k
 \quad&& \text{for short roots of}\quad B_n,\ C_n,\ F_4,&
\\
& \alpha^\vee_k=3\alpha_k \quad&& \text{for short roots of}\quad
G_2 .&
\end{alignat*}

The $\alpha$- and $\omega$-bases are related by the Cartan matrix
\eqref{Mmatrix} and by its inverse:
\begin{gather}\label{bases}
\alpha_j=\sum_{k=1}^nM_{jk}\,\omega_k,\qquad
\omega_j=\sum_{k=1}^n(M^{-1})_{jk} \alpha_k.
\end{gather}
For ranks 2 and 3 the inverse Cartan matrices are of the form
\begin{gather*}
 A_2 : \ \frac 13 \left(
 \begin{array}{cc}
 2&1\\ 1&2\
 \end{array} \right) ,\qquad
C_2 : \  \left(
 \begin{array}{cc}
 1&1/2\\ 1&1\
 \end{array} \right) ,\qquad
 G_2 : \
 \left(
 \begin{array}{cc}
 2&3\\ 1&2\
 \end{array} \right) ,
\\
 A_3 : \ \frac 14
\left(
 \begin{array}{ccc}
 3&2&1\\ 2&4&2\\ 1&2&3
 \end{array} \right) ,\qquad
 B_3 :  \ \frac 12
\left(
 \begin{array}{ccc}
 2&2&2\\ 2&4&4\\ 1&2&3
 \end{array} \right) ,\qquad
 C_3 : \ \frac 12
\left(
 \begin{array}{ccc}
 2&2&1\\ 2&4&2\\ 2&4&3
 \end{array} \right) .
\end{gather*}

Later on we need to calculate scalar products $\l x, y\r$ when $x$
and $y$ are given by coordina\-tes~$x_i$ and $y_i$ in
$\omega$-basis. It is given by the formula
\begin{gather}\label{matr}
\l x, y\r
     =\frac12\sum_{j,k=1}^n
                    x_jy_k(M^{-1})_{jk}\l\alpha_k\,|\,\alpha_k\r
    = xM^{-1}Dy^{T}=xSy^{T},
\end{gather}
where $D$ is the diagonal matrix ${\rm diag}\, (\frac 12 \l
\alpha_1,\alpha_1 \r,\dots ,\frac 12 \l \alpha_n,\alpha_n \r)$.
Matrices $S$, called `quadratic form matrices', are shown in
\cite{BMP} for all connected Coxeter--Dynkin diagrams.

The sets $P$ and $P_+$, def\/ined as
\[
P=\Z\omega_1+\cdots+\Z\omega_n \ \supset\
P_+=\Z^{\geq0}\,\omega_1+\cdots+\Z^{\geq0}\,\omega_n,
\]
are called respectively the {\it weight lattice} and the {\it cone
of dominant weights}. The set $P$ can be characterized as a set of
all $\lambda\in E_n$ such that
\[
\frac{2\langle \alpha_j,\lambda \rangle}{\langle \alpha_j,
\alpha_j \rangle}= \langle \alpha^\vee_j,\lambda \rangle\in \Z
\]
for all simple roots $\alpha_j$. Clearly, $Q\subset P$. Below we
shall need also the set $P^+_+$ of dominant weights of $P_+$,
which do not belong to any Weyl chamber (the set of integral
strictly dominant weights). Then $\lambda\in P^+_+$ means that
$\langle \lambda,\alpha_i \rangle >0$ for all simple roots
$\alpha_i$. We have
\[
P^+_+=\Z^{{}>0}\omega_1+\Z^{{}>0}\omega_2+\cdots
+\Z^{{}>0}\omega_n.
\]

The smallest dominant weights of $P_+$, dif\/ferent from zero,
coincide with the elements $\omega_1,\omega_2$, $\dots,\omega_n$
of the $\omega$-basis. They are called {\it fundamental weights}.
They are highest weights of fundamental irreducible
representations of the corresponding simple Lie algebra $L$.

Through the paper we often use the following notation for weights
in $\omega$-basis:
\[
z=\sum_{j=1}^n a_j\omega_j=(a_1\ a_2\ \dots\ a_n),\qquad
a_1,\dots,a_n\in\Z.
\]
If $x=\sum\limits_{j=1}^n b_j\alpha^\vee_j$, then
\begin{gather}\label{weight}
\l z,x\r =\sum_{j=1}^n a_jb_j.
\end{gather}

\subsection{Highest root}\label{section2.4}
There exists a unique highest (long) root $\xi$ and a unique
highest short root $\xi_s$. The highest (long) root can be written
as
\begin{gather}\label{highestroot}
\xi=\sum_{i=1}^nm_i\alpha_i=\sum_{i=1}^n m_i\frac{\langle
\alpha_i,\alpha_i\rangle}{2} \alpha_i^\vee\equiv \sum_{i=1}^n
q_i\alpha_i^\vee  .
\end{gather}
The coef\/f\/icients $m_i$ and $q_i$ can be viewed as attached to
the $i$-th node of the diagram. They are called {\it marks\/} and
{\it comarks\/} and are often listed in the literature (see, for
example, \cite{BMP}). In root systems with two lengths of roots,
namely in $B_n$, $C_n$, $F_4$ and $G_2$, the highest (long) root
$\xi$ is of the form
\begin{alignat}{3}
B_n\ &:&\quad \xi &=&\ (0\,1\,0\,\dots\,0)
               &=\alpha_1+2\alpha_2+2\alpha_3+\cdots+2\alpha_n ,\\
C_n\ &:&\ \xi     &=&\ (2\,0\,\dots\,0)
               &=2\alpha_1+2\alpha_2+\cdots+2\alpha_{n-1}+\alpha_n ,\\
F_4\ &:&\ \xi     &=&\ (1\,0\,0\,0)
               &=2\alpha_1+3\alpha_2+4\alpha_3+2\alpha_4 ,\\
G_2\ &:&\ \xi     &=&\ (1\,0)   &= 2\alpha_1+3\alpha_2.
\end{alignat}
For $A_n$, $D_n$, and $E_n$, all roots are of the same length,
hence $\xi_s=\xi$. We have
\begin{alignat}{3}
A_n\ &:&\quad \xi &=&\ (1\,0\,\dots\,0\,1)
               &=\alpha_1+\alpha_2+\cdots+\alpha_n ,\label{eq2.13}\\
D_n\ &:&\ \xi     &=&\ (0\,1\,0\,\dots\,0)
               &=\alpha_1+2\alpha_2+\cdots+2\alpha_{n-2}+
               \alpha_{n-1}+\alpha_n ,\label{eq2.14}\\
E_6\ &:&\ \xi     &=&\ (0\, 1\, 0\,\dots\,
0)&=\alpha_1+2\alpha_2+3\alpha_3+2\alpha_4+\alpha_5+2\alpha_6 ,
               \label{eq2.15}\\
E_7\ &:&\ \xi     &=&\ (1\, 0\,0\,\dots\,
0)&=2\alpha_1+3\alpha_2+4\alpha_3+3\alpha_4+2\alpha_5+\alpha_6+2\alpha_7,
      \label{eq2.16}\\
E_8\ &:&\ \xi     &=&\ (0\,0\,\dots\, 0\,
1)&=2\alpha_1+3\alpha_2+4\alpha_3+5\alpha_4+6\alpha_5+4\alpha_6+2\alpha_7+
3\alpha_8 .\label{eq2.17}
\end{alignat}

Note that for highest root $\xi$ we have
\begin{gather}\label{highest}
\xi^\vee =\xi .
\end{gather}
Moreover, if all simple roots are of the same length, then
 \[
\alpha_i^\vee =\alpha_i.
 \]
For this reason,
 \[
(q_1,q_2,\dots,q_n)=(m_1,m_2,\dots,m_n).
 \]
for $A_n$, $D_n$ and $E_n$. Formulas
\eqref{eq2.13}--\eqref{highest} determine these numbers. For short
roots $\alpha_i$ of $B_n$, $C_n$ and $F_4$ we have
$\alpha_i^\vee=2\alpha_i$.  For short root $\alpha_2$ of $G_2$ we
have $\alpha_2^\vee=3\alpha_2$. For this reason,
 \[
(q_1,q_2,\dots,q_n)=(1,2,\dots, 2,1)\ \ \ \ {\rm for}\ \ \ \  B_n,
 \]   \[
(q_1,q_2,\dots,q_n)=(1,1,\dots, 1,1)\ \ \ \ {\rm for}\ \ \ \  C_n,
 \]   \[
(q_1,q_2,q_3,q_4)=(2,3, 2,1)\ \ \ \ {\rm for}\ \ \ \  F_4,
 \]   \[
(q_1,q_2)=(2,1)\ \ \ \ {\rm for}\ \ \ \  G_2.
 \]

To each root system $R$ there corresponds an {\it extended
root system} (which is also called an {\it affine root system}).
It is constructed with the help of the highest root $\xi$ of $R$.
Namely, if $\alpha_1,\alpha_2,\dots, \alpha_n$ is a set of all
simple roots, then the roots
\[
\alpha_0:=-\xi,\alpha_1,\alpha_2,\dots, \alpha_n
\]
constitute a set of simple roots of the corresponding extended
root system. Taking into account the orthogonality (non-orthogonality) of
the root $\alpha_0$ to other simple roots, the diagram of an
extended root system can be constructed (which is an extension of
the corresponding Coxeter--Dynkin diagram; see, for example,
\cite{K}). Note that for all simple Lie algebras (except for
$A_n$) only one simple root is orthogonal to the root $\alpha_0$.
In the case of $A_n$, the two simple roots $\alpha_1$ and
$\alpha_n$ are not orthogonal to $\alpha_0$.

\subsection[Affine Weyl groups]{Af\/f\/ine Weyl groups}\label{section2.5}
We are interested in antisymmetric orbit functions which are given
on the Euclidean space~$E_n$. These functions are anti-invariant
with respect to action by elements of a Weyl group $W$, which is a
transformation group of $E_n$. However, $W$ does not describe all
symmetries of orbit functions corresponding to weights $\lambda\in
P^+_+$. A whole group of anti-invariances of antisymmetric orbit
functions is isomorphic to the af\/f\/ine Weyl group $W^{\rm aff}$
which is an extension of the Weyl group~$W$. This group is
def\/ined as follows.

Let $\alpha_1,\alpha_2,\dots ,\alpha_n$ be simple roots in the
Euclidean space $E_n$ and let $W$ be the corresponding Weyl group.
The group $W$ is generated by ref\/lections $r_{\alpha_i}$,
$i=1,2,\dots ,n$. In order to construct the af\/f\/ine Weyl group
$W^{\rm aff}$, corresponding to the group $W$, we have to add an
additional ref\/lection. This ref\/lection is constructed as
follows.

We consider the ref\/lection $r_{\xi}$ with respect to the
$(n-1)$-dimensional subspace (hyperplane) $X_{n-1}$ containing the
origin and orthogonal to the highest (long) root $\xi$, given in
\eqref{highestroot}:
\begin{gather} \label{refl-s}
r_{\xi}x=x-\frac{2\langle x,\xi\rangle}{\langle \xi,\xi\rangle}
\xi .
\end{gather}
Clearly, $r_{\xi} \in W$. We shift the hyperplane $X_{n-1}$ by the
vector $\xi^\vee/2$, where $\xi^\vee =2\xi/\langle\xi,\xi
\rangle$. (Note that by \eqref{highest} we have $\xi^\vee=\xi$.
However, it is convenient to use here  $\xi^\vee$.) The
ref\/lection with respect to the hyperplane $X_{n-1}+\xi^\vee/2$
will be denoted by $r_0$. Then in order to fulf\/ill the
transformation $r_0$ we have to fulf\/ill the transformation
$r_\xi$ and then to shift the result by $\xi^\vee$, that~is,
\[
r_0x=r_\xi x+\xi^\vee .
\]
We have $r_00=\xi^\vee$ and it follows from \eqref{refl-s} that
$r_0$ maps $x+\xi^\vee/2$ to
\[
r_\xi(x+ \xi^\vee/2)+\xi^\vee=x+\xi^\vee/2-\langle x,\xi^\vee
\rangle \xi.
\]
Therefore,
\begin{gather*}
r_0(x+\xi^\vee/2)=x+\xi^\vee/2 - \frac{2\langle
x,\xi\rangle}{\langle \xi,\xi\rangle} \xi =x+\xi^\vee/2
-\frac{2\langle x,\xi^\vee\rangle}{\langle
\xi^\vee,\xi^\vee\rangle} \xi^\vee
\notag\\
\phantom{r_0(x+\xi^\vee/2)}{}=x+\xi^\vee/2-\frac{2\langle
x+\xi^\vee/2,\xi^\vee\rangle}{\langle \xi^\vee,\xi^\vee\rangle}
\xi^\vee +\frac{2\langle \xi^\vee/2,\xi^\vee\rangle}{\langle
\xi^\vee,\xi^\vee\rangle} \xi^\vee . \notag
\end{gather*}
Denoting $x+\xi^\vee/2$ by $y$ we obtain that $r_0$ is given also
by the formula
 \begin{gather} \label{refl-0}
r_0y=y+\left( 1- \frac{2\langle y,\xi^\vee\rangle}{\langle
\xi^\vee,\xi^\vee\rangle}\right) \xi^\vee=\xi^\vee +r_\xi y .
 \end{gather}
The element $r_0$ does not belong to $W$ since elements of $W$ do
not move the point $0\in E_n$.

The hyperplane $X_{n-1}+\xi^\vee/2$ coincides with the set of
points $y$ such that $r_0y=y$. It follows from \eqref{refl-0} that
this hyperplane is given by the equation
 \begin{gather} \label{hyperpl}
1=\frac{2\langle y,\xi^\vee\rangle}{\langle
\xi^\vee,\xi^\vee\rangle}  =\langle y,\xi\rangle =\sum _{k=1}^n
a_kq_k,
 \end{gather}
where{\samepage
\[
y=\sum _{k=1}^n a_k\omega_k,\qquad \xi=\sum _{k=1}^n q_k
\alpha^\vee_k
\]
(see \eqref{weight}).}

A group of transformations of the Euclidean space $E_n$ generated
by ref\/lections $r_0,r_{\alpha_1},\dots ,r_{\alpha_n}$ is called
the {\it affine Weyl group} of the root system $R$ and is denoted
by $W^{\rm aff}$ or by $W^{\rm aff}_R$ (if is necessary to
indicate the initial root system), see \cite{K}.

Adjoining the ref\/lection $r_0$ to the Weyl group $W$
completely change properties of the group $W^{\rm aff}$.

If $r_{\xi}$ is the ref\/lection with respect to the hyperplane
$X_{n-1}$, then due to \eqref{refl-s} and \eqref{refl-0} for any
$x\in E_n$ we have
\[
r_0r_{\xi}x=r_0(r_{\xi}x)=\xi^\vee+r_\xi r_\xi x =x+\xi^\vee .
\]
Clearly, $(r_0r_{\xi})^kx=x+k\xi^\vee$, $k= 0,\pm 1,\pm 2,\dots $,
that is, the set of elements $(r_0r_{\xi})^k$, $k= 0,\pm 1,\pm
2,\dots $, is an inf\/inite commutative subgroup of $W^{\rm aff}$.
This means that (unlike to the Weyl group~$W$) $W^{\rm aff}$ {\it
is an infinite group}.

Since $r_00=\xi^\vee$, for any $w\in W$ we have
\[
wr_00=w\xi^\vee=\xi^\vee_w,
\]
where $\xi^\vee_w$ is a coroot of the same length as the coroot
$\xi^\vee$. For this reason, $wr_0$ is the ref\/lection with
respect to the $(n-1)$-hyperplane perpendicular to the root
$\xi^\vee_w$ and containing the point~$\xi^\vee_w/2$. Moreover,
\[
(wr_0)r_{\xi^\vee_w} x=x+\xi^\vee_w.
\]
We also have $((wr_0)r_{\xi^\vee_w})^kx=x+k\xi^\vee_w$, $k=0,\pm
1,\pm 2,\dots$. Since $w$ is any element of $W$, then the set
$w\xi^\vee$, $w\in W$, coincides with the set of coroots of
$R^\vee$, corresponding to all long roots of the root system $R$.
Thus, {\it the set $W^{\rm aff}\cdot 0$ coincides with the lattice
$Q^\vee_l$ generated by coroots $\alpha^\vee$ taken for all long
roots $\alpha$ from $R$.}

It is checked for each type of root systems that each coroot
$\xi_s^\vee$ for a short root $\xi_s$ of $R$ is a~linear
combination of coroots $w\xi^\vee\equiv \xi_w$, $w\in W$, with
integral coef\/f\/icients, that is, $Q^\vee =Q_l^\vee$. Therefore,
{\it The set $W^{\rm aff}\cdot 0$ coincides with the coroot
lattice $Q^\vee$ of $R$.}

Let $\hat Q^\vee$ be the subgroup of $W^{\rm aff}$ generated by
the elements
 \begin{gather} \label{ref-ow}
 (wr_0)r_w, \qquad\qquad w\in W,
 \end{gather}
where $r_w\equiv r_{\xi^\vee_w}$ for $w\in W$. Since elements
\eqref{ref-ow} pairwise commute with each other (since they are
shifts), $\hat Q^\vee$ is a commutative group. The subgroup $\hat
Q^\vee$ can be identif\/ied with the coroot lattice $Q^\vee$.
Namely, if for $g\in \hat Q^\vee$ we have $g\cdot 0=\gamma\in
Q^\vee$, then $g$ is identif\/ied with $\gamma$. This
correspondence is one-to-one.

The subgroups $W$ and $\hat Q^\vee$ generate $W^{\rm aff}$ since a
subgroup of $W^{\rm aff}$, generated by $W$ and $\hat Q^\vee$,
contains the element $r_0$. {\it The group $W^{\rm aff}$ is a
semidirect product of its subgroups $W$ and $\hat Q^\vee$, where
$\hat Q^\vee$ is an invariant subgroup} (see Section 5.2 in
\cite{KP06} for details).

\subsection{Fundamental domain}\label{section2.6}
An open connected simply connected set $D\subset E_n$ is called a
{\it fundamental domain} for the group $W^{\rm aff}$ (for the
group $W$) if it does not contains equivalent points (that is,
points $x$ and $x'$ such that $x=wx$) and if its closure contains
at least one point from each $W^{\rm aff}$-orbit (from each
$W$-orbit). It is evident that {\it the dominant Weyl chamber
(without walls of this chamber) is a fundamental domain for the
Weyl group $W$}. Recall that this domain consists of all points
$x=a_1\omega_1+\cdots +a_n\omega_n\in E_n$ for which
\[
a_i=\langle x,\alpha^\vee_i \rangle > 0,\qquad i=1,2,\dots,n.
\]
We wish to describe a fundamental domain for the group $W^{\rm
aff}$. Since $W\subset W^{\rm aff}$, it can be chosen as a subset
of the dominant Weyl chamber for $W$.

We have seen that the element $r_0\in W^{\rm aff}$ is a
ref\/lection with respect to the hyperplane $X_{n-1}+\xi^\vee/2$,
orthogonal to the root $\xi$ and containing the point
$\xi^\vee/2$. This hyperplane is given by the equation
\eqref{hyperpl}. This equation shows that the hyperplane
$X_{n-1}+\xi^\vee/2$ intersects the axices, determined by the
vectors $\omega_i$, in the points $\omega_i/q_i$, $i=1,2,\dots
,n$, where $q_i$ are such as in~\eqref{hyperpl}. We create the
simplex with $n+1$ vertices in the points
 \begin{gather}\label{sympl}
0,\ \frac{\omega_1}{q_1},\ \dots , \ \frac{\omega_n}{q_n} .
 \end{gather}
By the def\/inition of this simplex and by \eqref{hyperpl}, this
simplex consists of all points $y$ of the dominant Weyl chamber
for which $\langle y,\xi \rangle \le 1$. Clearly, the interior $F$
of this simplex belongs to the dominant Weyl chamber. The
following theorem is true (see, for example, \cite{KP06}):

\begin{theorem}
 The set $F$ is a fundamental domain for the
affine Weyl group $W^{\rm aff}$.
\end{theorem}

For the rank 2 cases the fundamental domain is the interior of the
simplex with the following vertices:
 \begin{alignat}{2} A_2\
&:&\quad& \{ 0,\ \omega_1,\ \omega_2\} ,
\notag\\
C_2\ &:&\quad& \{ 0,\ \omega_1,\ \omega_2\} , \notag
\\
G_2\ &:&\quad&\{ 0,\ \tfrac{\omega_1}2,\ \omega_2\} .\notag
\end{alignat}

\section{Weyl group signed orbits}\label{section3}

\subsection{Signed orbits}\label{section3.1}
As we have seen, the $(n-1)$-dimensional linear subspaces
$X_\alpha$ of $E_n$, orthogonal to positive roots~$\alpha$ and
containing the origin,  divide the space $E_n$ into connected
parts, which are called {\it Weyl chambers}. A number of such
chambers is equal to an order of the corresponding Weyl group~$W$.
Elements of the Weyl group permute these chambers. There exists a
single chamber~$D_+$ such that $\l\alpha_i, x \r \ge 0$, $x\in
D_+$, $i=1,2,\dots ,n$. It is the {\it dominant Weyl chamber}.

Clearly, the cone of dominant weights $P_+$ belongs to the
dominant Weyl chamber $D_+$. (Note that it is not a case for the
set $Q_+$.) We have $P\cap D_+=P_+$.

Let $y$ be an arbitrary dominant element of the Euclidean space
$E_n$, which {\it does not lie on some Weyl chamber}. We act upon
$y$ by all elements of the Weyl group $W$. As a result, we obtain
a set of elements $wy$, $w\in W$, which is called {\it Weyl group
orbit}. All these elements are pairwise dif\/ferent. We attach to
each point $wy$ a sign coinciding with a sign of $\det w$. The set
of all points $wy$, $w\in W$, together with their signs is called
a {\it signed orbit of the point} $y$ with respect the Weyl group
(or a Weyl group signed orbit, containing $y$). Points of signed
orbits will be denoted by $x^+$ or $x^-$, depending on a sign.
Sometimes, we denote points $wy$, $y\in D_+$, of the signed orbit,
containing the point $y$, as $wy^{\det w}$, where instead of a
sign we have $+1$ or $-1$, respectively.

An orbit (where points do not have signs) of a point $y\in D_+$ is
denoted by $O(y)$ or $O_W(y)$. A~size of an orbit $O(y)$ is a
number $|O(y)|$ of its elements. Each Weyl chamber contains only
one point of a f\/ixed orbit $Q(y)$. A signed orbit of a strictly
dominant point $y\in E_n$ is denoted by~$O^\pm(y)$
or~$O^\pm_W(y)$.

Note that orbits $O(y)$ are def\/ined for any dominant elements
$y\in E_n$. Signed orbits $O^\pm(y)$ can be def\/ined only for
strictly dominant $y\in E_n$.

\subsection[Signed orbits of $A_1$, $A_1\times A_1$, $A_2$,
$C_2$, $G_2$]{Signed orbits of $\boldsymbol{A_1}$,
$\boldsymbol{A_1\times A_1}$, $\boldsymbol{A_2}$,
$\boldsymbol{C_2}$, $\boldsymbol{G_2}$}\label{section3.2}

Assuming that $a> 0$ and $b> 0$, we list the contents of signed
orbits in $\omega$-basis:
\begin{align}
A_1:\quad
  &O^\pm(a)\ni(a)^+,\ (-a)^-\\
A_1\times A_1:\quad
    &O^\pm(a\ b)\ni(a\ b)^+,\ ({-}a\ b)^-,\ (a\ {-}b)^-,\ ({-}a\ {-}b)^+\\
A_2:\quad
           &O^\pm(a\ b)\ni(a\ b)^+,\ ({-}a\ a{+}b)^-,\ (a{+}b\ {-}b)^-,\notag\\
           &\qquad\qquad ({-}b\ {-}a)^-,\ ( {-}a{-}b\ a)^+,\ (b\ {-}a{-}b)^+.
\end{align}
In the cases of $C_2$ and $G_2$ (where the second simple root is
the longer one for $C_2$ and the shorter one for $G_2$) we have
\begin{align}
C_2:\quad
   &O^\pm(a\ b)\ni (a\ b)^+,\ ({-}a\ a{+}b)^-,\ (a{+}2b\ {-}b)^-,\
           (a{+}2b\ {-}a{-}b)^+,\notag \\
&\qquad\quad    ({-}a\ {-}b)^+,\ ({-}a\ {-}a{-}b)^-,\ ({-}a{-}2b\
b)^-,\
           ({-}a{-}2b\ a{+}b)^+,\\
G_2:\quad
  &O^\pm(a\ b)\ni\pm(a\ b)^+,\ \pm({-}a\ 3a{+}b)^-,\ \pm(a{+}b\ {-}b)^-,\notag\\
  &\qquad\quad\pm(2a{+}b\ {-}3a{-}b)^+,\ \pm({-}a{-}b\ 3a{+}2b)^+,\ \pm({-}2a{-}b\
  3a{+}2b)^- ,
\end{align}
where $\pm (c,d)^+$ means two signed points $(c,d)^+$ and
$(-c,-d)^+$.

As we see, for each point $(c\ d)$ of a signed orbit of $C_2$ or
$G_2$ there exists in the orbit the point $({-}c\ {-}d)$ with the
same sign.

\subsection[The case of $A_n$]{The case of $\boldsymbol{A_n}$}\label{section3.3}
In the cases of Coxeter--Dynkin diagrams $A_{n-1}$, $B_n$, $C_n$,
$D_n$, root and weight lattices, Weyl groups and signed orbits are
described in a simple way by using the orthogonal coordinate
system in~$E_n$. In particular, this coordinate system is useful
under practical work with signed orbits.

In the case $A_n$ it is convenient to describe root and weight
lattices, Weyl group and antisymmetric orbit functions in the
subspace of the Euclidean space $E_{n+1}$, given by the equation
\[
x_1+x_2+\cdots +x_{n+1}=0,
\]
where $x_1,x_2,\dots ,x_{n+1}$ are orthogonal coordinates of a
point $x\in E_{n+1}$. The unit vectors in directions of these
coordinates are denoted by ${\bf e}_j$, respectively. Clearly,
${\bf e}_i\bot {\bf e}_j$, $i\ne j$. The set of roots of $A_n$ is
given by the vectors
\[
\alpha_{ij}={\bf e}_i-{\bf e}_j, \qquad i\ne j.
\]
The roots
\[
\alpha_{ij}={\bf e}_i-{\bf e}_j, \qquad i< j,
\]
are positive and the roots
\[
\alpha_i\equiv \alpha_{i,i+1}={\bf e}_i-{\bf e}_{i+1},\qquad
i=1,2,\dots ,n,
\]
constitute the system of simple roots.

If $x=\sum\limits_{i=1}^{n+1} x_i{\bf e}_i$, $x_1+x_2+\cdots
+x_{n+1}=0$, is a point of $E_{n+1}$, then this point belongs to
the dominant Weyl chamber $D_+$ if and only if
\[
x_1\ge x_2\ge \cdots \ge x_{n+1}.
\]
Indeed, if this condition is fulf\/illed, then $\langle
x,\alpha_i\rangle =x_i-x_{i+1}\geq 0$, $i=1,2,\dots, n$, and $x$
is dominant. Conversely, if $x$ is dominant, then $\langle
x,\alpha_i\rangle \geq 0$ and this condition is fulf\/illed. {\it
The point $x$ is strictly dominant if and only if}
\[
x_1> x_2> \cdots > x_{n+1}.
\]

If $\lambda =\sum\limits_{i=1}^n \lambda_i \omega_i$, then the
coordinates $\lambda_i$ in the $\omega$-coordinates are connected
with the orthogonal coordinates $m_j$ of
$\lambda=\sum\limits_{i=1}^{n+1} m_i{\bf e}_i$ by the formulas
 \begin{gather*}
 m_1  = \frac{n}{n+1} \lambda_1
  + \frac{n-1}{n+1}\lambda_2 + \frac{n-2}{n+1}\lambda_3
 + \cdots  + \frac{2}{n+1}\lambda_{n-1} + \frac{1}{n+1}\lambda_n,\\
 m_2  = -\frac{1}{n+1} \lambda_1
  + \frac{n-1}{n+1}\lambda_2 + \frac{n-2}{n+1}\lambda_3
 + \cdots  + \frac{2}{n+1}\lambda_{n-1} + \frac{1}{n+1}\lambda_n,\\
 m_3  = -\frac{1}{n+1} \lambda_1
  - \frac{2}{n+1}\lambda_2 + \frac{n-2}{n+1}\lambda_3
 + \cdots  + \frac{2}{n+1}\lambda_{n-1} + \frac{1}{n+1}\lambda_n,\\
 \cdots    \cdots    \cdots      \cdots    \cdots    \cdots
  \cdots    \cdots    \cdots      \cdots    \cdots    \cdots
\cdots    \cdots    \cdots      \cdots    \cdots    \cdots
  \cdots    \cdots    \cdots      \cdots    \\
 m_{n}  = -\frac{1}{n+1} \lambda_1
  - \frac{2}{n+1}\lambda_2 - \frac{3}{n+1}\lambda_3
 - \cdots  - \frac{n-1}{n+1}\lambda_{n-1} + \frac{1}{n+1}\lambda_n,\\
 m_{n+1}  = -\frac{1}{n+1} \lambda_1
  - \frac{2}{n+1}\lambda_2 - \frac{3}{n+1}\lambda_3
 - \cdots
  - \frac{n-1}{n+1}\lambda_{n-1} - \frac{n}{n+1}\lambda_n.
 \end{gather*}
 The inverse formulas are
  \begin{gather}\label{***}
 \lambda_i=m_i-m_{i+1},\qquad i=1,2,\dots ,n.
 \end{gather}

By means of the formula
 \begin{gather}\label{refl}
 r_\alpha \lambda=\lambda -\frac{2\l \lambda,\alpha\r}{\l \alpha,
 \alpha \r}\alpha \qquad
  \end{gather}
for the ref\/lection with respect to the hyperplane, orthogonal to
a root $\alpha$, we can f\/ind that the ref\/lection
$r_{\alpha_{ij}}$ acts upon the vector
$\lambda=\sum\limits_{i=1}^{n+1} m_i{\bf e}_i$, given by
orthogonal coordinates, by permuting the coordinates $m_i$ and
$m_j$. Since for each $i$ and $j$, $1\le i,j\le n+1$, there exists
a root $\alpha_{ij}$, {\it the Weyl group $W(A_n)$ consists of all
permutations of the orthogonal coordinates $m_1,m_2,\dots
,m_{n+1}$ of a point $\lambda$, that is, $W(A_n)$ coincides with
the symmetric group $S_{n+1}$}.

Sometimes (for example, if we wish that coordinates would be
integers or non-negative integers), it is convenient to introduce
orthogonal coordinates $x_1,x_2,\dots,$ $x_{n+1}$ for $A_n$ in
such a way that
\[
x_1+x_2+\cdots +x_{n+1}=m,
\]
where $m$ is some f\/ixed real number. They are obtained from the
previous orthogonal coordinates by adding the same number
$m/(n+1)$ to each coordinate. Then, as one can see from
\eqref{***}, $\omega$-coordinates $\lambda_i=x_i-x_{i+1}$ and the
Weyl group $W$ do not change. Sometimes, it is natural to use
orthogonal coordinates $x_1,x_2,\dots,x_{n+1}$ for which all $x_i$
are non-negative.

We need below the half-sum $\rho$ of the positive roots of $A_n$,
$\rho=\frac12 \sum\limits_{\alpha>0} \alpha$. It is easy to see
that up to a common constant we have
\[
  \rho=n{\bf e}_1+ (n-1){\bf e}_2+\cdots
+{\bf e}_n,
 \]
that is, in orthogonal coordinates we have
 \begin{gather}\label{pho-A}
 \rho=(n,n-1,\dots,1,0).
\end{gather}
In the non-orthogonal $\omega$-coordinates we have
\[
\rho=\omega_1+\omega_2+\cdots +\omega_n.
\]

The signed orbit $O^\pm(\lambda)$,
$\lambda=(m_1,m_2,\dots,m_{n+1})$, $m_1> m_2> \cdots > m_{n+1}$,
consists of all points
\[
(m_{i_1},m_{i_2},\dots,m_{i_{n+1}})^{{\rm sgn}\, (\det w)}
\]
obtained from $(m_1,m_2,\dots,m_{n+1})$ by permutations $w\in
W\equiv S_{n+1}$. Below instead of ${\rm sgn}\, (\det w)$ we write
simply $\det w$.

\subsection[The case of $B_n$]{The case of $\boldsymbol{B_n}$}\label{section3.4}

Orthogonal coordinates of a point $x\in E_n$ are denoted by
$x_1,x_2,\dots ,x_n$. We denote by ${\bf e}_i$ the corresponding
unit vectors. Then the set of roots of $B_n$ is given by the
vectors
\[
\alpha_{\pm i,\pm j}=\pm {\bf e}_i\pm {\bf e}_j, \quad i\ne j,
\qquad  \alpha_{\pm i}=\pm {\bf e}_i,\quad i=1,2,\dots ,n
\]
(all combinations of signs must be taken). The roots
\[
\alpha_{i,\pm j}={\bf e}_i\pm {\bf e}_j, \quad i< j,\qquad
\alpha_{i}={\bf e}_i,\quad i=1,2,\dots ,n,
\]
are positive and $n$ roots
\[
\alpha_i:={\bf e}_i-{\bf e}_{i+1},\quad i=1,2,\dots ,n-1, \qquad
\alpha_n={\bf e}_n
\]
constitute the system of simple roots.

It is easy to see that if $\lambda=\sum\limits_{i=1}^{n} m_i{\bf
e}_i$ is a point of $E_{n}$, then this point belongs to the
dominant Weyl chamber $D_+$ if and only if
\[
m_1\ge m_2\ge \cdots \ge m_{n}\ge 0.
\]
Moreover, {\it this point is strictly dominant if and only if}
\[
m_1> m_2> \cdots > m_{n}> 0.
\]

If $\lambda =\sum\limits_{i=1}^n \lambda_i \omega_i$, then the
coordinates $\lambda_i$ in the $\omega$-coordinates are connected
with the coordinates $m_j$ of $\lambda=\sum\limits_{i=1}^{n}
m_i{\bf e}_i$ by the formulas
 \begin{alignat*}{11}
 m_1 &=&\lambda_1&+&\lambda_2&+&\cdots &+&\lambda_{n-1}&+&\tfrac 12
 \lambda_n, \\
 m_2&=&         && \lambda_2&+&\cdots &+&\lambda_{n-1}&+&\tfrac 12
 \lambda_n, \\
 \cdots &&\cdots && \cdots && \cdots   && \cdots  && \cdots \\
 m_{n}&=&         &&    &&   &&  &&\tfrac 12  \lambda_n,
 \end{alignat*}
 The inverse formulas are
 \[
 \lambda_i=m_i-m_{i+1},\quad i=1,2,\dots ,n-1,\qquad
 \lambda_n=2m_n.
 \]
It is easy to see that if $\lambda\in P_+$, then the coordinates
$m_1,m_1,\dots ,m_n$ are all integers or all half-integers.

The half-sum $\rho$ of positive roots of $B_n$, $\rho=\frac12
\sum\limits_{\alpha>0} \alpha$, in orthogonal coordinates has the
form{\samepage
 \begin{gather}\label{pho-B}
 \rho=(n-\tfrac 12,n-\tfrac 32,\dots, \tfrac 12).
\end{gather}
In $\omega$-coordinates we have $\rho=\omega_1+\omega_2+\cdots
+\omega_n$.}

By means of the formula \eqref{refl} we f\/ind that the
ref\/lection $r_{\alpha}$ acts upon orthogonal coordinates of the
vector $\lambda=\sum\limits_{i=1}^{n} m_i{\bf e}_i$ by permuting
$i$-th and $j$-th coordinates if $\alpha=\pm ({\bf e}_i - {\bf
e}_j)$, as the permutation of $i$-th and $j$-th coordinates and
the change of their signs if $\alpha=\pm ({\bf e}_i + {\bf e}_j)$,
and as the change of a sign of $i$-th coordinate if $\alpha=\pm
{\bf e}_i$. Thus, the Weyl group $W(B_n)$ consists of all
permutations of the orthogonal coordinates $m_1,m_2,\dots ,m_{n}$
of a point $\lambda$ with possible sign alternations of any number
of them.

The signed orbit $O^\pm(\lambda)$,
$\lambda=(m_1,m_2,\dots,m_{n})$, $m_1> m_2> \cdots > m_{n}> 0$,
consists of all points
  \begin{gather}\label{B-altern}
(\pm m_{i_1}, \pm m_{i_2},\dots,\pm m_{i_{n}})^{\det w}
  \end{gather}
(each combination of signs is possible) obtained from $(m_1,m_2,
\dots,m_{n})$ by permutations and alternations of signs which
constitute an element $w$ of the Weyl group $W(B_n)$. Moreover,
$\det w$ is equal to $\pm 1$ depending on whether $w$ consists of
even or odd number of ref\/lections and alternations of signs. A
sign of $\det w$ can be determined as follows. We represent $w$ as
a product $w=\epsilon s$, where $s$ is a permutation of
$(m_1,m_2,\dots,m_n)$ and $\epsilon$ is an alternation of signs of
coordinates. Then $\det w=(\det
s)\epsilon_{i_1}\epsilon_{i_2}\cdots \epsilon_{i_n}$, where $\det
s$ is def\/ined as in the previous subsection and $\epsilon_{i_j}$
is a sign of $i_j$-th coordinate.

\subsection[The case of $C_n$]{The case of $\boldsymbol{C_n}$}\label{section3.5}
In the orthogonal system of coordinates of the Euclidean space
$E_{n}$ the set of roots of $C_n$ is given by the vectors
\[
\alpha_{\pm i,\pm j}=\pm {\bf e}_i\pm {\bf e}_j, \quad i\ne j,
\qquad  \alpha_{\pm i}=\pm 2{\bf e}_i,\quad i=1,2,\dots ,n,
\]
where ${\bf e}_i$ is the unit vector in the direction of $i$-th
coordinate $x_i$ (all combinations of signs must be taken). The
roots
\[
\alpha_{i,\pm j}={\bf e}_i\pm {\bf e}_j, \quad i< j,\qquad
\alpha_{i}=2{\bf e}_i,\quad i=1,2,\dots ,n,
\]
are positive and $n$ roots
\[
\alpha_i:={\bf e}_i-{\bf e}_{i+1},\quad i=1,2,\dots ,n-1, \qquad
\alpha_n=2{\bf e}_n
\]
constitute the system of simple roots.

It is easy to see that a point $\lambda=\sum\limits_{i=1}^{n}
m_i{\bf e}_i\in E_n$ belongs to the dominant Weyl chamber~$D_+$ if
and only if
\[
m_1\ge m_2\ge \cdots \ge m_{n}\ge 0.
\]
This point is strictly dominant if and only if
\[
m_1> m_2> \cdots > m_{n}> 0.
\]

If $\lambda =\sum\limits_{i=1}^n \lambda_i \omega_i$, then the
coordinates $\lambda_i$ in the $\omega$-coordinates are connected
with the coordinates $m_j$ of $\lambda=\sum\limits_{i=1}^{n}
m_i{\bf e}_i$ by the formulas
 \begin{alignat*}{11}
 m_1 &=&\lambda_1&+&\lambda_2&+&\cdots &+&\lambda_{n-1}&+&
 \lambda_n,\\
 m_2&=&  &&\lambda_2&+&\cdots &+&\lambda_{n-1}&+& \lambda_n,\\
 \cdots &&\cdots && \cdots && \cdots   && \cdots  && \cdots \\
 m_{n}&=&        &&      &&        &&     && \lambda_n .
 \end{alignat*}
 The inverse formulas are
 \[
 \lambda_i=m_i-m_{i+1},\qquad i=1,2,\dots ,n-1,\qquad
 \lambda_n=m_n.
 \]
If $\lambda\in P_+$, then all coordinates $m_i$ are integers.

The half-sum $\rho$ of positive roots of $C_n$, $\rho=\frac12
\sum\limits_{\alpha>0} \alpha$, in orthogonal coordinates has the
form
 \begin{gather}\label{pho-C}
 \rho=(n,n-1,\dots, 2,1).
\end{gather}

By means of the formula \eqref{refl} we f\/ind that the
ref\/lection $r_{\alpha}$ acts upon orthogonal coordinates of the
vector $\lambda=\sum\limits_{i=1}^{n} m_i{\bf e}_i$ by permuting
$i$-th and $j$-th coordinates if $\alpha=\pm ({\bf e}_i - {\bf
e}_j)$, as the permutation of $i$-th and $j$-th coordinates and
the change of their signs if $\alpha=\pm ({\bf e}_i + {\bf e}_j)$,
and as the change of a sign of $i$-th coordinate if $\alpha=\pm
2{\bf e}_i$. Thus, the Weyl group $W(C_n)$ consists of all
permutations of the orthogonal coordinates $m_1,m_2,\dots ,m_{n}$
of a point $\lambda$ with sign alternations of some of them, that
is, this Weyl group acts on orthogonal coordinates exactly in the
same way as the Weyl group $W(B_n)$ does.

The signed orbit $O^\pm(\lambda)$,
$\lambda=(m_1,m_2,\dots,m_{n})$, $m_1> m_2> \dots > m_{n}> 0$,
consists of all points
\[
(\pm m_{i_1}, \pm m_{i_2},\dots,\pm m_{i_{n+1}})^{\det w}
\]
(each combination of signs is possible) obtained from $(m_1,m_2,
\dots,m_{n})$ by permutations and alternations of signs which
constitute an element $w$ of the Weyl group $W(C_n)$. Moreover,
$\det w$ is equal to $\pm 1$ depending on whether $w$ consists of
even or odd numbers of ref\/lections and alternations of signs.
Since $W(C_n)=W(B_n)$, then a sign of $\det w$ is determined as in
the case~$B_n$.

As we see, in the orthogonal coordinates signed orbits for $C_n$
coincide with signed orbits of~$B_n$.

\subsection[The case of $D_n$]{The case of $\boldsymbol{D_n}$} \label{section3.6}

In the orthogonal system of coordinates of the Euclidean space
$E_{n}$ the set of roots of $D_n$ is given by the vectors
\[
\alpha_{\pm i,\pm j}=\pm {\bf e}_i\pm {\bf e}_j, \quad i\ne j,
\]
where ${\bf e}_i$ is the unit vector in the direction of $i$-th
coordinate (all combinations of signs must be taken). The roots
\[
\alpha_{i,\pm j}={\bf e}_i\pm {\bf e}_j, \quad i< j,
\]
are positive and $n$ roots
\[
\alpha_i:={\bf e}_i-{\bf e}_{i+1},\quad i=1,2,\dots ,n-1, \qquad
\alpha_n={\bf e}_{n-1}+ {\bf e}_n
\]
constitute the system of simple roots.

It is easy to see that if $\lambda=\sum\limits_{i=1}^{n} m_i{\bf
e}_i$ is a point of $E_{n}$, then this point belongs to the
dominant Weyl chamber $D_+$ if and only if
\[
m_1\ge m_2\ge \cdots \ge m_{n-1}\ge |m_n|.
\]
{\it This point is strictly dominant if and only if}
\[
m_1> m_2> \cdots > m_{n-1}> |m_n|
\]
(in particular, $m_n$ can take the value 0).

If $\lambda =\sum\limits_{i=1}^n \lambda_i \omega_i$, then the
coordinates $\lambda_i$ in the $\omega$-coordinates are connected
with the coordinates $m_j$ of $\lambda=\sum\limits_{i=1}^{n}
m_i{\bf e}_i$ by the formulas
  \begin{alignat*}{13}
 m_1&=&\lambda_1&+&\lambda_2&+&\cdots &+&\lambda_{n-2}&+&\tfrac 12
 (\lambda_{n-1} &+&\lambda_n),\\
 m_2&=&   &&  \lambda_2&+&\cdots &+&\lambda_{n-2}&+&\tfrac 12
 (\lambda_{n-1}&+&  \lambda_n),\\
 \cdots &&\cdots && \cdots && \cdots && \cdots  && \cdots &&\cdots\\
 m_{n-1}&=&     &&         &&        &&        &&
 \tfrac 12 (\lambda_{n-1}&+& \lambda_n),\\
 m_{n}&=&     &&         &&        &&        &&
 \tfrac 12 (\lambda_{n-1}&-& \lambda_n),
  \end{alignat*}
 The inverse formulas are
 \[
 \lambda_i=m_i-m_{i+1},\quad i=1,2,\dots ,n-2,\qquad
 \lambda_{n-1}=m_{n-1}+m_n,\qquad \lambda_{n}=m_{n-1}-m_n .
 \]
If $\lambda\in P_+$, then the coordinates $m_1,m_2,\dots,m_n$ are
all integers or all half-integers.

The half-sum $\rho$ of positive roots of $D_n$, $\rho=\frac12
\sum\limits_{\alpha>0} \alpha$, in orthogonal coordinates has the
form
 \begin{gather}\label{pho-D}
 \rho=(n-1,n-2,\dots, 1,0).
\end{gather}

By means of the formula \eqref{refl}  for the ref\/lection
$r_\alpha$ we f\/ind that $r_{\alpha}$ acts upon orthogonal
coordinates of the vector $\lambda=\sum\limits_{i=1}^{n} m_i{\bf
e}_i$ by permuting $i$-th and $j$-th coordinates if $\alpha=\pm
({\bf e}_i - {\bf e}_j)$, and as the permutation of $i$-th and
$j$-th coordinates and the change of their signs if $\alpha=\pm
({\bf e}_i + {\bf e}_j)$. Thus, the Weyl group $W(D_n)$ consists
of all permutations of the orthogonal coordinates $m_1,m_2,\dots
,m_{n}$ of a point $\lambda$ with sign alternations of even number
of them.

Since an alternation of signs of two coordinates $x_i$ and $x_j$
is a product of two ref\/lections $r_\alpha$ with $\alpha=({\bf
e}_i + {\bf e}_j)$ and with $\alpha=({\bf e}_i - {\bf e}_j)$, a
sign of the determinant of this alternation is plus. Note that
$|W(D_n)|=\frac 12 |W(B_n)|$.

The signed orbit $O^\pm(\lambda)$,
$\lambda=(m_1,m_2,\dots,m_{n})$, $m_1> m_2> \cdots > m_{n}> 0$,
consists of all points
\[
(\pm m_{i_1}, \pm m_{i_2},\dots,\pm m_{i_{n+1}})^{\det w}
\]
obtained from $(m_1,m_2, \dots,m_{n})$ by permutations and
alternations of even number of signs which constitute an element
$w$ of the Weyl group $W(D_n)$. Moreover, $\det w$ is equal to
$\pm 1$ and a sign of $\det w$ is determined as follows. The
element $w\in W(D_n)$ can be represented as a product $w=\tau s$,
where $s$ is a permutation from $S_n$ and $\tau$ is an alternation
of even number of coordinates. Then $\det w=\det s$. Indeed, a
determinant of a transform, given by an element of $W$ which is an
alternation of two signs, is equal to $+1$ (since this element can
be represented as a product of two ref\/lections).

\subsection[Signed orbits of $A_3$]{Signed orbits of $\boldsymbol{A_3}$}\label{section3.7}
Signed orbits for $A_3$, $B_3$ and $C_3$ can be calculated by
using the orthogonal coordinates in the corresponding Euclidean
space, described above, and the description of the Weyl groups
$W(A_3)$, $W(B_3)$ and $W(C_3)$ in the orthogonal coordinate
systems. Below we give results of such calculations. Points
$\lambda$ of signed orbits are given in the $\omega$-coordinates
as $(a\,b\,c)$, where $\lambda=a\omega_1+b\omega_2+c\omega_3$.

The signed orbit $O^\pm(a\ b\ c)$, $a>0$, $b>0$, $c>0$, of $A_3$
contains the points
\begin{gather*}
O^\pm(a\ b\ c)\ni (a\ b\ c)^+, (a{+}b\ {-}b\ b{+}c)^-, (a{+}b\ c\
{-}b{-}c)^+,
   (a\ b{+}c\ {-}c)^-,
\\
\qquad\qquad (a{+}b{+}c\ {-}c\ {-}b)^-, (a{+}b{+}c\ {-}b{-}c\
b)^+, ({-}a\ a{+}b\ c)^-, ({-}a\ a{+}b{+}c\ {-}c)^+,
\\
\qquad\qquad (b\ {-}a{-}b\ a{+}b{+}c)^+, (b{+}c\ {-}a{-}b{-}c\
a{+}b)^-, ({-}a{-}b\ a\ b{+}c)^+, ({-}b\ {-}a\ a{+}b{+}c)^-
\end{gather*}
and the points, contragredient to these points, where the
contragredient of the point $(a'\ b'\ c')^+$ is $({-}c'\ {-}b'\
{-}a')^+$ and the contragredient of the point $(a'\ b'\ c')^-$ is
$({-}c'\ {-}b'\ {-}a')^-$.

\subsection[Signed orbits of $B_3$]{Signed orbits of $\boldsymbol{B_3}$}
\label{section3.8}
As in the previous case, points $\lambda$ of signed orbits are
given by the $\omega$-coordinates $(a\ b\ c)$, where
$\lambda=a\omega_1+b\omega_2+c\omega_3$. The signed orbit
$O^\pm(a\ b\ c)$,\ $a>0,\ b>0,\ c>0$, of $B_3$ contains the points
\begin{gather*}
O^\pm(a\ b\ c)\ni  (a\ b\ c)^+,  (a{+}b\ {-}b\ 2b{+}c)^-,
    ({-}a\ a{+}b\ c)^-, (b\ {-}a{-}b\ 2a{+}2b{+}c)^+,
\\
\qquad\qquad ({-}a{-}b\ a\ 2b{+}c)^+, ({-}b\ {-}a\ 2a{+}2b{+}c)^-,
(a\ b{+}c\ {-}c)^-,   (a{+}b{+}c\ {-}b{-}c\ 2b{+}c)^+,
\\
\qquad\qquad ({-}a\ a{+}b{+}c\ {-}c)^+,  (b{+}c\ {-}a{-}b{-}c\
2a{+}2b{+}c)^-,
   ({-}a{-}b{-}c\ a\ 2b{+}c)^-,
\\
\qquad\qquad ({-}b{-}c\ {-}a\ 2a{+}2b{+}c)^+, ({-}a{-}2b{-}c\ b\
c)^-, ({-}a{-}b{-}c\ {-}b\ 2b{+}c)^+,
\\
\qquad\qquad (a{+}2b{+}c\ {-}a{-}b{-}c\ c)^+, (b\ a{+}b{+}c\
{-}2a{-}2b{-}c)^-, (a{+}b{+}c\ {-}a{-}2b{-}c\ 2b{+}c)^-,
\\
\qquad\qquad ({-}b\ a{+}2b{+}c\ {-}2a{-}2b{-}c)^+, ({-}a{-}2b{-}c\
b{+}c\ {-}c)^+, ({-}a{-}b\ {-}b{-}c\ 2b{+}c)^-,
\\
\qquad\qquad (a{+}2b{+}c\ {-}a{-}b\ {-}c)^-,(b{+}c\ a{+}b\
{-}2a{-}2b{-}c)^+,
   (a{+}b\ {-}a{-}2b{-}c\ 2b{+}c)^+,
\\
\qquad\qquad ({-}b{-}c\ a{+}2b{+}c\ {-}2a{-}2b{-}c)^-
\end{gather*}
and also all these points taken with opposite signs of
coordinates, signs of these points are also opposite.

\subsection[Signed orbits of $C_3$]{Signed orbits of $\boldsymbol{C_3}$}\label{section3.9}
As in the previous cases, points $\lambda$ of signed orbits are
given by the $\omega$-coordinates $(a\ b\ c)$, where
$\lambda=a\omega_1+b\omega_2+c\omega_3$. The signed orbit
$O^\pm(a\ b\ c)$, $a>0$, $b>0$, $c>0$, of $C_3$ contains the
points
\begin{gather*}
O^\pm(a\ b\ c)\ni  (a\ b\ c)^+, (a{+}b\ {-}b\ b{+}c)^-,  ({-}a\
a{+}b\ c)^-, (b\ {-}a{-}b\ a{+}b{+}c)^+,
\\
\qquad\qquad ({-}a{-}b\ a\ b{+}c)^+, ({-}b\ {-}a\ a{+}b{+}c)^-,
(a\ b{+}2c\ {-}c)^-,
 (a{+}b{+}2c\ {-}b{-}2c\ b{+}c)^+,
\\
\qquad\qquad ({-}a\ a{+}b{+}2c\ {-}c)^+,  (b{+}2c\ {-}a{-}b{-}2c\
a{+}b{+}c)^-, ({-}a{-}b{-}2c\ a\ b{+}c)^-,
\\
\qquad\qquad ({-}b{-}2c\ {-}a\ a{+}b{+}c)^+, ({-}a{-}2b{-}2c\ b\
c)^-, ({-}a{-}b{-}2c\ {-}b\ b{+}c)^+,
\\
\qquad\qquad (a{+}2b{+}2c\ {-}a{-}b{-}2c\ c)^+, (b\ a{+}b{+}2c\
{-}a{-}b{-}c)^-, (a{+}b{+}2c\ {-}a{-}2b{-}2c\ b{+}c)^-,
\\
\qquad\qquad ({-}b\ a{+}2b{+}2c\ {-}a{-}b{-}c)^+,({-}a{-}2b{-}2c\
b{+}2c\ {-}c)^+, ({-}a{-}b\ {-}b{-}2c\ b{+}c)^-,
\\
\qquad\qquad (a{+}2b{+}2c\ {-}a{-}b\ {-}c)^-,(b{+}2c\ a{+}b\
{-}a{-}b{-}c)^+, (a{+}b\ {-}a{-}2b{-}2c\ b{+}c)^+,
\\
\qquad\qquad ({-}b{-}2c\ a{+}2b{+}2c\ {-}a{-}b{-}c)^-
\end{gather*}
and also all these points taken with opposite signs of
coordinates, signs of these points are also opposite.

\section{Antisymmetric orbit functions}\label{section4} 

\subsection[Definition]{Def\/inition}\label{section4.1}
The exponential functions $e^{2\pi{\rm i}\langle m,x\rangle}$,
$x\in E_n$, with f\/ixed $m=(m_1,m_2,\dots ,m_n)$ determine the
Fourier transform on $E_n$. Antisymmetric orbit functions are an
antisymmetrized (with respect to a Weyl group) version of
exponential functions. Correspondingly, they determine an
antisymmetrized version of the Fourier transform.

First we def\/ine symmetric orbit functions, studied in
\cite{KP06}. Let $W$ be a Weyl group of transformations of the
Euclidean space $E_n$. To each element $\lambda\in E_n$ from the
dominant Weyl chamber (that is, $\langle \lambda, \alpha_i\rangle
\ge 0$ for all simple roots $\alpha_i$) there corresponds
a {\it symmetric orbit function} $\phi_\lambda$ on~$E_n$, which is
given by the formula
 \begin{gather}\label{orb-s}
\phi_\lambda(x)=\sum_{\mu\in O(\lambda)} e^{2\pi{\rm i}\langle
\mu,x\rangle}, \qquad x\in E_n,
 \end{gather}
where $O(\lambda)$ is the $W$-orbit of the element $\lambda$. The
number of summands is equal to the size $|O(\lambda)|$ of the
orbit $O(\lambda)$ and we have $\phi_\lambda(0)=|O(\lambda)|$.

Sometimes (see, for example, \cite{Pat-Z-1} and \cite{Pat-Z-2}),
it is convenient to use a modif\/ied def\/inition of orbit
functions:
\begin{gather}\label{orb-2}
\hat \phi_\lambda(x)=|W_\lambda| \phi_\lambda (x),
\end{gather}
where $W_\lambda$ is a subgroup in $W$ whose elements leave
$\lambda$ f\/ixed. Then for all orbit functions $\hat\phi_\lambda$
we have $\hat \phi_\lambda (0)=|W|$.

Antisymmetric orbit functions are def\/ined (see \cite{PZ-06} and
\cite{P-SIG-05}) for dominant elements $\lambda$, which do not
belong to a wall of the dominant Weyl chamber (that is, for
strictly dominant elements~$\lambda$). The {\it antisymmetric
orbit function}, corresponding to such an element, is def\/ined as
 \begin{gather}\label{orb-a}
\varphi_\lambda(x)=\sum_{w\in W} (\det w) e^{2\pi{\rm i}\langle
w\lambda,x\rangle}, \qquad x\in E_n.
 \end{gather}
A number of summands in \eqref{orb-a} is equal to the size $|W|$
of the Weyl group $W$. We have $\varphi_\lambda(0)=0$.

Symmetric orbit functions $\phi_\lambda$ for which $\lambda\in
P_+$ and antisymmetric orbit functions $\varphi_\lambda(x)$ for
which $\lambda\in P^+_+$ are of special interest for
representation theory.
\medskip

\noindent {\bf Example.} {\it Antisymmetric orbit functions for
$A_1$}. In this case, there exists only one simple (positive) root
$\alpha$. We have $\langle \alpha,\alpha \rangle =2$. Then the
relation $2\langle \omega,\alpha \rangle / \langle \alpha,\alpha
\rangle =1$ means that $\langle \omega,\alpha \rangle =1$. This
means that $\omega=\alpha/2$ and $\langle \omega,\omega \rangle
=1/2$. Elements of $P^+_+$ coincide with $m\omega$, $m\in {\mathbb
Z}_+$. We identify points $x$ of $E_1\equiv {\mathbb R}$ with
$\theta \omega$. Since the Weyl group $W(A_1)$ consists of two
elements 1 and $r_\alpha$, and
\[
r_\alpha x=x-\frac{2\langle \theta\omega,\alpha \rangle}{\langle
\alpha,\alpha \rangle}\alpha=x-\theta\alpha =x-2x=-x ,
\]
antisymmetric orbit functions $\varphi_\lambda(x)$,
$\lambda=m\omega$, $m> 0$, are given by the formula
\[
\varphi_\lambda(x)=e^{2\pi{\rm i}\langle m\omega,\theta\omega
\rangle}-e^{-2\pi{\rm i}\langle m\omega,\theta\omega \rangle}
=e^{\pi{\rm i}m\theta}-e^{-\pi{\rm i}m\theta}=2{\rm i}\sin (\pi
m\theta).
\]
Note that for the symmetric orbit function $\phi_\lambda(x)$ we
have
$\phi_\lambda(x)=2\cos (\pi m\theta)$. 

\subsection[Antisymmetric orbit functions of $A_2$]{Antisymmetric orbit functions of $\boldsymbol{A_2}$}
\label{section4.2}
Antisymmetric orbit functions for the Coxeter--Dynkin diagrams of
rank 2 were given in \cite{PZ-06}. In this subsection we give
these functions for $A_2$.

Put $\lambda=a\omega_1+b\omega_2\equiv (a\ b)$ with $a>0$, $b>0$.
Then for $\varphi_\lambda(x)\equiv \varphi_{(a\ b)}(x)$ we have
from \eqref{orb-a} that
\begin{gather*}
 \varphi_{(a\ b)}(x)
   =    e^{2\pi i\l(a\ b), x\r}
       - e^{2\pi i\l(-a\ a+b), x\r}
       - e^{2\pi i\l(a+b\ -b), x\r} \notag\\
\phantom{\varphi_{(a\ b)}(x)=}{} + e^{2\pi i\l(b\ -a-b), x\r}
       + e^{2\pi i\l(-a-b\ a), x\r}
       - e^{2\pi i\l(-b\ -a), x\r} . \notag
\end{gather*}
Using the representation $x=\psi_1\alpha_1+\psi_2\alpha_2$, one
obtains
\begin{gather}
  \varphi_{(a\ b)}(x)
   =   e^{2\pi i(a\psi_1+b\psi_2)}
      - e^{2\pi i(-a\psi_1+(a+b)\psi_2)}
      - e^{2\pi i((a+b)\psi_1-b\psi_2)} \notag\\
 \phantom{\varphi_{(a\ b)}(x)=}{} + e^{2\pi i(b\psi_1-(a+b)\psi_2)}
      + e^{2\pi i((-a-b)\psi_1+a\psi_2)}
      - e^{2\pi i(-b\psi_1-a\psi_2)}.
\end{gather}
The actual expression for $ \varphi_{(a\ b)}(x)$ depends on a
choice of coordinate systems for $\lambda$ and $x$. Setting
$x=\theta_1\omega_1+\theta_2\omega_2$ and $\lambda$ as before, we
get
\begin{gather}
  \varphi_{(a\ b)}(x)
   =   e^{\tfrac{2\pi i}3((2a+b)\theta_1+(a+2b)\theta_2)}
      - e^{\tfrac{2\pi i}3((-a+b)\theta_1+(a+2b)\theta_2)}\notag\\
\phantom{ \varphi_{(a\ b)}(x)=}{}- e^{\tfrac{2\pi
i}3((2a+b)\theta_1+(a- b)\theta_2)}
      + e^{-\tfrac{2\pi i}3((a-b)\theta_1+(2a+b)\theta_2)}\\
\phantom{ \varphi_{(a\ b)}(x)=}{} + e^{-\tfrac{2\pi
i}3((a+2b)\theta_1+(-a+b)\theta_2)}
      - e^{-\tfrac{2\pi i}3((a+2b)\theta_1+(2a+b)\theta_2)}.\notag
\end{gather}
Note that $ \phi_{(a\ a)}(x)$ are pure imaginary for all $a>0$ and
\begin{gather}
 \varphi_{(a\ a)}(x) = 2{\rm i}\left\{\sin 2\pi a( \psi_1+ \psi_2)
                +\sin 2\pi a(\psi_1-2\psi_2)
                -\sin 2\pi a(2\psi_1- \psi_2) \right\}\notag\\
\phantom{\varphi_{(a\ a)}(x)}{} = 2{\rm i}\left\{\sin 2\pi
a(\theta_1+\theta_2)
                -\sin 2\pi a\theta_1-\sin 2\pi a\theta_2 \right\}.
\end{gather}
The pairs $ \varphi_{(a\ b)}(x)+ \varphi_{(b\ a)}(x)$ are always
pure imaginary functions.

\subsection[Antisymmetric orbit functions of $C_2$ and $G_2$]{Antisymmetric orbit functions of
 $\boldsymbol{C_2}$ and $\boldsymbol{G_2}$}\label{section4.3}
Putting again $\lambda=a\omega_1+b\omega_2=(a\,b)$,
$x=\theta_1\omega_1+\theta_2\omega_2$ and using the matrices $S$
from \eqref{matr}, which are of the form
\[
S(C_2)=\tfrac12\begin{pmatrix}1&1\\1&2\end{pmatrix},\qquad
S(G_2)=\tfrac16\begin{pmatrix}6&3\\3&2\end{pmatrix},
\]
we f\/ind the orbit functions for $C_2$ and $G_2$:
\begin{gather}
C_2 :\quad  \varphi_{(a\ b)}(x)  = 2\cos\pi
((a+b)\theta_1+(a+2b)\theta_2)
   -2\cos\pi (b\theta_1+(a+2b)\theta_2)\notag \\
\phantom{C_2 :\quad  \varphi_{(a\ b)}(x)  =}{} -2\cos\pi
((a+b)\theta_1+a\theta_2)
   +2\cos\pi(b\theta_1-a\theta_2), \\
 G_2 :\quad  \varphi_{(a\ b)}(x)   =2\cos\pi ((2a+b)\theta_1+(a+\tfrac23b)\theta_2)
        -2\cos\pi ((a+b)\theta_1+(a+\tfrac23 b)\theta_2)\notag\\
\phantom{G_2 :\quad  \varphi_{(a\ b)}(x)   =}{}
-2\cos\pi((2a+b)\theta_1+(a+\tfrac13 b)\theta_2)
+2\cos\pi((a+b)\theta_1+\tfrac13 b\theta_2)\notag\\
\phantom{G_2 :\quad  \varphi_{(a\ b)}(x)   =}{}
+2\cos\pi(a\theta_1+(a+\tfrac13b)\theta_2)
               -2\cos\pi(a\theta_1-\tfrac13b\theta_2).
\end{gather}
As we see, orbit functions for $C_2$ and $G_2$ are real.

\subsection[Antisymmetric orbit functions of $A_n$]{Antisymmetric orbit functions of $\boldsymbol{A_n}$}
\label{section4.4}
It is dif\/f\/icult to write down an explicit form of orbit
functions for $A_n$, $B_n$, $C_n$ and $D_n$ in coordinates with
respect to the $\omega$- or $\alpha$-bases. For this reason, for
these cases we use the orthogonal coordinate systems, described in
Section~\ref{section3}.

Let $\lambda=(m_1,m_2,\dots ,m_{n+1})$ be a strictly dominant
element for $A_n$ in orthogonal coordinates described in
Subsection~\ref{section3.3}. Then $m_1> m_2> \cdots > m_{n+1}$.
The Weyl group in this case coincides with the symmetric group
$S_{n+1}$. Then the signed orbit $O^\pm(\lambda)$ consists of
points $(w\lambda)^{\det w}$, $w\in W\equiv S_{n+1}$. Representing
points $x\in E_{n+1}$ in the orthogonal coordinate system,
$x=(x_1,x_2,\dots ,x_{n+1})$, and using formula \eqref{orb-a} we
f\/ind that
\begin{gather}
 \varphi_\lambda(x)  =
 \sum _{w\in S_{n+1}} (\det w) e^{2\pi{\rm i}\langle w(m_1,
 \dots ,m_{n+1}),(x_1,\dots,x_{n+1})\rangle} \notag \\
  \phantom{\varphi_\lambda(x)}{} =
 \sum _{w\in S_{n+1}} (\det w) e^{2\pi{\rm i}((w\lambda)_1x_1+
 \cdots +(w\lambda)_{n+1}x_{n+1})} ,\label{orb-f}
\end{gather}
where $((w\lambda)_{1},(w\lambda)_{2},\dots ,(w\lambda)_{n+1})$
are the coordinates of the point $w\lambda$.

Note that the element $-(m_{n+1},m_n,\dots ,m_1)$ is strictly
dominant if the element $(m_1,m_2,\dots$, $m_{n+1})$ is strictly
dominant. In the Weyl group $W(A_n)$ there exists an element $w_0$
such that
\[
w_0(m_{1},m_2,\dots ,m_{n+1})= (m_{n+1},m_n,\dots ,m_1).
\]
Moreover, we have
\begin{gather*}
\det w_0=1\qquad {\rm for}\qquad A_{4k-1}\qquad {\rm and}\qquad
A_{4k},
\\
\det w_0=-1\qquad {\rm for}\qquad A_{4k+1}\qquad {\rm and}\qquad
A_{4k+2}.
\end{gather*}
It follows from here that in the expressions for the orbit
functions $\varphi_{(m_{1},m_2,\dots ,m_{n+1})}(x)$ and
$\varphi_{-(m_{n+1},m_n,\cdots ,m_{1})}(x)$ there are summands
\begin{gather}\label{w_0}
e^{2\pi{\rm i}\langle w_0\lambda ,x\rangle}= e^{2\pi{\rm
i}(m_{n+1}x_1+
 \cdots +m_{1}x_{n+1})}\qquad
{\rm and}\qquad e^{-2\pi{\rm i}(m_{n+1}x_1+
 \cdots +m_{1}x_{n+1})} ,
\end{gather}
respectively, which are complex conjugate to each other. Moreover,
the f\/irst expression is contained with the sign $(\det w_0)$ in
$\varphi_{(m_{1},m_2,\dots ,m_{n+1})}(x)$, that is, the
expressions \eqref{w_0} are contained in
$\varphi_{(m_{1},m_2,\dots ,m_{n+1})}(x)$ and
$\varphi_{-(m_{n+1},m_n,\dots ,m_{1})}(x)$ with the same sign for
$n=4k-1, 4k$ and with opposite signs for $n=4k+1, 4k+2$, $k\in
\mathbb{Z}_+$.

 Similarly, in the expressions \eqref{orb-f} for the function
$\varphi_{(m_{1},m_2,\dots ,m_{n+1})}(x)$ and for the function
$\varphi_{-(m_{n+1},m_n,\dots ,m_{1})}(x)$ all other summands are
(up to a sign, which depends on a value of $n$) pairwise complex
conjugate. Therefore,
\begin{gather}\label{compl}
\varphi_{(m_{1},m_2,\dots
,m_{n+1})}(x)=\overline{\varphi_{-(m_{n+1},m_n,\dots ,m_{1})}(x)}
\end{gather}
for $n=4k-1,4k$ and
\begin{gather}\label{compl-n}
\varphi_{(m_{1},m_2,\dots
,m_{n+1})}(x)=-\overline{\varphi_{-(m_{n+1},m_n,\dots ,m_{1})}(x)}
\end{gather}
for $n=4k+1,4k+2$.

If we use for $\lambda$ the coordinates $\lambda_i=\langle
\lambda,\alpha^\vee_i\rangle$ in the $\omega$-basis instead of the
orthogonal coordinates~$m_j$, then these equations can be written
as
\[
\varphi_{(\lambda_{1},\dots
,\lambda_{n})}(x)=\overline{\varphi_{(\lambda_{n},\dots
,\lambda_{1})}(x)}, \qquad \varphi_{(\lambda_{1},\dots
,\lambda_{n})}(x)=-\overline{\varphi_{(\lambda_{n},\dots
,\lambda_{1})}(x)},
\]
respectively. According to \eqref{compl} and \eqref{compl-n}, if
\begin{gather}\label{m-A}
(m_1,m_2,\dots ,m_{n+1})=-(m_{n+1},m_n,\dots ,m_{1})
\end{gather}
(that is, the element $\lambda$ has in the $\omega$-basis the
coordinates $(\lambda_1,\lambda_2,\dots ,\lambda_2,\lambda_1)$),
then {\it the orbit function $\varphi_\lambda$ is real for
$n=4k-1,4k$ and pure imaginary for $n=4k+1,4k+2$.} In the f\/irst
case the antisymmetric orbit function can be represented as a sum
of cosines of angles  and in the second case as a sum of sines of
angles multiplied by ${\rm i}=\sqrt{-1}$.

\begin{proposition}\label{prop2}
In the orthogonal coordinates, antisymmetric orbit functions of
$A_n$ can be represented as determinants of certain matrices:
\begin{gather}
\varphi_{(m_{1},m_2,\dots ,m_{n+1})}(x)= \det \left( e^{2\pi{\rm
i}m_ix_j}\right)_{i,j=1}^{n+1}
\notag\\
\phantom{\varphi_{(m_{1},m_2,\dots ,m_{n+1})}(x)}{} \equiv  \det
\left(
 \begin{array}{cccc}
 e^{2\pi{\rm i}m_1x_1}& e^{2\pi{\rm i}m_1x_2}&\cdots & e^{2\pi{\rm
i}m_1x_{n+1}}\\
  e^{2\pi{\rm i}m_2x_1}& e^{2\pi{\rm i}m_2x_2}&\cdots & e^{2\pi{\rm
i}m_2x_{n+1}}\\
  \cdots & \cdots &
\cdots & \cdots \\\
 e^{2\pi{\rm i}m_{n+1}x_1}& e^{2\pi{\rm i}m_{n+1}x_2}&\cdots & e^{2\pi{\rm
i}m_{n+1}x_{n+1}} \end{array} \right) .\label{det-A}
 \end{gather}
 \end{proposition}

\begin{proof} A proof of this formula follows from the fact that this expression
for $\varphi_{(m_{1},m_2,\dots ,m_{n+1})}(x)$ coincides with the
expression given by the formula \eqref{orb-f}; see \cite{Weyl}.
\end{proof}

Taking into account the form of the half-sum of positive roots
$\rho$ for $A_n$, we can write down the orbit function
$\varphi_\rho(x)$, corresponding to the weight $\rho=\frac 12
\sum\limits_{\alpha>0} \alpha$, in the form of the Vandermonde
determinant,
 \begin{gather} \label{a-pho-A}
\varphi_\rho(x)=\det \left( e^{2\pi{\rm i}ix_j}
\right)_{i,j=1}^{n+1}
 =\prod_{k<l} (e^{2\pi{\rm i}x_k}-e^{-2\pi{\rm i}x_l}).
 \end{gather}
The last equality follows from the expression for the Vandermonde
determinant.

\subsection[Antisymmetric orbit functions of $B_n$]{Antisymmetric orbit functions of $\boldsymbol{B_n}$}
\label{section4.5}
Let $\lambda=(m_1,m_2,\dots ,m_{n})$ be a strictly dominant
element for $B_n$ in orthogonal coordinates described in
Subsection~\ref{section3.4}. Then $m_1> m_2> \cdots > m_{n}> 0$.
The Weyl group $W(B_n)$ consists of permutations of the
coordinates $m_i$ with sign alternations of some of them.
Representing points $x\in E_{n}$ also in the orthogonal coordinate
system, $x=(x_1,x_2,\dots ,x_{n})$, and using
formula~\eqref{orb-a} we f\/ind that
\begin{gather}
 \varphi_\lambda(x)  =
\sum _{\varepsilon_i=\pm 1}\sum _{w\in S_{n}} (\det
w)\varepsilon_1 \varepsilon_2\dots \varepsilon_n e^{2\pi{\rm
i}\langle w(\varepsilon_1m_1, \dots ,\varepsilon_nm_{n}),
(x_1,\dots,x_{n})\rangle} \notag \\
\phantom{\varphi_\lambda(x)}{} =\sum _{\varepsilon_i=\pm 1}\sum
_{w\in S_{n}} (\det w)\varepsilon_1 \varepsilon_2\cdots
\varepsilon_n e^{2\pi{\rm i}((w(\varepsilon \lambda))_1x_1+
 \cdots +(w(\varepsilon\lambda))_{n}x_{n})} ,\label{orb-B}
\end{gather}
where $(w(\varepsilon \lambda))_1,
 \dots ,(w(\varepsilon\lambda))_{n}$ are the orthogonal
 coordinates of the
 points $w(\varepsilon \lambda)$ if
$\varepsilon \lambda=(\varepsilon_1m_1$, $\dots
,\varepsilon_nm_{n})$.

Since in $W(B_n)$ there exists an element which changes signs of
all coordinates $m_i$, then for each summand $e^{2\pi{\rm
i}((w(\varepsilon \lambda))_1x_1+  \cdots +(w(\varepsilon
\lambda))_{n}x_{n})}$ in the expressions \eqref{orb-B} for the
antisymmetric orbit function $\varphi_{(m_{1},m_2,\dots
,m_{n})}(x)$ there exists exactly one summand complex conjugate to
it, that is, the summand $e^{-2\pi{\rm i}(((w(\varepsilon
\lambda))_1x_1+  \cdots +(w(\varepsilon \lambda))_{n}x_{n})}$.
This summand is with sign $(-1)^n=(\det w')$, where $w'$ changes
signs of all coordinates. Therefore, $(\det w')=1$ if $n=2k$ and
$(\det w')=-1$ if $n=2k+1$. This means that {\it antisymmetric
orbit functions of $B_n$ are real if $n=2k$ and pure imaginary if
$n=2k+1$.} Each antisymmetric orbit function of $B_n$ can be
represented as a sum of cosines of the corresponding angles if
$n=2k$ and as a sum of sines, multiplied by ${\rm i}=\sqrt{-1}$,
if $n=2k+1$. The following proposition is true \cite{Weyl}:

\begin{proposition}\label{prop3}
Antisymmetric orbit functions of $B_n$ can be represented in the
form
 \begin{gather}\label{det-B}
\varphi_{(m_{1},m_2,\dots ,m_{n})}(x)= \det \left( e^{2\pi{\rm
i}m_ix_j}-e^{-2\pi{\rm i}m_ix_j}   \right)_{i,j=1}^{n}= (2{\rm
i})^n \det \left(
 \sin 2\pi m_ix_j  \right)_{i,j=1}^{n}.
 \end{gather}
\end{proposition}

\begin{proof} Let us take on the right hand side of \eqref{orb-B}
the sum of terms with f\/ixed $w\in S_n$. It can be written down
as
\begin{gather*}
(\det w)\sum _{\varepsilon_i=\pm 1} \varepsilon_1
\varepsilon_2\cdots \varepsilon_n e^{2\pi{\rm i}((w(\varepsilon
\lambda))_1x_1+
 \cdots +(w(\varepsilon\lambda))_{n}x_{n})}
\\
\qquad  {}=(\det w) (e^{2\pi{\rm i}(w\lambda)_1x_1}-e^{-2\pi{\rm
i}(w\lambda)_1x_1}) \cdots (e^{2\pi{\rm
i}(w\lambda)_nx_n}-e^{-2\pi{\rm i}(w\lambda)_nx_n})
\\
\qquad {}=(\det w)(2{\rm i})^n \sin 2\pi (w\lambda)_1x_1\cdot
{\dots} \cdot \sin 2\pi (w\lambda)_nx_n.
\end{gather*}
Then for $\varphi_\lambda(x)$ we have
\begin{gather*}
 \varphi_\lambda(x)= (2{\rm i})^n\sum_{w\in S_n} (\det w)
\sin 2\pi (w\lambda)_1x_1\cdot {\dots} \cdot \sin 2\pi
(w\lambda)_nx_n
\\
\phantom{\varphi_\lambda(x)}{}= (2{\rm i})^n \det \left(
 \sin 2\pi m_ix_j  \right)_{i,j=1}^{n},
\end{gather*}
where $\lambda=(m_1,m_2,\dots ,m_{n})$. Proposition is proved.
\end{proof}

For the antisymmetric orbit function $\varphi_\rho$, corresponding
to the half-sum $\rho$ of positive roots of~$B_n$, one has
\[
\varphi_\rho(x)=(2{\rm i})^n \det \left(  \sin 2\pi\rho_ix_j
\right)_{i,j=1}^{n},
\]
where $\rho=(\rho_1,\rho_2,\dots,\rho_n)=
(n-\frac12,n-\frac32,\dots, \frac12)$.

\subsection[Antisymmetric orbit functions of $C_n$]{Antisymmetric orbit functions of $\boldsymbol{C_n}$}
\label{section4.6}

Let $\lambda=(m_1,m_2,\dots ,m_{n})$ be a strictly dominant
element for $C_n$ in the orthogonal coordinates described in
Subsection~\ref{section3.5}. Then $m_1> m_2> \cdots > m_{n}> 0$.
The Weyl group $W(C_n)$ consists of permutations of the
coordinates with sign alternations of some of them. Representing
points $x\in E_{n}$ also in the orthogonal coordinate system,
$x=(x_1,x_2,\dots ,x_{n})$, we f\/ind that
\begin{gather}
 \varphi_\lambda(x)  =
\sum _{\varepsilon_i=\pm 1}\sum _{w\in S_{n}} (\det w)
\varepsilon_1\varepsilon_2\dots \varepsilon_n e^{2\pi{\rm
i}\langle w(\varepsilon_1m_1, \dots ,\varepsilon_nm_{n}),
(x_1,\cdots,x_{n})\rangle} \notag \\
\phantom{\varphi_\lambda(x)}{} =\sum _{\varepsilon_i=\pm 1}\sum
_{w\in S_{n}} (\det w) \varepsilon_1\varepsilon_2\cdots
\varepsilon_n
 e^{2\pi{\rm i}((w(\varepsilon \lambda))_1x_1+
 \cdots +(w(\varepsilon\lambda))_{n}x_{n})} ,\label{orb-C}
\end{gather}
where, as above, $(w(\varepsilon \lambda))_1, \dots ,
(w(\varepsilon\lambda))_{n}$ are the orthogonal coordinates of the
points $w(\varepsilon \lambda)$ if $\varepsilon
\lambda=(\varepsilon_1m_1$, $\dots ,\varepsilon_nm_{n})$.

As in the case of $B_n$, in the expressions \eqref{orb-C} for the
functions $\varphi_{(m_{1},m_2,\dots ,m_{n})}(x)$ for each summand
$e^{2\pi{\rm i}((w(\varepsilon \lambda))_1x_1+  \cdots
+(w(\varepsilon \lambda))_{n}x_{n})}$ there exists exactly one
summand complex conjugate to it, that is, the summand
$e^{-2\pi{\rm i}((w(\varepsilon \lambda))_1x_1+  \cdots
+(w(\varepsilon \lambda))_{n}x_{n})}$. Moreover, this summand is
with sign ``+'' if $n=2k$ and with sign ``$-$'' if $n=2k+1$.
Therefore, {\it antisymmetric orbit functions of $C_n$ are real if
$n=2k$ and pure imaginary if $n=2k+1$.}

Note that in the orthogonal coordinates the antisymmetric orbit
functions $\varphi_{(m_{1},m_2,\dots ,m_{n})}(x)$ of $C_n$
coincides with the antisymmetric orbit functions
$\varphi_{(m_{1},m_2,\dots ,m_{n})}(x)$ of $B_n$, that is,
antisymmetric orbit functions \eqref{orb-B} and \eqref{orb-C}
coincide. However, $\alpha$-coordinates of the element
$(m_1,m_2,\dots,m_n)$ for $C_n$ do not coincide with
$\alpha$-coordinates of the element $(m_1,m_2,\dots,m_n)$
for~$B_n$, that is, in $\alpha$-coordinates the corresponding
antisymmetric orbit functions of $B_n$ and $C_n$ are dif\/ferent.

\begin{proposition}\label{prop4}
Antisymmetric orbit functions of $C_n$ can be represented in the
form
 \begin{gather}\label{det-C}
\varphi_{(m_{1},m_2,\dots ,m_{n})}(x)=\det \left( e^{2\pi{\rm
i}m_ix_j}-e^{-2\pi{\rm i}m_ix_j}   \right)_{i,j=1}^{n} = (2{\rm
i})^n \det \left(
 \sin 2\pi m_ix_j  \right)_{i,j=1}^{n}.
 \end{gather}
\end{proposition}

\begin{proof} This proposition follows from Proposition~\ref{prop3} if we take into account
that antisymmetric orbit functions of~$C_n$ and of $B_n$ coincide
in the orthogonal coordinate systems.
\end{proof}

For the antisymmetric orbit function $\varphi_\rho$, corresponding
to the half-sum of positive roots of~$C_n$, one has
\[
\varphi_\rho(x)=(2{\rm i})^n \det \left(  \sin 2\pi\rho_ix_j
\right)_{i,j=1}^{n},
\]
where $\rho=(\rho_1,\rho_2,\dots,\rho_n)= (n,n-1,\dots,1)$.

\subsection[Antisymmetric orbit functions of $D_n$]{Antisymmetric orbit functions of $\boldsymbol{D_n}$}
\label{section4.7}
Let $\lambda=(m_1,m_2,\dots ,m_{n})$ be a strictly dominant
element for $D_n$ in the orthogonal coordinates described in
Subsection~\ref{section3.6}. Then $m_1> m_2> \cdots > m_{n-1}>
|m_n|$. The Weyl group $W(D_n)$ consists of permutations of the
coordinates with sign alternations of even number of them.
Representing points $x\in E_{n}$ also in the orthogonal coordinate
system, $x=(x_1,x_2,\dots ,x_{n})$, and using formula
\eqref{orb-a} we f\/ind that
\begin{gather}
 \varphi_\lambda(x)  =
{\sum_{\varepsilon_i=\pm 1}}' \sum _{w\in S_{n}} (\det w)
  e^{2\pi{\rm i}\langle
w(\varepsilon_1m_1, \dots ,\varepsilon_nm_{n}),
(x_1,\dots,x_{n})\rangle} \notag \\
\phantom{\varphi_\lambda(x)}{} ={\sum_{\varepsilon_i =\pm 1}}'\;
\sum _{w\in S_{n}}  (\det w)   e^{2\pi{\rm
i}((w(\varepsilon\lambda))_1x_1+
 \cdots +(w(\varepsilon\lambda))_nx_{n})} ,\label{orb-D}
\end{gather}
where $(w(\varepsilon \lambda))_1, \dots ,
(w(\varepsilon\lambda))_{n}$ are the orthogonal coordinates of the
points $w(\varepsilon \lambda)$ and the prime at the sum sign
means that the summation is over values of $\varepsilon_i$ with
even number of sign minus. We have taken into account that an
alternation of coordinates without any permutation does not change
a determinant.

Let $m_n\ne 0$. Then in the expressions \eqref{orb-D} for the
orbit function $\varphi_{(m_{1},m_2,\dots ,m_{n})}(x)$ of
$D_{n=2k}$ for each summand $ e^{2\pi{\rm
i}((w(\varepsilon\lambda))_1x_1+
 \cdots +(w(\varepsilon\lambda))_nx_{n})}$
there exists exactly one summand (with the same sign) complex
conjugate to it. This means that these {\it antisymmetric orbit
functions of $D_{2k}$ are real.} Each orbit function of $D_{2k}$
can be represented as a sum of cosines of the corresponding
angles.

It is also proved by using the formula \eqref{orb-D} that for
$m_n\ne 0$ {\it the antisymmetric orbit functions
$\varphi_{(m_{1},\dots, m_{2k},m_{2k+1})}(x)$ and
$\varphi_{(m_{1},\dots ,m_{2k},-m_{2k+1})}(x)$ of $D_{2k+1}$ are
complex conjugate.}

If $m_n=0$, then it follows from \eqref{orb-D} that antisymmetric
orbit functions of $D_n$ are real and can be represented as a sum
of cosines of certain angles.

Explicit forms of antisymmetric orbit functions of $D_n$ are
described by the following proposition \cite{Weyl}:

\begin{proposition}\label{prop5}
Antisymmetric orbit functions of $D_n$ are representable in the
form
\begin{gather}
\varphi_{(m_{1},m_2,\dots ,m_{n})}(x)= \tfrac12 \det \left(
e^{2\pi{\rm i}m_ix_j}{-}e^{-2\pi{\rm i}m_ix_j}
\right)_{i,j=1}^{n}{+} \tfrac12 \det \left( e^{2\pi{\rm
i}m_ix_j}{+}e^{-2\pi{\rm i}m_ix_j} \right)_{i,j=1}^{n}
\notag \\
 \phantom{\varphi_{(m_{1},m_2,\dots ,m_{n})}(x)}{}
 = \tfrac12 (2{\rm i})^n \det \left(  \sin 2\pi m_ix_j  \right)_{i,j=1}^{n}
+2^{n-1} \det \left(  \cos 2\pi m_ix_j
\right)_{i,j=1}^{n}\label{det-D-1}
 \end{gather}
if $m_n\ne 0$, and in the form
 \begin{gather}
\varphi_{(m_{1},m_2,\dots ,m_{n})}(x)= \tfrac12  \det \left(
e^{2\pi{\rm i}m_ix_j}+e^{-2\pi{\rm i}m_ix_j} \right)_{i,j=1}^{n} =
2^{n-1} \det \left(  \cos 2\pi m_ix_j
\right)_{i,j=1}^{n}\label{det-D-2}
 \end{gather}
if $m_n=0$.
\end{proposition}

 \begin{proof} Let $m_n\ne 0$. We take on the right hand side of \eqref{orb-D}
a sum of terms with f\/ixed $w\in S_n$. It can be written as
\[
I_w\equiv  (\det w){\sum_{\varepsilon_i=\pm 1}}' \varepsilon_1
\varepsilon_2\cdots \varepsilon_n e^{2\pi{\rm i}((w(\varepsilon
\lambda))_1x_1+
 \cdots +(w(\varepsilon\lambda))_{n}x_{n})},
\]
where $\varepsilon_1 \varepsilon_2\cdots \varepsilon_n =1$ (since
there is an even number of $\varepsilon_i$ with
$\varepsilon_i=-1$). A value of $I_w$ does not change if we add
and subtract the same term to it:
\begin{gather*}
I_w = (\det w){\sum _{\varepsilon_i=\pm 1}}' \varepsilon_1
\varepsilon_2\cdots \varepsilon_n e^{2\pi{\rm i}((w(\varepsilon
\lambda))_1x_1+
 \cdots +(w(\varepsilon\lambda))_{n}x_{n})}  \\
\phantom{I_w=}{}  +\tfrac 12 (\det w){\sum _{\varepsilon_i=\pm
1}}'' \varepsilon_1 \varepsilon_2\cdots \varepsilon_n e^{2\pi{\rm
i}((w(\varepsilon \lambda))_1x_1+
 \cdots +(w(\varepsilon\lambda))_{n}x_{n})}  \\
\phantom{I_w=}{}  -\tfrac 12 (\det w){\sum _{\varepsilon_i=\pm
1}}'' \varepsilon_1 \varepsilon_2\cdots \varepsilon_n e^{2\pi{\rm
i}((w(\varepsilon \lambda))_1x_1+
 \cdots +(w(\varepsilon\lambda))_{n}x_{n})} ,
\end{gather*}
the sum with two primes here means that the summation is over
values of $\varepsilon_i$ with an odd number of negative
$\varepsilon_i$. We split down the right hand side into two parts,
\begin{gather*}
I_w = \left[ \tfrac 12 (\det w){\sum _{\varepsilon_i=\pm 1}}'
\!\cdots -
 \tfrac 12 (\det w){\sum _{\varepsilon_i=\pm 1}}'' \!\cdots \right]\! +
 \left[ \tfrac 12 (\det w){\sum _{\varepsilon_i=\pm 1}}'\! \cdots +
 \tfrac 12 (\det w){\sum _{\varepsilon_i=\pm 1}}'' \!\cdots \right]
\end{gather*}
and repeat the reasoning of the proof of Proposition~\ref{prop3}.
As a result, we have
\begin{gather*}
I_w  = \tfrac 12 (\det w) (e^{2\pi{\rm
i}(w\lambda)_1x_1}-e^{-2\pi{\rm i}(w\lambda)_1x_1}) \cdots
(e^{2\pi{\rm i}(w\lambda)_nx_n}-e^{-2\pi{\rm i}(w\lambda)_nx_n}) \\
\phantom{I_w  =}{}  +\tfrac 12 (\det w) (e^{2\pi{\rm
i}(w\lambda)_1x_1}+e^{-2\pi{\rm i}(w\lambda)_1x_1}) \cdots
(e^{2\pi{\rm i}(w\lambda)_nx_n}+e^{-2\pi{\rm i}(w\lambda)_nx_n}) \\
\phantom{I_w }{} =  \tfrac 12 (\det w)(2{\rm i})^n \sin 2\pi
(w\lambda)_1x_1\cdot \,{\cdots}\, \cdot
\sin 2\pi (w\lambda)_nx_n \\
\phantom{I_w =}{} +\tfrac 12  (\det w) 2^n \cos 2\pi
(w\lambda)_1x_1\cdot \,{\cdots}\, \cdot \cos 2\pi (w\lambda)_nx_n
.
\end{gather*}
Then for $\varphi_\lambda(x)$ we have
\begin{gather*}
 \varphi_\lambda(x)= \tfrac12 (2{\rm i})^n\sum_{w\in S_n} (\det w)
\sin 2\pi (w\lambda)_1x_1\cdot \,{\cdots}\, \cdot \sin 2\pi
(w\lambda)_nx_n
\\
\phantom{\varphi_\lambda(x)=}{} + \tfrac12 2^n \sum_{w\in S_n}
(\det w) \cos 2\pi (w\lambda)_1x_1\cdot \,{\cdots}\, \cdot \cos
2\pi (w\lambda)_nx_n
\\
\phantom{\varphi_\lambda(x)}{}= \tfrac12 (2{\rm i})^n \det \left(
 \sin 2\pi m_ix_j  \right)_{i,j=1}^{n}+2^{n-1} \det \left(
 \cos 2\pi m_ix_j  \right)_{i,j=1}^{n} ,
\end{gather*}
where $\lambda=(m_1,m_2,\dots ,m_{n})$. Thus, the proposition is
proved for the case $m_n\ne 0$.

Let $m_n=0$. Then, in the expression \eqref{orb-D} for
$\varphi_\lambda(x)$, in each term of the sum there exists the
multiplier $e^{2\pi{\rm i} \varepsilon_n m_n x_i}$ with some $i$.
Since $m_n=0$ we have $e^{2\pi{\rm i} \varepsilon_n m_n x_i}=1$
for $\varepsilon_n=1$ and for $\varepsilon_n=-1$. The case
$\varepsilon_n=1$ gives an even number of negative $\varepsilon_i$
in the set $(\varepsilon_1,\varepsilon_2,\dots,\varepsilon_{n-1})$
and the case $\varepsilon_n=-1$ gives an odd number of negative
$\varepsilon_i$ in this set. Therefore, we may throw out the sum
over $\varepsilon_n$ and consider that summation in \eqref{orb-D}
runs over $\varepsilon_i=\pm 1$, $i=1,2,\dots n-1$ (with even or
odd number of negative $\varepsilon_i$):
\begin{gather}\label{orb-D-o}
 \varphi_\lambda(x)  =
\sum_{\varepsilon_i =\pm 1}\; \sum _{w\in S_{n}}  (\det w)
e^{2\pi{\rm i}((w(\varepsilon\lambda))_1x_1+
 \cdots +(w(\varepsilon\lambda))_nx_{n})} ,
\end{gather}
where in the sum the multipliers, containing $m_n$, are removed
and the f\/irst sum does not contain summation over
$\varepsilon_n$. As in the case of the proof of
Proposition~\ref{prop3}, we take in \eqref{orb-D-o} terms with
f\/ixed $w$ and write down it as
\begin{gather*}
    (\det w) \sum_{\varepsilon_i =\pm 1}
e^{2\pi{\rm i}((w(\varepsilon\lambda))_1x_1+
 \cdots +(w(\varepsilon\lambda))_nx_{n})} \\
\qquad{}  = \tfrac12 (\det w) (e^{2\pi{\rm
i}(w\lambda)_1x_1}+e^{-2\pi{\rm i} (w\lambda)_1x_1}) \cdots
(e^{2\pi{\rm i}(w\lambda)_nx_n}+e^{-2\pi{\rm i}(w\lambda)_nx_n}) \\
 \qquad{} =  2^{n-1} (\det w) \cos 2\pi (w\lambda)_1x_1\cdot\, {\cdots} \,\cdot
\cos 2\pi (w\lambda)_nx_n ,
\end{gather*}
where, as before, the multipliers, containing $m_n$, are omitted.
Since $e^{2\pi{\rm i}m_nx_i}+e^{-2\pi{\rm i}m_nx_i}=2,$ this leads
to the formula \eqref{det-D-2} for $\varphi_\lambda(x)$ when
$m_n=0$. The proposition is proved.  \end{proof}

For the antisymmetric orbit function $\varphi_\rho$, $\rho=\frac
12 \sum\limits_{\alpha>0} \alpha$, one has
\[
\varphi_\rho(x)=2^{n-1}\det \left( \cos 2\pi\rho_ix_j
\right)_{i,j=1}^{n},
\]
where $\rho=(\rho_1,\rho_2,\dots,\rho_n)= (n-1,n-2,\dots,1,0)$.

\subsection[Symmetric orbit functions of $B_n$, $C_n$ and $D_n$]{Symmetric orbit functions
of $\boldsymbol{B_n}$, $\boldsymbol{C_n}$ and
$\boldsymbol{D_n}$}\label{section4.8}
In Subsections~\ref{section4.5}--\ref{section4.7} we have derived
expressions for antisymmetric orbit functions $\varphi_\lambda
(x)$ of~$B_n$, $C_n$ and $D_n$ in orthogonal coordinates as
determinants of sine and cosine functions. Similar expressions can
be derived for symmetric orbit functions of $B_n$, $C_n$ and
$D_n$. Since in the def\/ining expressions for symmetric orbit
functions $\phi_\lambda(x)$ (see formula \eqref{orb-s}) there are
no multipliers $(\det w)$, then $\phi_\lambda(x)$ is expressed in
terms of products of sine and cosine functions (instead of
determinants).

As before, we express elements $\lambda=(m_1,m_2,\dots,m_n)$ and
$x=(x_1,x_2,\dots,x_n)$ in the corresponding orthogonal coordinate
systems. In Propositions \ref{prop6} and \ref{prop7} below we
suppose that $\lambda$ is an element of the dominant Weyl chamber
and is not obligatory integral.

\begin{proposition}\label{prop6}
Symmetric orbit functions of $B_n$ and $C_n$ can be represented in
the form
 \begin{gather}\label{det-B-symm}
\phi_{(m_{1},m_2,\dots ,m_{n})}(x)=
2^n \sum_{w\in S_n}
\cos 2\pi m_{w(1)}x_1 \cdot \,{\cdots}\,\cdot \cos 2\pi
m_{w(n)}x_n,
 \end{gather}
where $w(1),w(2),\dots,w(n)$ is the set of numbers $1,2,\dots,n$
obtained by acting by the permutation $w\in S_n$.
\end{proposition}

This proposition is proved in the same way as
Proposition~\ref{prop3} and we omit it.

\begin{proposition}\label{prop7}
Symmetric orbit functions of $D_n$ can be represented in the form
\begin{gather}
\phi_{(m_{1},m_2,\dots ,m_{n})}(x)=
2^{n-1} \sum_{w\in S_n}
\cos 2\pi m_{w(1)}x_1 \cdot\, {\cdots}\,\cdot \cos 2\pi m_{w(n)}x_n\notag\\
\phantom{\phi_{(m_{1},m_2,\dots ,m_{n})}(x)=}{}
 +\tfrac12 (2{\rm i})^n \sum_{w\in S_n} \sin 2\pi m_{w(1)}x_1 \cdot\, {\cdots}\,\cdot
\sin 2\pi m_{w(n)}x_n\label{det-D-symm}
\end{gather}
if $m_n\ne 0$ and in the form
 \begin{gather}\label{det-D-2-sym}
\varphi_{(m_{1},m_2,\dots ,m_{n})}(x)=
 2^{n-1} \sum_{w\in S_n}
 \cos 2\pi m_{w(1)}x_1 \cdot \,{\cdots}\,\cdot
\cos 2\pi m_{w(n)}x_n
 \end{gather}
if $m_n=0$, where $w(1),w(2),\dots,w(n)$ is the set of numbers
$1,2,\dots,n$ obtained by acting by the permutation $s\in S_n$.
\end{proposition}

A proof is similar to the proof of Proposition~\ref{prop5}.

\section{Properties of antisymmetric orbit functions}\label{section5}

\subsection{Anti-invariance with respect to the Weyl group}\label{section5.1}
Since the scalar product $\langle\cdot ,\cdot \rangle$ in $E_n$ is
invariant with respect to the Weyl group $W$, that is,
\[
\langle wx,wy\rangle =\langle x,y\rangle, \qquad
 w\in W,\quad x,y\in E_n,
\]
antisymmetric orbit functions $\varphi_\lambda$ for strictly
positive elements $\lambda$ (not obligatory belonging to~$P^+_+$)
are anti-invariant with respect to $W$, that is,
\[
\varphi_\lambda (w'x)=(\det w')\varphi_\lambda (x), \qquad w'\in
W.
\]
Indeed,
\begin{gather*}
\varphi_\lambda (w'x)= \sum_{w\in W} (\det w) e^{2\pi{\rm
i}\langle w\lambda,w'x\rangle} = \sum_{w\in W}(\det w) e^{2\pi{\rm
i}\langle {w'}^{-1}w\lambda,x\rangle}
 \notag\\
\phantom{\varphi_\lambda (w'x)}{}=  (\det w')\sum_{w\in W} (\det
w) e^{2\pi{\rm i}\langle w\lambda,x\rangle} =(\det w')
\varphi_\lambda (x) \notag
\end{gather*}
since ${w'}^{-1}w$ runs over the whole group $W$ when $w$ runs
over~$W$.

\subsection[Anti-invariance with respect to the affine Weyl group]{Anti-invariance
with respect to the af\/f\/ine Weyl group}\label{section5.2}

The af\/f\/ine Weyl group $W^{\rm aff}$ is generated by
ref\/lections $r_0, r_{\alpha_1},\dots ,r_{\alpha_n}$ (see
Subsection~\ref{section2.5}). We say that an antisymmetric orbit
function $\varphi_\lambda$, $\lambda\in P^+_+$, is anti-invariant
with respect to the af\/f\/ine Weyl group $W^{\rm aff}$ if
$\varphi_\lambda (r_0x)=- \varphi_\lambda (x)$. Let us show that
$\varphi_\lambda (x)$ satisf\/ies this relation. Since $r_0x=r_\xi
x+\xi^\vee$, where $\xi$ is the highest root (see
Subsection~\ref{section2.4}), for $\mu\in P$ we have
\[
\langle\mu,r_0x\rangle=\langle \mu ,\xi^\vee + r_{\xi}x\rangle
=\frac{2\langle \mu ,\xi\rangle}{\langle \xi,\xi \rangle} +\langle
\mu, r_{\xi}x\rangle = {\rm integer}+\langle r_{\xi}\mu ,x\rangle
\]
since $r^2_\xi=1$. Hence,
\begin{gather*}
\varphi_\lambda(r_0x)= \sum_{w\in W} (\det w) e^{2\pi{\rm
i}\langle w\lambda,r_0x\rangle}=\sum_{w\in W} (\det w) e^{2\pi{\rm
i}\langle r_{\xi}w\lambda,x\rangle}
\notag\\
\phantom{\varphi_\lambda(r_0x)}{}= (\det r_\xi)\sum_{w\in W}(\det
w) e^{2\pi{\rm i}\langle w\lambda,x\rangle}=-\varphi_\lambda(x)
\notag
\end{gather*}
since $r_{\xi}$ is a ref\/lection belonging to $W$.

If $\lambda\ne P$, then $\varphi_\lambda$ is not anti-invariant
with respect to $r_0$. It is anti-invariant only under action by
elements of the Weyl group $W$.

Due to the anti-invariance of antisymmetric orbit functions
$\varphi_\lambda$, $\lambda\in P^+_+$, with respect to the group
$W^{\rm aff}$, it is enough to consider them only on the
fundamental domain $F\equiv F(W^{\rm aff})$ of~$W^{\rm aff}$.
Values of $\varphi_\lambda$ on other points of $E_n$ are
determined by using the action of $W^{\rm aff}$ on $F$ or taking a
limit. In particular, {\it functions $\varphi_\lambda$,
$\lambda\in P^+_+$, are anti-invariant under a reflection with
respect to any $(n-1)$-dimensional wall of the fundamental domain
$F$.}

\subsection{Continuity and vanishing}\label{section5.3}
An antisymmetric orbit function $\varphi_\lambda$ is a f\/inite
sum of exponential functions. Therefore it is continuous and has
continuous derivatives of all orders in $E_n$.

Due to anti-invariance of the orbit functions $\varphi_\lambda$,
$\lambda\in D_+^+$, with respect to the Weyl group~$W$,
$\varphi_\lambda$ vanishes on all walls of the Weyl chambers. The
anti-invariance $\varphi_\lambda (r_0x)=-\varphi_\lambda(x)$ for
$\lambda\in P^+_+$ shows that $\varphi_\lambda$, $\lambda\in
P^+_+$, {\it vanishes on the boundary of the fundamental domain
$F$ of the affine Weyl group $W^{\rm aff}$.}

\subsection{Realness and complex conjugation}\label{section5.4}
The results, formulated below and concerning antisymmetric orbit
functions of the Coxeter--Dynkin diagrams $A_n$, $B_n$, $C_n$,
$D_n$ and $G_2$, were proved in the previous section.

Antisymmetric orbit functions of the following Coxeter--Dynkin
diagrams are real:
\[
 B_{2k},\quad C_{2k},\quad   G_2.
\]
Antisymmetric orbit functions of the Coxeter--Dynkin diagrams
$B_{2k+1}$ and $C_{2k+1}$ are purely imaginary. The antisymmetric
orbit functions $\varphi_\lambda$ of the Coxeter--Dynkin diagrams
$A_{4k-1}$ and $A_{4k}$ satisfy the condition
\[
\varphi_{(\lambda_1,\lambda_2,\dots,\lambda_n)}(x)=
\overline{\varphi_{(\lambda_n,\lambda_{n-1},\dots,\lambda_1)}(x)}
\]
and antisymmetric orbit functions $\varphi_\lambda$ of the
Coxeter--Dynkin diagrams $A_{4k+1}$ and $A_{4k+2}$ satisfy the
condition
\[
\varphi_{(\lambda_1,\lambda_2,\dots,\lambda_n)}(x)=-
\overline{\varphi_{(\lambda_n,\lambda_{n-1},\dots,\lambda_1)}(x)}.
\]

Antisymmetric orbit functions
$\varphi_{(\lambda_1,\lambda_2,\dots,\lambda_n)}$ of $D_n$ are
real if $\lambda_{n-1}=\lambda_n$. If $\lambda_{n-1}\ne
\lambda_n$, then orbit functions
$\varphi_{(\lambda_1,\lambda_2,\dots,\lambda_n)}$ are real for
$n=2k$ and satisfy the condition
\[
\varphi_{(\lambda_1,\dots,\lambda_{n-2},\lambda_{n-1}, \lambda_n)}
=\overline{\varphi_{(\lambda_1,\dots,\lambda_{n-2},\lambda_{n},
\lambda_{n-1})}}
\]
if $n=2k+1$.

\subsection{Scaling symmetry}\label{section5.5}
Let $O(\lambda)$ be an orbit of $\lambda$, $\lambda\in D^+_+$.
Since $w(c\lambda)=cw(\lambda)$ for any $c\in {\mathbb R}$ and for
any $w\in W$, then the orbit $O(c\lambda)$ is an orbit consisting
of the points $cw\lambda$, $w\in W$. Moreover, points $w\lambda$
and $cw\lambda$ of the signed orbits $O^\pm(\lambda)$ and $O^\pm
(c\lambda)$, respectively, have the same sign. Let
$\varphi_\lambda=\sum\limits_{w\in W}(\det w) e^{2\pi{\rm
i}w\lambda}$ be the antisymmetric orbit function for $\lambda\in
D^+_+$. Then
\[
\varphi_{c\lambda}(x)=\sum_{w\in W} (\det w) e^{2\pi{\rm i}\langle
cw\lambda ,x\rangle} = \sum_{w\in W} (\det w) e^{2\pi{\rm
i}\langle w\lambda ,cx\rangle} = \varphi_{\lambda}(cx).
\]
The equality $\varphi_{c\lambda}(x)=\varphi_{\lambda}(cx)$
expresses the {\it scaling symmetry of orbit functions.}

If we consider only antisymmetric orbit functions
$\varphi_\lambda$, corresponding to $\lambda\in P^+_+$, then the
scaling symmetry $\varphi_{c\lambda}(x)=\varphi_{\lambda}(cx)$
holds for those values $c\in {\mathbb R}\backslash \{ 0\}$ for
which $c\lambda\in P^+_+$.

\subsection{Duality}\label{section5.6}
Due to the invariance of the scalar product $\langle \cdot, \cdot
\rangle$ with respect to the Weyl group $W$, $\langle w\mu ,wy
\rangle= \langle \mu ,y \rangle$, for $x\in E_n$  not lying on a
wall of some Weyl chamber we have
 \[
 \varphi_\lambda(x)=\sum_{w\in W}(\det w) e^{2\pi {\rm i}\langle
\lambda,w^{-1}x \rangle} =\sum_{w\in W}(\det w) e^{2\pi {\rm
i}\langle \lambda,wx \rangle}=  \varphi_x(\lambda).
 \]
The relation $ \varphi_\lambda(x)=\varphi_x(\lambda)$ expresses
the {\it duality} of antisymmetric orbit functions.

\subsection{Orthogonality}\label{section5.7}
Antisymmetric orbit functions $\varphi_\lambda$, $\lambda\in
P^+_+$, are orthogonal on $F$ with respect to the Euclidean
measure \cite{MP06}:
 \begin{gather}\label{ortog}
|F|^{-1}\int_F
\varphi_\lambda(x)\overline{\varphi_{\lambda'}(x)}dx=
|W|\delta_{\lambda\lambda'} ,
 \end{gather}
where the overbar means complex conjugation. This relation
directly follows from  orthogonality of the exponential functions
$e^{2\pi{\rm i}\langle \mu,x \rangle}$ (entering into the
def\/inition of orbit functions) for dif\/ferent weights $\mu$ and
from the fact that a given weight $\nu\in P$ belongs to precisely
one orbit function. In \eqref{ortog}, $|F|$ means an area of the
fundamental domain $F$.

Sometimes, it is dif\/f\/icult to f\/ind the area $|F|$. In this
case it is useful the following form of the formula \eqref{ortog}:
\[
\int_{\sf T} \varphi_\lambda(x)\overline{\varphi_{\lambda'}(x)}dx=
|W|\delta_{\lambda\lambda'} ,
 \]
where ${\sf T}$ is the torus in $E_n$ described in
Subsection~\ref{section9.1} below. If to assume that an area of
${\sf T}$ is equal to 1, $|{\sf T}|=1$, then $|F|=|W|^{-1}$ and
formula \eqref{ortog} takes the form
\begin{gather}\label{ortog-tor}
\int_F \varphi_\lambda(x)\overline{\varphi_{\lambda'}(x)}dx=
\delta_{\lambda\lambda'} .
 \end{gather}

The formula \eqref{ortog-tor} gives the orthogonality relation for
the antisymmetric multivariate sine function \eqref{det-C}. We
have
\begin{gather}\label{ortog-ant-sin}
2^{2n} \int_F  \det \left(  \sin 2\pi m_ix_j  \right)_{i,j=1}^{n}
\det \left(  \sin 2\pi m'_ix_j  \right)_{i,j=1}^{n}     dx=
\delta_{{\bf m},{\bf m}'} ,
 \end{gather}
where ${\bf m}=(m_1,m_2,\dots,m_n)$ and ${\bf
m}'=(m'_1,m'_2,\dots,m'_n)$ are strictly dominant and integral
(that is, $m_i,m'_j\in {\mathbb Z}^{>0}$, $m_1>m_2>\cdots>m_n>0$),
and the domain $F$ consists of points $x=(x_1,x_2, \dots,x_n)\in
E_n$ such that
\[
 \frac12 \ge x_1\ge x_2\ge\cdots \ge x_n\ge 0
\]
(see Subsection~\ref{section5.10} below).

A similar orthogonality relation can be written down for the
symmetric multivariate cosine function~\eqref{det-B-symm}.
Introducing the notation
\begin{gather}\label{symm-cos}
{\det}^+(\cos 2\pi m_ix_j)_{i,j=1}^n:=  \sum_{w\in S_n}
\cos 2\pi m_{w(1)}x_1 \cdot \,{\cdots}\,\cdot \cos 2\pi
m_{w(n)}x_n
\end{gather}
we have
\begin{gather}\label{ortog-ant-cos}
2^{2n} \int_F  {\det}^+ \left(  \cos 2\pi m_ix_j
\right)_{i,j=1}^{n} {\det}^+ \left(  \cos 2\pi m'_ix_j
\right)_{i,j=1}^{n}     dx= |W_{\bf m}|\delta_{{\bf m},{\bf m}'} ,
 \end{gather}
where ${\bf m}$ and ${\bf m}'$ are such that $m_1\ge m_2\ge
\cdots\ge m_n$, $m'_1\ge m'_2\ge \cdots\ge m'_n$, and $|W_{\bf
m}|$ is an order of the subgroup $|W_{\bf m}|\subset S_n$
consisting of elements leaving ${\bf m}$ invariant.

\subsection{Orthogonality to symmetric orbit functions}\label{section5.8}
Let $\alpha_i$ be a simple root. We create the domain $F^{\rm
ext}=F\cup r_{\alpha_i}F$, where $r_{\alpha_i}$ is the
ref\/lection corresponding to the root $\alpha_i$ and $F$ is the
fundamental domain. Since for $\mu\in P^+_+$ we have
\[
\phi_\mu(r_{\alpha_i}x)=\phi_\mu(x),\qquad
\varphi_\mu(r_{\alpha_i}x)=-\varphi_\mu(x) ,
\]
where $\phi_\mu$ is a symmetric orbit function, then
 \begin{gather}\label{ortog-ant}
 \int_{F^{\rm ext}} \phi_\mu(x)\overline{\varphi_\mu(x)}dx=0.
 \end{gather}
Indeed, due to symmetry and antisymmetry of symmetric and
antisymmetric orbit functions we have
\begin{gather*}
\int_{F^{\rm ext}} \phi_\mu(x)\overline{\varphi_\mu(x)}dx=
\int_{F} \phi_\mu(x)\overline{\varphi_\mu(x)}dx+
\int_{r_{\alpha_i}F} \phi_\mu(x)\overline{\varphi_\mu(x)}dx
\notag\\
\phantom{\int_{F^{\rm ext}}
\phi_\mu(x)\overline{\varphi_\mu(x)}dx}{} = \int_{F}
\phi_\mu(x)\overline{\varphi_\mu(x)}dx+ \int_{F}
\phi_\mu(x)\overline{(-\varphi_\mu(x)})dx=0.
\end{gather*}
For the case of $A_1$ the orthogonality \eqref{ortog-ant} means
the orthogonality of the functions sine and cosine on the interval
$(0,2\pi)$.

The formula \eqref{ortog-ant} determines orthogonality of
multivariate sine and cosine functions \eqref{det-C} and
\eqref{symm-cos}:
\[
 \int_{F^{\rm ext}} \det \left(  \sin 2\pi m_ix_j  \right)_{i,j=1}^{n}
{\det}^+(\cos 2\pi m_ix_j)_{i,j=1}^n dx=0,
\]
where the notations are such as in \eqref{ortog-ant-sin} and
$F^{\rm ext}$ consists of points $x\in E_n$ such that
\[
\tfrac12 \ge x_1\ge x_2\ge\cdots \ge x_n\ge 0\qquad {\rm or}\qquad
\tfrac12 \ge x_2\ge x_1\ge x_3\ge x_4\ge\cdots \ge x_n\ge 0.
\]

\subsection[Antisymmetric orbit functions $\varphi_\rho$]{Antisymmetric orbit functions $\boldsymbol{\varphi_\rho}$}
\label{section5.9}
Let $\rho$ be the half-sum of all positive roots of a root system:
 \begin{gather}\label{rho}
\rho=\frac12 \sum_{\alpha>0} \alpha .
 \end{gather}
It is well-known that for all simple Lie algebras in
$\omega$-coordinate we have
\[
\rho=\omega_1+\omega_2+\cdots +\omega_n\equiv (1\ 1\ \cdots\ 1).
\]
The antisymmetric orbit function $\varphi_\rho$ is important in
the theory of characters of group representations. We have
\begin{gather}\label{rho-for}
\varphi_\rho(x)=\sum_{w\in W} (\det w)e^{2\pi{\rm i}\langle
w\rho,x \rangle}.
\end{gather}

\begin{proposition}\label{prop8}
 The antisymmetric orbit function $\varphi_\rho$
can be represented as
\begin{gather}
\varphi_\rho(x)= \prod_{\alpha>0} ( e^{\pi{\rm i}\langle\alpha,x
\rangle}-e^{-\pi{\rm i}\langle\alpha,x \rangle})
 =e^{2\pi{\rm i}\langle\rho,x \rangle} \prod_{\alpha>0}
 (1 - e^{-2\pi{\rm i}\langle\alpha,x \rangle})
\notag\\
\phantom{\varphi_\rho(x)}{}= (2{\rm i})^r \prod_{\alpha>0} \sin
\pi \langle\alpha,x \rangle,\label{rho-sin}
 \end{gather}
where $r$ is a number of positive roots in the corresponding root
system.
\end{proposition}

\begin{proof} The expression $g(x)=\prod\limits_{\alpha>0}
( e^{\pi{\rm i}\langle\alpha,x \rangle}-e^{-\pi{\rm
i}\langle\alpha,x \rangle})$ is antisymmetric with respect to the
Weyl group $W$. Indeed, if $r_i$ is a ref\/lection, corresponding
to the simple root $\alpha_i$, then due to Proposition~\ref{prop1}
we have
\begin{gather*}
g(r_ix)= \prod_{\alpha>0} ( e^{\pi{\rm i}\langle\alpha,r_ix
\rangle}-e^{-\pi{\rm i}\langle\alpha,r_ix \rangle})
=\prod_{\alpha>0} ( e^{\pi{\rm i}\langle r_i\alpha,x
\rangle}-e^{-\pi{\rm i}\langle r_i\alpha,x \rangle})
\notag\\
\phantom{g(r_ix)}{}= ( e^{-\pi{\rm i}\langle\alpha_i,x
\rangle}-e^{\pi{\rm i}\langle\alpha_i,x \rangle})
\prod_{\alpha>0,\alpha\ne \alpha_i} ( e^{\pi{\rm i}\langle\alpha,x
\rangle}-e^{-\pi{\rm i}\langle\alpha,x \rangle}) =-g(x).
\end{gather*}
Since $\alpha_i$ is an arbitrary simple root, $g(x)$ is
antisymmetric with respect to $W$. Then $g(x)$ is a~sum of
antisymmetric orbit functions. Representing $\prod_{\alpha>0} (
e^{\pi{\rm i}\langle\alpha,x \rangle}-e^{-\pi{\rm
i}\langle\alpha,x \rangle})$ in the form~\eqref{orb-a}, we see
that in this form there exists only one term $e^{\pi {\rm
i}\langle \lambda,x \rangle}$ with a strictly dominant weight
$\lambda$ and this weight coincides with $\rho$. Therefore, $g(x)=
\varphi_\rho(x)$. Proposition is proved.
\end{proof}

An invariant measure of the compact Lie group $G$, associated with
the Weyl group $W$, is expressed in terms of $|\varphi_\rho|^2$.
Taking into account an explicit form of positive roots in the
orthogonal coordinate systems, it is easy to derive from
\eqref{rho-sin} that in these coordinates we have
\begin{gather*}
\varphi_\rho(x)=(2{\rm i})^{n(n+1)/2} \prod_{1\le i<j\le n+1} \sin
\pi (x_i-x_j)\qquad \mbox{for $A_n$},
\\
\varphi_\rho(x)=(2{\rm i})^{n^2} \prod_{1\le i<j\le n} \sin  \pi
(x_i-x_j) \sin  \pi (x_i+x_j)
 \prod_{1\le i\le n} \sin \pi x_i \qquad \mbox{for $B_n$},
\\
\varphi_\rho(x)=(2{\rm i})^{n^2} \prod_{1\le i<j\le n} \sin  \pi
(x_i-x_j) \sin  \pi (x_i+x_j)
 \prod_{1\le i\le n} \sin 2\pi x_i\qquad \mbox{for $C_n$},
\\
\varphi_\rho(x)=(2{\rm i})^{n(n-1)} \prod_{1\le i<j\le n} \sin
\pi (x_i-x_j) \sin  \pi (x_i+x_j) \qquad\mbox{for $D_n$}.
\end{gather*}
These formulas give other expressions for $|\varphi_\rho(x)|$ with
respect to formulas derived in
Subsections~\ref{section4.4}--\ref{section4.7}.

\begin{proposition}\label{prop9}
 The antisymmetric orbit function $\varphi_\rho$
vanishes on the boundary of the fundamental domain $F$. It does
not vanish on intrinsic points of $F$.
\end{proposition}

\begin{proof}  Since $\rho\in P^+_+$, then $\varphi_\rho$ vanishes on the
boundary of $F$ due to the results of Subsection~\ref{section5.3}.
From the other side, it is easy to see from \eqref{rho-sin} that
the set of points, on which $\varphi_\rho(x)$ vanishes, coincides
with the set of all points $x\in E_n$ for which $\langle \alpha,x
\rangle \in \mathbb{Z}$ for some root $\alpha$. No of these points
is an intrinsic point of $F$. The proposition is proved.
\end{proof}

From Proposition~\ref{prop9} and from the above formulas for
$\varphi_\rho(x)$ we easily derive explicit forms of the
fundamental domains for the cases $A_n$, $B_n$, $C_n$, $D_n$ in
the orthogonal coordinates.

 \medskip

(a) The fundamental domain $F(A_n)$ is contained in the domain of
real points $x=(x_1,x_2,\dots$, $x_{n+1})$ such that
\[
 x_1>x_2>\cdots>x_{n+1},\qquad x_1+x_2+\cdots +x_{n+1}=0.
\]
Moreover, a point $x$ of this domain belongs to $F(A_n)$ if and
only if $x_1+|x_{n+1}|<1$. The last condition means in fact the
relation $\langle x,\xi \rangle<1$ for the highest root $\xi$ of
the root system $A_n$.
\medskip

(b) The fundamental domain $F(B_n)$ is contained in the domain of
points $x=(x_1,x_2,\dots$, $x_{n})$ such that
\[
1> x_1>x_2>\cdots>x_{n}>0.
\]
Moreover, a point $x$ of this domain belongs to $F(B_n)$ if and
only if $x_1+x_2<1$.
\medskip

(c) The fundamental domain $F(C_n)$ consists of all points
$x=(x_1,x_2,\dots,x_{n})$ such that
\[
\tfrac12 > x_1>x_2>\cdots>x_{n}>0.
\]

(d) The fundamental domain $F(D_n)$ is contained in the domain of
points $x=(x_1,x_2,\dots$, $x_{n})$ such that
\[
1> x_1>x_2>\cdots>x_{n-1}>|x_n|.
\]
Moreover, a point $x$ of this domain belongs to $F(D_n)$ if and
only if $x_1+x_2<1$.

\subsection[Symmetric orbit functions $\phi_\rho$]{Symmetric orbit functions $\boldsymbol{\phi_\rho}$}
\label{section5.10}
Let us consider a symmetric counterpart of the antisymmetric orbit
function $\varphi_\rho(x)$:
\begin{gather}\label{rho-symm}
\phi_\rho(x):=\sum_{w\in W} e^{2\pi{\rm i}\langle w\rho,x
\rangle},
\end{gather}
where, as before, $\rho=\frac 12 \sum\limits_{\alpha>0} \alpha$.

\begin{proposition}\label{prop10}
 The symmetric orbit function $\phi_\rho$
can be represented as
\begin{gather}
\varphi_\rho(x)= \prod_{\alpha>0} ( e^{\pi{\rm i}\langle\alpha,x
\rangle}+e^{-\pi{\rm i}\langle\alpha,x \rangle})
 =e^{2\pi{\rm i}\langle\rho,x \rangle} \prod_{\alpha>0}
 (1 + e^{-2\pi{\rm i}\langle\alpha,x \rangle})
 \notag\\
\phantom{\varphi_\rho(x)}{} = 2^r \prod_{\alpha>0} \cos \pi
\langle\alpha,x \rangle,\label{rho-cos}
\end{gather}
where $r$ is a number of positive roots in the corresponding root
system.
\end{proposition}

This proposition is proved in the same way as
Proposition~\ref{prop6}.

 \medskip

It is easy to derive from \eqref{rho-cos} that in the orthogonal
coordinate systems we have
\begin{gather*}
\phi_\rho(x)=2^{n(n+1)/2} \prod_{1\le i<j\le n+1} \cos \pi
(x_i-x_j)\qquad \mbox{for $A_n$},
\\
\phi_\rho(x)=2^{n^2} \prod_{1\le i<j\le n} \cos  \pi (x_i-x_j)
\cos  \pi (x_i+x_j)
 \prod_{1\le i\le n} \cos \pi x_i\qquad \mbox{for $B_n$},
\\
\phi_\rho(x)=2^{n^2} \prod_{1\le i<j\le n} \cos  \pi (x_i-x_j)
\cos  \pi (x_i+x_j)
 \prod_{1\le i\le n} \cos 2\pi x_i
\qquad \mbox{for $C_n$},
\\
\phi_\rho(x)=2^{n(n-1)} \prod_{1\le i<j\le n} \cos  \pi (x_i-x_j)
\cos  \pi (x_i+x_j)\qquad \mbox{for $D_n$}.
\end{gather*}

\section[Properties of antisymmetric orbit functions of $A_n$]{Properties of antisymmetric orbit functions
of $\boldsymbol{A_n}$}\label{section6}

By using results on decomposition of certain reducible
representations of the group $GL(n,\mathbb{C})$ into irreducible
constituents, properties of antisymmetric orbit functions of $A_n$
can be derived. We use in this section the results of~\cite{Weyl}.

\subsection{Decomposition of symmetric powers of representations}\label{section6.1}
We need some formulas for decomposition of symmetric powers of
f\/inite dimensional irreducible representations of the group
$GL(n_1,\mathbb{C})\times GL(n_2,\mathbb{C})$. Let us f\/irst
def\/ine symmetric powers of representations.

Let $T$ be a f\/inite dimensional representation of a group $G$ on
a linear space $X$. It induces a~representation on the space
$\mathcal{P}_m(X)$, which is a subspace of the space
$\mathcal{P}(X)$ of polynomials on $X$, consisting of all
homogeneous polynomials of power $m$. In order to determine this
representation we note that the formula
\[
(Q(g)p)(x)=p(T(g^{-1})x),\qquad x\in X,\quad g\in G,
\]
gives a representation of $G$ on $\mathcal{P}(X)$ which is denoted
by $Q$. The subspace $\mathcal{P}_m(X)$ is invariant with respect
to the representation $Q$. The restriction of $Q$ onto
$\mathcal{P}_m(X)$ is called an $m$-th {\it symmetric power of the
representation} $T$ and is denoted as $\sigma_m(T)$.

Let $G=GL(n_1,\mathbb{C})\times GL(n_2,\mathbb{C})$, $n_1\le n_2$.
Then a f\/inite dimensional irreducible representation of $G$ is
given (in the orthogonal coordinate system) by
\[
(\lambda|\mu)=(m_1,m_2,\dots ,m_{n_1}\,|\,r_1,r_2,\dots ,r_{n_2}),
\]
where
\[
m_1\ge m_2\ge\cdots \ge m_{n_1},\qquad r_1\ge r_2\ge\cdots \ge
r_{n_2}.
\]
We assume that $m_{n_1}\ge 0$ and $r_{n_2}\ge 0$.

One can consider three representations of
$G=GL(n_1,\mathbb{C})\times GL(n_2,\mathbb{C})$ on the space
$\mathfrak{M} _{n_1n_2}(\mathbb{C})$ of $n_1\times n_2$ complex
matrices. They are given by the formulas
 \begin{gather}\label{T}
 T(g_1,g_2)X=g_1Xg_2^t,
 \\
 \label{T'}
 T'(g_1,g_2)X=g_1Xg_2^*,
 \\
 \label{T''}
 T''(g_1,g_2)X=g_1Xg_2^{-1},
 \end{gather}
where the index $t$ means a transposition and $*$ means a
transposition together with complex conjugation. Then for
symmetric powers of these representations we have the following
decompositions into irreducible representations of $G$ (a proof
see in~\cite{Weyl}):
 \begin{gather}\label{s-T}
 \sigma_m(T)=\sum_{\sum_i s_i=m} (s_1,s_2,\dots,s_{n_1}\,|\,
 s_1,s_2,\dots,s_{n_1},0,\dots,0),
 \\
 \label{s-T'}
 \sigma_m(T')=\sum_{\sum_i s_i=m} (s_1,s_2,\dots,s_{n_1}\,|\,\overline{
 s_1,s_2,\dots,s_{n_1},0,\cdots,0}),
 \\
 \label{s-T''}
 \sigma_m(T'')=\sum_{\sum_i s_i=m} ( s_1,s_2,\dots,s_{n_1}\,|\,0,\ldots,0,
 -s_{n_1},-s_{n_2},\dots,-s_{1}),
 \end{gather}
where summations are over all $s_1,s_2,\dots,s_{n_1}$ such that
\[
s_1\ge s_2\ge \cdots \ge s_{n_1}\ge 0,\qquad s_1+s_2+\cdots
+s_{n_1}=m
\]
and the overbar means that a representation is anti-analytic.

If $n_1=n_2=n$, then the formulas \eqref{T}--\eqref{T''} determine
the following tensor product representations of
$G=GL(n,\mathbb{C})$ on the space $\mathfrak{M}_n(\mathbb{C})$ of
complex $n\times n$ matrices:
\begin{gather*}
(T_1\otimes T_1)(g)X=gXg^t,
\\
(T_1\otimes \bar{T}_1)(g)X=gXg^*,
\\
(T_1\otimes \hat T_1)(g)X=gXg^{-1},
\end{gather*}
where $T_1$ is the f\/irst fundamental representation (with
highest weight $(1,0,\dots,0)$) of $G=GL(n,\mathbb{C})$,  ${\bar
T}_1$ and ${\hat T}_1$ are the complex conjugate and
contragredient representations to the representation $T_1$,
respectively. Then according to \eqref{s-T}--\eqref{s-T''} we have
 \begin{gather}\label{s-TT}
 \sigma_m(T_1\otimes T_1)=\sum_{{\bf m}} T_{\bf m}\otimes T_{\bf m},
 \\
 \label{s-TT'}
 \sigma_m(T_1\otimes \bar T_1)=\sum_{{\bf m}} T_{\bf m}\otimes \bar T_{\bf m},
 \\
 \label{s-TT''}
 \sigma_m(T_1\otimes T_1)=\sum_{{\bf m}} T_{\bf m}\otimes \hat T_{\bf m},
 \end{gather}
where $T_{\bf m}$ is the irreducible representation of the group
$GL(n,\mathbb{C})$ with highest weight ${\bf m}\equiv
(m_1,m_2,\dots ,m_n)$, $m_1\ge m_2\ge \cdots \ge m_n\ge 0$, and
the summation is over those ${\bf m}$ for which $m_1+m_2+\cdots
+m_n=m$.

Replacing the space $\mathfrak{M}_n(\mathbb{C})$ by the subspaces
of all symmetric or all antisymmetric matrices from
$\mathfrak{M}_n(\mathbb{C})$ we obtain the following
decompositions of symmetric powers of the irreducible
representations of $GL(n,\mathbb{C})$ with highest weights
$(2,0,\dots,0)$ and $(1,1,0,\dots,0)$, respectively:
 \begin{gather}\label{s-T2}
 \sigma_m(T_{(2,0,\dots,0)})=\sum_{{\bf m}} T_{(2m_1,2m_2,\dots,2m_n)},
 \\
 \label{s-T11}
 \sigma_m(T_{(1,1,0,\dots,0)})=\sum_{{\bf m}} T_{(m_1,m_1,\dots,m_k,m_k)},
 \end{gather}
where the summations are over those ${\bf m}=(m_1,m_2,\dots ,m_n)$
for which $m_1\ge m_2\ge \cdots \ge m_n\ge 0$ and $m_1+m_2+\dots
+m_n=m$. In the second formula $n=2k$; if $n=2k+1$, then on the
right hand side we have to replace $T_{(m_1,m_1,\dots,m_k,m_k)}$
by $T_{(m_1,m_1,\dots,m_k,m_k,0)}$.


\subsection[Properties of antisymmetric orbit functions of $A_n$]{Properties of
antisymmetric orbit functions of
$\boldsymbol{A_n}$}\label{section6.2}
We represent antisymmetric orbit functions of $A_n$ in the
orthogonal coordinate system as in formula \eqref{orb-f}. Let
$\lambda=(m_1,m_2,\dots ,m_{n+1})$ and $\lambda
+r=(m_1+r,m_2+r,\dots , m_{n+1}+r)$, where $r$ is a f\/ixed real
number. If $x=(x_1,x_2,\dots, x_{n+1})$, $x_1+x_2+\cdots
+x_{n+1}=0$, and $w\in W$, then we have
\[
e^{2\pi {\rm i}\langle \lambda+r,wx \rangle}= e^{2\pi {\rm
i}\langle \lambda,wx \rangle} e^{2\pi {\rm i}\langle 0+r,wx
\rangle}=e^{2\pi {\rm i}\langle \lambda,wx \rangle} .
\]
It follows from this equality that
\begin{gather}
\varphi_\lambda(x)=\varphi_{\lambda+r}(x),\qquad
\phi_\lambda(x)=\phi_{\lambda+r}(x)
 \end{gather}
where $\lambda=(m_1,m_2,\dots ,m_{n+1})$ is given in the
orthogonal coordinate system. This means that instead of $m_i$,
$i=1,2,\dots, n+1$, determined by formulas of
Subsection~\ref{section3.3}, we may assume that $m_1,m_2,\dots
,m_{n+1}$ are integers such that $m_1\ge m_2\ge \cdots \ge
m_{n+1}\ge 0$. We adopt this assumption in this subsection.

For simplicity we introduce the following notations:
\[
e^{2\pi {\rm i}x_j}=y_j,\qquad j=1,2,\dots ,n+1.
\]
In order to receive relations for antisymmetric orbit functions we
have to take characters of representations on both sides of
relations of the previous subsection  and to substitute the
expression
\[
\chi_\lambda(x)=\frac{\varphi_{\lambda+\rho}(x)}{\varphi_\rho(x)}
\]
for characters (see Subsection~\ref{section9.1} below).

The relation \eqref{s-T} gives the equality
\[
\chi_{\sigma_m(T)}(g_1,g_2)=\sum_{\sum_i m_i=m}
\chi_s(g_1)\chi_{s'}(g_2)
\]
for characters of representations, where
$s'=(s_1,s_2,\dots,s_{n_1},0,\dots,0)$ and the sum is such as
in~\eqref{s-T}. Multiply both sides by $t^m$ and sum up over all
non-negative integral values of $m$. We get
 \begin{gather}\label{ch-1}
\sum_{m=0}^\infty t^m\chi_{\sigma_m(T)}(g_1,g_2)=\sum_{m=0}^\infty
t^m \sum_{\sum\limits_i m_i=m} \chi_s(g_1)\chi_{s'}(g_2).
  \end{gather}
Take this relation for $g_1=\diag (y_1,y_2,\dots ,y_{n_1})$ and
$g_2=\diag (z_1,z_2,\dots ,z_{n_2})$. For such $g_1$ and~$g_2$ the
left hand side takes the form
\[
\sum_{m=0}^\infty t^m\chi_{\sigma_m(T)}(g_1,g_2)=\sum_{m=0}^\infty
t^m \sum y_1^{r_1}y_2^{r_2}\cdots y_{n_1}^{r_{n_1}}
z_1^{p_1}z_2^{p_2}\cdots z_{n_2}^{p_{n_2}},
\]
where the second summation on the right hand side is over integral
$r_i$ and $p_j$ such that $\sum\limits_{i=1}^{n_1}r_i+
\sum\limits_{j=1}^{n_2}p_j=m$. Therefore, the relation
\eqref{ch-1} can be written as{\samepage
\[
\sum_{m=0}^\infty t^m \sum_{\sum_i m_i=m}
\chi_s(g_1)\chi_{s'}(g_2) =\prod_{i=1}^{n_1} \prod_{j=1}^{n_2}
(1-ty_iz_j)^{-1},
\]
where $g_1=\diag (y_1,y_2,\dots ,y_{n_1})$, $g_2=\diag
(z_1,z_2,\dots ,z_{n_2})$ and $|t|<1$, $|y_i|<1$, $|z_j|<1$.}

Further, we represent characters in \eqref{ch-1} in terms of
antisymmetric orbit functions and set $t=1$. As a result, we
receive the following relation for antisymmetric orbit functions:
 \begin{gather}\label{R-1}
\varphi^{(n_1)}_{\rho_1}(y) \varphi^{(n_2)}_{\rho_2}(z)
\prod_{i=1}^{n_1} \prod_{j=1}^{n_2} (1-y_iz_j)^{-1}= \sum_m
\varphi^{(n_1)}_{s+\rho_1}(y) \varphi^{(n_2)}_{\hat s+\rho_2}(z),
 \end{gather}
where $y=(y_1,y_2,\dots,y_{n_1})$, $z=(z_1,z_2,\dots,z_{n_2})$,
and $\rho_1$ and $\rho_2$ are half-sums of positive roots for
$A_{n_1-1}$ and $A_{n_2-1}$, respectively. In particular, if
$n_1=n_2=n$, then
\[
\varphi_{\rho}(y) \varphi_{\rho}(z) \prod_{i,j=1}^{n}
(1-y_iz_j)^{-1}= \sum_m \varphi_{s+\rho}(y) \varphi_{s+\rho}(z).
\]
Applying to this relation the Cauchy lemma, which states that
\[
\det \left( (1-y_iz_j)^{-1}\right)_{i,j=1}^n = \varphi_{\rho}(y)
\varphi_{\rho}(z) \prod_{i,j=1}^{n} (1-y_iz_j)^{-1} ,
\]
we obtain the identity
\[
\sum_\lambda \varphi_{\lambda+\rho}(y) \varphi_{\lambda+\rho}(z)=
\det \left( (1-y_iz_j)^{-1}\right)_{i,j=1}^n.
\]

Now we use the relation \eqref{s-T2} for representations of the
group $GL(n,\mathbb{C})$ in order to derive the equality
\[
\chi_{(2,0,\dots,0),m} (g)=\sum_{|{\bf m}|=m} \chi_{2{\bf
m}}(g),\qquad g\in GL(n,\mathbb{C}),
\]
where on the left hand side is the character of the representation
$\sigma_m(T_{(2,0,\dots,0)})$ from \eqref{s-T2}, $2{\bf
m}=(2m_1,2m_2,\dots,2m_n)$, $m_1\ge m_2\ge\cdots \ge m_n\ge 0$,
$|{\bf m}|=m_1+m_2+\cdots +m_n$. We derive from this equality that
 \begin{gather}\label{cha-2}
\sum_{m=0}^\infty \chi_{(2,0,\dots,0),m} (g)t^m=\sum_{m=0}^\infty
t^m \sum_{|{\bf m}|=m} \chi_{2{\bf m}}(g).
 \end{gather}
Setting $g=\diag (y_1,y_2,\dots,y_n)$, for the left hand side we
have the expression
\[
\sum_{m=0}^\infty \chi_{(2,0,\dots,0),m} (g)t^m =\prod_{1\le i\le
j\le n} (1-ty_iy_j)^{-1},
\]
where $|y_i|<1$ and $|t|<1$. Now using the Weyl character formula
one receives
 \begin{gather}\label{cha-3}
\sum_{\bf m} \varphi_{2 {\bf m}+\rho}(y)t^{|{\bf m}|}=
\varphi_\rho(y) \prod_{1\le i\le j\le n} (1-ty_iy_j)^{-1}.
 \end{gather}

Substituting into \eqref{cha-3} the expression for $\varphi_{2
{\bf m}+\rho}(y)$, setting $t=1$ and using the evident relation
\[
\sum_{\bf m} a_1^{m_1}a_2^{m_2}\cdots a_n^{m_n}=(1-a_1)^{-1}
(1-a_1a_2)^{-1}\cdots (1-a_1a_2\cdots a_n)^{-1},
\]
we get a non-trivial identity
\[
\sum_{(i_1,i_2,\dots,i_n)} \frac{{\rm sign} (i_1,i_2,\dots,i_n)\,
y_{i_1}^{n-1} y_{i_2}^{n-2}\cdots y_{i_{n-1}}}{(1-y^2_{i_1})
(1-y^2_{i_1}y^2_{i_2})\cdots (1-y^2_{i_1}\cdots y^2_{i_n})}
=\varphi_\rho(y) \prod_{1\le i\le j\le n} (1-y_iy_j)^{-1},
\]
where summation is over all permutations $(i_1,i_2,\dots,i_n)$ of
the natural numbers $1,2,\dots ,n$ and ${\rm sign}\,
(i_1,i_2,\dots,i_n)$ means a sign of the permutation
$(i_1,i_2,\dots,i_n)$.

In the same way, from the decomposition \eqref{s-T11} we derive
 \begin{gather}\label{cha-4}
\sum_{\bf m} \varphi_{ {\tilde {\bf m}}+\rho}(y)t^{|{\bf m}|}=
\varphi_\rho(y) \prod_{1\le i< j\le n} (1-ty_iy_j)^{-1},
 \end{gather}
where summation is over ${\bf m}=(m_1,m_2,\dots,m_k)$, $k=[n/2]$,
$m_1\ge m_2\ge\cdots \ge m_k\ge 0$, ${\tilde {\bf
m}}=(m_1,m_1,m_2,m_2,\dots,m_k,m_k)$ (here one has to add 0 at the
end if $n=2k+1$). As above, we receive from here the identity
\[
\sum_{(i_1,i_2,\dots,i_n)} \frac{{\rm sign} (i_1,i_2,\dots,i_n)\,
y_{i_1}^{n-1} y_{i_2}^{n-2}\cdots y_{i_{n-1}}} {(1-y_{i_1}y_{i_2})
(1-y_{i_1}y_{i_2}y_{i_3}y_{i_4})\cdots (1-y_{i_1}\cdots
y_{i_{2\nu}})}
 =\varphi_\rho(y) \prod_{1\le i\le j\le n} (1-y_iy_j)^{-1}.
\]

In order to derive other identities we note that for
$\theta=(\theta_1,\theta_2,\dots,\theta_n)$ one has
\[
\lim_{\theta\to 0} \varphi_{\lambda+\rho}(e^{{2\pi\rm i}\theta})
=\varphi_\rho(\lambda+\rho).
\]
Taking into account this relation and substituting $y=(1,\dots,1)$
and $z=(1,\dots,1)$ into \eqref{R-1}, we derive a relation for the
antisymmetric orbit functions $\varphi_\rho$:
  \begin{gather}\label{cha-5}
\sum_{|{\bf m}|=m} \varphi^{(s)}_{\rho_s}({\bf m}+\rho_s)
\varphi^{(r)}_{\rho_r}(\hat {\bf
m}+\rho_r)=\frac{(sr+m-1)!}{(sr-1)!m!} \varphi^{(s)}_{\rho_s}(
\rho_s) \varphi^{(r)}_{\rho_r}(\rho_r),
  \end{gather}
where $r\ge s\ge 0$, ${\bf m}=(m_1,m_2,\dots,m_s)$, $\hat {\bf m}=
(m_1,m_2,\dots,m_s,0,\dots,0)$, $|{\bf m}|=m_1+m_2+\cdots+m_s$ and
$\rho_k=(k-1,k-2,\dots,1,0)$.

From the relations \eqref{cha-3} and \eqref{cha-4} one similarly
obtains the identities
  \begin{gather}\label{cha-6}
\sum_{|{\bf m}|}\varphi_\rho(2{\bf m}+\rho)=\frac{(m+\frac12
n(n+1)-1)!}{(\frac12 n(n+1)-1)!m!} \varphi_\rho(\rho),
 \\
 \label{cha-7}
\sum_{|{\bf m}|}\varphi_\rho(\tilde{\bf m}+\rho)=\frac{(m+\frac12
n(n-1)-1)!}{(\frac12 n(n-1)-1)!m!} \varphi_\rho(\rho),
 \end{gather}
where ${\bf m}=(m_1,m_2,\dots,m_\nu)$, $\tilde{\bf
m}=(m_1,m_1,m_2,m_2,\dots,m_\nu,m_\nu)$, $\nu=[n/2]$.

From \eqref{cha-5}--\eqref{cha-7} we derive the relations
\begin{gather*}
\sum_{\bf m} \varphi^{(s)}_{\rho_s}({\bf m}+\rho_s)
\varphi^{(r)}_{\rho_r}(\hat {\bf m}+\rho_r)t^{|{\bf m}|}=
(1-t)^{-sr} \varphi^{(s)}_{\rho_s}( \rho_s)
\varphi^{(r)}_{\rho_r}(\rho_r),
\\
\sum_{\bf m}\varphi_\rho(2{\bf m}+\rho)t^{|{\bf m}|}=
(1-t)^{-n(n+1)/2}\varphi_\rho(\rho),
\\
\sum_{\bf m}\varphi_\rho(2\tilde{\bf m}+\rho)t^{|{\bf m}|}=
(1-t)^{-n(n-1)/2}\varphi_\rho(\rho).
\end{gather*}

Other decompositions of the previous subsection lead to new
relations for antisymmetric orbit functions of $A_n$.

\section{Decomposition of products of (anti)symmetric orbit functions}\label{section7}
The aim of this section is to derive how to decompose products of
(anti)symmetric orbit functions into sums of (anti)symmetric orbit
functions. Such operations are fulf\/illed by means of the
corresponding decompositions of (signed) orbits.

\subsection{Products of symmetric and antisymmetric orbit functions}\label{section7.1}
Invariance of symmetric orbit functions $\phi_\lambda$ and
anti-invariance of antisymmetric orbit functions~$\varphi_\lambda$
with respect to the Weyl group $W$ lead to the following
statements:

\begin{proposition}\label{prop11}
{\rm (a)} A product of symmetric orbit functions expands into a
sum of symmetric orbit functions:
 \begin{gather}\label{dec-1}
 \phi_\lambda \phi_\mu =\sum _\nu n_\nu
\phi_\nu,
 \end{gather}
where an integer $n_\nu$ shows how many times the orbit function
$\phi_\nu$ is contained in the product~$ \phi_\lambda \phi_\mu$.

{\rm (b)} A product of antisymmetric orbit functions expands into
a sum of symmetric orbit functions:
 \begin{gather}\label{dec-2}
 \varphi_\lambda \varphi_\mu =\sum _\nu n_\nu
\phi_\nu ,
 \end{gather}
where $n_\nu$ are positive or negative integral coefficients.

{\rm (c)} A product of symmetric and antisymmetric orbit functions
expands into a sum of antisymmetric orbit functions:
 \begin{gather}\label{dec-3}
 \phi_\lambda \varphi_\mu =\sum _\nu n_\nu
\varphi_\nu.
 \end{gather}
where $n_\nu$ are positive or negative integral coefficients.
\end{proposition}

\begin{proof} In the case (b) we have
\[
 \varphi_\lambda(wx) \varphi_\mu(wx)=(\det w)^2 \varphi_\lambda(x)
 \varphi_\mu(x)= \varphi_\lambda(x) \varphi_\mu(x).
\]
Therefore, the product $ \varphi_\lambda \varphi_\mu $ is
invariant with respect to $W$. Hence, it can be expanded into
symmetric orbit functions (see Subsection 7.8 in \cite{KP06}).
Since antisymmetric orbit functions $\varphi_\lambda$ are sums of
exponential functions $e^{2\pi{\rm i}\langle \sigma,x\rangle}$
with coef\/f\/icients $\pm 1$, the coef\/f\/icients $n_\nu$ in
\eqref{dec-2} are integers. In the case (c) we have a similar
situation. The case (a) is considered in \cite{KP06}.
\end{proof}

In order to fulf\/ill expansions (a)--(c) in an explicit form it
is necessary to fulf\/ill the corresponding decompositions of
products of (signed) orbit, considering usual (not signed) orbits
as signed orbits in which to all points the sign ``+'' is
assigned, and to take into account that multiplication of (singed)
orbit functions are reduced to multiplication of exponential
functions.

A product $O^\pm(\lambda)\otimes O^\pm(\lambda')$ of two signed
orbits $O^\pm(\lambda)$ and $O^\pm(\lambda')$ (one or two of them
can be replaced by usual orbits, that is, to the
corresponding points the sign ``+'' is assigned) is the set of all
points of the form $\lambda_1+\lambda_2$ (where $\lambda_1\in
O^\pm(\lambda)$ and $\lambda_2\in O^\pm(\lambda')$) with a sign
which is a product of signs of $\lambda$ and $\lambda'$. Since a
set of points $\lambda_1+\lambda_2$ (without signs), $\lambda_1\in
O(\lambda)$, $\lambda_2\in O(\lambda')$, is invariant with respect
to action of the corresponding Weyl group, each product of orbits
is decomposable into a sum of orbits. Then it follows from
assertions (a)--(c) of Proposition~\ref{prop11} that, considering
points $\lambda_1+\lambda_2$ with signs, we obtain decomposition
of $O^\pm(\lambda)\otimes O^\pm(\lambda')$ into usual orbits,
where to each point a sign is assigned  (the same sign for points
of a f\/ixed orbit). Moreover, a product of a signed orbit with a
usual orbit decomposes into signed orbits (not into usual orbits).

Under product of (signed) orbits we may receive an orbit with
signed points in which   the sign ``$-$'' corresponds to
a~dominant weight. This means that we obtain a signed orbit with
opposite signs for its points. In this case we say that a product
of (signed) orbits contains the corresponding signed orbit with
sign ``$-$'' and denote it by $-O^\pm (\mu)$. That is why in
\eqref{dec-2} and~\eqref{dec-3} negative coef\/f\/icients can
appear.
 \medskip

\noindent {\bf Example.} {\it Orbits of $A_1$.} If $a\in E_1$ is
strictly positive, then the signed orbit of this point $O^\pm(a)$
consists of two signed points $a^+$ and $-a^-$. It is easy to see
that for the product $O^\pm(a)\otimes O^\pm(b)$ we have
\begin{gather*}
O^\pm(a)\otimes O^\pm(b)
\equiv\{a^+,-a^-\}\otimes\{b^+,-b^-\}\\
\phantom{O^\pm(a)\otimes O^\pm(b)}{}=\{(a+b)^+,(-a-b)^+\}
  \cup\{(|a-b|)^-,(-|a-b|)^-\}\\
\phantom{O^\pm(a)\otimes O^\pm(b)}{}=O(a+b)\cup -O(|a-b|),
\end{gather*}
where $O(a+b)$ and $O(|a-b|)$ are usual (not signed) orbits.
Therefore,
\[
\varphi_a(x)\varphi_b(x)=\phi_{a+b}(x)-\phi_{|a-b|}(x).
\]
Similarly, we have
\begin{gather*}
O^\pm(a)\otimes O(b)= O^\pm(a+b)\cup O^\pm(a-b),\qquad a>b>0,
\\
O^\pm(a)\otimes O(b)= O^\pm(a+b)\cup -O^\pm(b-a),\qquad b>a>0.
\end{gather*}
Thus,
\begin{gather*}
\varphi_a(x)\varphi_b(x)=\varphi_{a+b}(x)+\varphi_{a-b}(x),\qquad
a>b>0,
\\
\varphi_a(x)\phi_b(x)=\varphi_{a+b}(x)-\varphi_{|a-b|}(x), \qquad
b>a>0.
\end{gather*}
For the corresponding decompositions for $O(a)\otimes O(b)$ see~\cite{KP06}.

\medskip

Decomposition of products of orbits in higher dimension of the
Euclidean space is not a simple task. In the next subsection we
consider some general results on the decomposition.

\subsection{Products of symmetric and antisymmetric orbits}\label{section7.2}
Let $O(\lambda)=\{ w\lambda | w\in W/W_\lambda\}$ be a usual orbit
and $O^\pm(\mu)=\{ w\mu | w\in W\}$ be a signed orbit. Then
\begin{gather}
O(\lambda)\otimes O^\pm(\mu)= \{ (w\lambda +w'\mu)^{\det w'} \,|\,
w\in W/W_\lambda,w'\in W\}
\notag\\
\phantom{O(\lambda)\otimes O^\pm(\mu)}{}=  \{
(w\lambda{+}w_1\mu)^{\det w_1} \,|\, w{\in} W/W_\lambda\}\cup \{
(w\lambda{+}w_2\mu)^{\det w_2} \,| \,w{\in} W/W_\lambda\}\cup
\cdots
 \notag\\
  \phantom{O(\lambda)\otimes O^\pm(\mu)}{}
\quad \cup \{ (w\lambda{+}w_s\mu)^{\det w_s}\, | \, w{\in}
W/W_\lambda\},\label{decom-00}
\end{gather}
where $w_1,w_2,\dots ,w_s$ is the set of elements of $W$. Since a
product of an orbit and a signed orbit decomposes into signed
orbits, for decomposition of the product $O(\lambda)\otimes
O^\pm(\mu)$ into separate signed orbits it is necessary to take
dominant elements from each term of the right hand side
of~\eqref{decom-00}. That is, $O(\lambda)\otimes O^\pm(\mu)$ is a
union of the signed orbits, corresponding to points from
\begin{gather}
 D(\{ (w\lambda+w_1\mu)^{\det w_1} \,|\, w\in W/W_\lambda\}),\quad D(\{
(w\lambda+w_2\mu)^{\det w_2} \,|\, w\in W/W_\lambda\}), \quad
\dots ,
\nonumber\\
 \label{decom-11}
 D(\{ (w\lambda+w_s\mu)^{\det w_s} \,|\, w\in W/W_\lambda\}),
 \end{gather}
where $D(\{ (w\lambda+w_i\mu)^{\det w_i} \,|\, w\in
W/W_\lambda\})$ means the set of dominant signed elements in $\{
(w\lambda+w_i\mu)^{\det w_i} \,|\, w\in W/W_\lambda\}$.

 \begin{proposition}\label{prop12}
  The product $O(\lambda)\otimes O^\pm(\mu)$
consists only of signed orbits of the form
$O^\pm(|w\lambda+\mu|)$, $w\in W/W_\lambda$, where
$|w\lambda+\mu|$ is a dominant weight of the orbit containing
$w\lambda+\mu$. Moreover, each such orbit $O^\pm(|w\lambda+\mu|)$,
$w\in W/W_\lambda$, except for those of them, for which
$|w\lambda+\mu|$ lies on some wall of the dominant Weyl chamber,
belongs to the product $O(\lambda)\otimes O^\pm(\mu)$.
\end{proposition}

\begin{proof} For each dominant element $w\lambda+w_i\mu$ from
\eqref{decom-11} there exists an element $w''\in W$ such that
$w''( w\lambda+w_i\mu)=w'\lambda+\mu$. It means that
$w\lambda+w_i\mu$ is of the form $|w'\lambda+\mu|$, $w'\in
W/W_\lambda$. It is clear that a sign of this dominant element is
$\det w_iw''$. Conversely, take any element $w\lambda+\mu$, $w\in
W/W_\lambda$. It belongs (with some sign) to the product
$O(\lambda)\otimes O^\pm(\mu)$. This means that $|w\lambda+\mu|$
also belongs to this product. Therefore, the signed orbit
$O^\pm(|w\lambda+\mu|)$ is contained in $O(\lambda)\otimes
O^\pm(\mu)$ with some coef\/f\/icient if $|w\lambda+\mu|$ does not
lie on some wall of the dominant Weyl chamber.  This
coef\/f\/icient cannot vanish since if it vanishes, then in the
product $O(\lambda)\otimes O^\pm(\mu)$ there are contained points
of $O^\pm(|w\lambda+\mu|)$, but taken with an opposite signs. In
this case there exists another element $w'\lambda+\mu$ such that
$w\lambda+\mu=w''(w'\lambda+\mu)$. Since $\mu$ does not lie on a
wall, this is not possible. Proposition is proved.
\end{proof}

It follows from Proposition~\ref{prop12} that for decomposition of
the product $O(\lambda)\otimes O^\pm(\mu)$ into separate signed
orbits we have to take all elements $w\lambda+\mu$, $w\in
W/W_\lambda$, and to f\/ind the corresponding strictly dominant
elements $|w\lambda+\mu |$, $w\in W/W_\lambda$.

 \medskip

\noindent {\bf Corollary.} {\it For the product $O(\lambda)\otimes
O^\pm(\mu)$ we have
 \begin{gather} \label{decom02}
O(\lambda)\otimes O^\pm(\mu)={\bigcup}'_{w\in
W/W_\lambda}\varepsilon_w O^\pm(|w\lambda+\mu |),
 \end{gather}
where the prime means that the term with $|w\lambda+\mu |$ lying
on a wall must be omitted, and $\varepsilon_w$ is equal to $+1$ or
$-1$.}

\begin{proof} This corollary is similar to Proposition 4 in \cite{KP06}.
A proof is also similar. We only have to take into account that
signed orbits can be contained in $O(\lambda)\otimes O^\pm(\mu)$
with the sign ``$-$''. The corollary is proved.
\end{proof}

Note that in the case of product $O(\lambda)\otimes O(\mu)$,
$\lambda\in P_+$, $\mu\in P_+$, of usual orbits the orbits
$O(\nu)$ with multiplicities $m_\nu>1$ may appear in the
decomposition . As we see from Corollary, all coef\/f\/icients in
the decomposition \eqref{decom02} are modulo 1. The only problem
which appears here is to f\/ind signs of the coef\/f\/icients
$\varepsilon_\nu$.

According to Corollary we have{\samepage
\[
\phi_\lambda(x)\varphi_\mu(x)= {\sum}'_{w\in
W/W_\lambda}\varepsilon_w \varphi_{|w\lambda+\mu |}(x),
\]
where summation is such as in \eqref{decom02} and $\varepsilon_w$
are equal to $+1$ or $-1$.}

\begin{proposition}\label{prop13}
Let $O(\lambda)$, $\lambda\ne 0$, be an orbit and let $O^\pm
(\mu)$ be a signed orbit. If all elements $w\lambda+\mu$, $w\in
W/W_\lambda$, are dominant, then
\[
O(\lambda)\otimes O^\pm(\mu)={\bigcup}'_{w\in W/W_\lambda}
O^\pm(w\lambda+\mu),
\]
where the prime means that terms corresponding to $w\lambda+\mu$
lying on a wall must be omitted.
\end{proposition}

\begin{proof} The statement of this proposition follows from the above
corollary.
\end{proof}

At the end of this subsection we formulate the following method
for decomposition of products $O(\lambda)\otimes O^\pm(\mu)$,
which follows from statement of Proposition~\ref{prop12}. On the
f\/irst step we shift all points of the orbit $O(\lambda)$ by
$\mu$. As a result, we obtain the set of points $w\lambda+\mu$,
$w\in W/W_\lambda$. On the second step, we map non-dominant
elements of this set by elements of the Weyl group $W$ to the
dominant Weyl chamber. On this step we obtain the set
$|w\lambda+\mu|$, $w\in W/W_\lambda$. Then according to
Proposition~\ref{prop12}, $O(\lambda)\otimes O^\pm(\mu)$ consists
of the signed orbits $O^\pm(|w\lambda+\mu|)$ for which
$|w\lambda+\mu|$ do not lie on a wall of the dominant Weyl
chamber. On the third step, we determine signs of these orbits,
taking into account the above propositions or making a direct
calculation.

\subsection{Products of antisymmetric orbits}\label{section7.3}
Let $O^\pm(\lambda)=\{ (w\lambda)^{\det w} | w\in W\}$ and
$O^\pm(\mu)=\{ (w\mu)^{\det w} | w\in W\}$ be two signed orbits,
where $\lambda$ and $\mu$ are strictly dominant elements of $E_n$.
Then
\begin{gather}
O^\pm(\lambda)\otimes O^\pm(\mu)= \{ (w\lambda +w'\mu)^{\det ww'}
\,|\, w\in W,w'\in W\}
\notag\\
\phantom{O^\pm(\lambda)\otimes O^\pm(\mu)}{}=\{
(w\lambda{+}w_1\mu)^{\det ww_1} \,|\, w{\in} W\}\cup \{
(w\lambda{+}w_2\mu)^{\det ww_2} | w{\in} W\}\cup \cdots
 \notag\\
\phantom{O^\pm(\lambda)\otimes O^\pm(\mu)=}{} \cup \{
(w\lambda{+}w_s\mu)^{\det ww_s} \,| \,w{\in} W\},\label{decom}
 \end{gather}
where $w_1,w_2,\dots ,w_s$ is the set of all elements of $W$.
Since a product of two signed orbits decomposes into usual orbits
(all points of some of these orbits are taken with the sign
``$-$''), for decomposition of the product $O^\pm(\lambda)\otimes
O^\pm(\mu)$ into separate orbits it is necessary to take dominant
elements from each term of the right hand side of \eqref{decom}.
That is, $O^\pm(\lambda)\otimes O^\pm(\mu)$ is a union of the
orbits (some of them with the sign ``$-$'') corresponding to
points from
\begin{gather}
 D(\{ (w\lambda+w_1\mu)^{\det ww_1} \,| \,w\in W\}),\quad D(\{
(w\lambda+w_2\mu)^{\det ww_2} | w\in W\}), \quad\dots ,
\nonumber\\
 \label{decom1}
  D(\{ (w\lambda+w_s\mu)^{\det ww_s} \,| \,w\in W\}),
 \end{gather}
where $D(\{ (w\lambda+w_i\mu)^{\det ww_i} | w\in W\})$ means the
set of dominant signed elements in $\{ (w\lambda+w_i\mu)^{\det
ww_i} | w\in W\}$.

\begin{proposition}\label{prop14}
The product $O^\pm(\lambda)\otimes O^\pm(\mu)$ consists only of
orbits $($with signs ``$+$'' or ``$-$''$)$ of the form
$O(|w\lambda+\mu|)$, $w\in W$, where $|w\lambda+\mu|$ is a
dominant weight of the orbit containing $w\lambda+\mu$. Moreover,
each such orbit $O(|w\lambda+\mu|)$, $w\in W$, for which
$|w\lambda+\mu|$ does not lie on some wall of the dominant Weyl
chamber, belongs to the product $O^\pm(\lambda)\otimes
O^\pm(\mu)$.
\end{proposition}

\begin{proof} For each dominant element $w\lambda+w_i\mu$ from
\eqref{decom1} there exists an element $w''\in W$ such that $w''(
w\lambda+w_i\mu)=w'\lambda+\mu$. It means that $w\lambda+w_i\mu$
is of the form $|w'\lambda+\mu|$, $w'\in W$, and the f\/irst part
of the proposition follows.

Take any element $w\lambda+\mu$, $w\in W$, which does not lie on a
wall. Then $|w\lambda+\mu|$ does not lie on a wall. Then
$w\lambda+\mu$ belongs (with some sign) to the product
$O^\pm(\lambda)\otimes O^\pm(\mu)$. This means that
$|w\lambda+\mu|$ also belongs to this product. Therefore, the
orbit $O(|w\lambda+\mu|)$ is contained in $O^\pm(\lambda)\otimes
O^\pm(\mu)$ with some coef\/f\/icient. This coef\/f\/icient cannot
vanish since if it vanishes, then in the product $O^\pm
(\lambda)\otimes O^\pm(\mu)$ there are contained points of
$O^\pm(|w\lambda+\mu|)$, but taken with an opposite signs. In this
case there exists another element $w'\lambda+\mu$ such that
$w\lambda+\mu=w''(w'\lambda+\mu)$. Since $\mu$ does not lie on a
wall, this is not possible. Proposition is proved.
\end{proof}

It follows from Proposition~\ref{prop14} that for decomposition of
the product $O^\pm(\lambda)\otimes O^\pm(\mu)$ into orbits we have
to take all elements $w\lambda+\mu$, $w\in W$, and to f\/ind the
corresponding dominant elements $|w\lambda+\mu |$, $w\in W$.
 \medskip

\noindent {\bf Corollary.} {\it For the product
$O^\pm(\lambda)\otimes O^\pm(\mu)$ we have
 \begin{gather} \label{decom2}
O^\pm(\lambda)\otimes O^\pm(\mu)={\bigcup}_{w\in W}\varepsilon_w
O(|w\lambda+\mu |),
 \end{gather}
where $\varepsilon_w$ is equal to $+1$ or $-1$ if $|w\lambda+\mu
|$ does not lie on some wall.}
 \medskip

In the case of product $O^\pm(\lambda)\otimes O^\pm(\mu)$,
$\lambda\in P^+_+$, $\mu\in P^+_+$, of sighed orbits in the
decomposition may appear orbits $O(\nu)$ with integral
coef\/f\/icients $m_\nu$ such that $m_\nu>1$. Such
coef\/f\/icients may appear only for $\nu$ lying on a wall.

 \begin{proposition}\label{prop15}
 Let $O^\pm(\lambda)$ and  $O^\pm(\mu)$ be two
signed orbits. If all elements $w\lambda+\mu$, $w\in W$, are
dominant, then
\[
O^\pm(\lambda)\otimes O^\pm(\mu)=\bigcup_{w\in W}\varepsilon_w
O(w\lambda+\mu),
\]
where $\varepsilon_w$ is equal to $+1$ or $-1$, if $|w\lambda+\mu
|$ does not lie on some wall.
\end{proposition}

This proposition is proved in the same way as
Proposition~\ref{prop14} and we omit it.

\begin{proposition}\label{prop16}
Let $O^\pm(\lambda)$ and $O^\pm(\mu)$ be sighed orbits, and let
all elements $w\lambda+\mu$, $w\in W$, be strictly dominant (that
is, they are dominant and do not belong to any wall of the
dominant Weyl chamber). Then
 \[
O^\pm(\lambda)\otimes O^\pm(\mu)=\bigcup_{w\in W} \varepsilon_w
O(w\lambda+\mu),
 \]
where $\varepsilon_w$ are equal to $+1$ or $-1$.
\end{proposition}

This proposition is proved in the same way as Proposition~2
in~\cite{KP06}.

\subsection{Decomposition of products for rank 2}\label{section7.4}
We give here examples of decompositions of products of (signed)
orbits for the cases $A_2$ and~$C_2$. Orbits for these cases are
placed on the plane. Therefore, decompositions can be done by
geometrical calculations on this plane. These cases can be easily
considered by using for orbit points the orthogonal coordinates
from Section~\ref{section3}.  The corresponding Weyl groups have a
simple description in these coordinates and this gives a
possibility to make calculations in a simple manner.

For the case of $A_2$ at $a\ne b$ we have
\begin{alignat*}{3}
&A_2\,:\quad &O^\pm(a\ b)\otimes O(c\ 0) &= O^\pm(a{+}c\ b)\cup
O^\pm(a{-}c\ b{+}c) &&\\
        & & &\qquad \cup -O^\pm(a{+}b{-}c\ c{-}b)     \quad &&(a>c>b),\\
            &&O^\pm(a\ b)\otimes O(c\ 0) &= O^\pm(a{+}c\ b)\cup O^\pm(a{-}c\ b{+}c)&&\\
        & & &\qquad \cup O^\pm(a\ b{-}c)     \quad &&(a>c, b>c),\\
            &&O^\pm(a\ b)\otimes O(c\ 0) &= O^\pm(a{+}c\ b)&& (a=b=c),\\
            &&O^\pm(a\ b)\otimes O(c\ 0) &= O^\pm(a{+}c\ b)\cup
                  -O^\pm(a{+}b{-}c\ c{-}b)) \quad &&(a=c>b),\\
            &&O^\pm(a\ b)\otimes O(c\ 0) &= O^\pm(a{+}c\ b)\cup
                  O^\pm(a\ b{-}c)   && (b>a=c),\\
            &&O^\pm(a\ b)\otimes O(c\ 0) &= O^\pm(a{+}c\ b)\cup
                  -O^\pm(c{-}a\ a{+}b)  && (a<b=c),\\
            &&O^\pm(a\ b)\otimes O(c\ 0) &= O^\pm(a{+}c\ b)\cup -O^\pm(c{-}a\
            a{+}b)&&\\
        & & &\qquad \cup O^\pm(c{-}a{-}b\ a)     \quad &&(c>a+b),\\
            &&O^\pm(a\ b)\otimes O(c\ 0) &= O^\pm(a{+}c\ b)\cup -O^\pm(c{-}a\
            a{+}b)&&\\
        & & &\qquad \cup -O^\pm(a{+}b{-}c\ c{-}a)     \quad &&(a+b>c>b),\\
            &&O^\pm(a\ b)\otimes O(c\ 0) &= O^\pm(a{+}c\ b)\cup -O^\pm(c{-}a\
            a{+}b)&&\\
        & & &\qquad \cup O^\pm(a\ b{-}c)     \quad &&(a+b>b>c).
\end{alignat*}

If $a=b$, then we get
\begin{alignat*}{3}
&\qquad &O^\pm(a\ a)\otimes O(c\ 0) &= O^\pm(a{+}c\ a)\cup
O^\pm(c{-}2a\ a) &&\\
        & & &\qquad \cup -O^\pm(c{-}a\ 2a)     \quad &&(c>2a),\\
            &&O^\pm(a\ a)\otimes O(c\ 0) &= O^\pm(a{+}c\ a)\cup -O^\pm(2a{-}c\ c{-}a)&&\\
        & & &\qquad \cup -O^\pm(c{-}a\ 2a)     \quad &&(2a>c>a),\\
            &&O^\pm(a\ a)\otimes O(c\ 0) &= O^\pm(a{+}c\ a)\cup O^\pm(a{-}c\ a{+}c)&&\\
        & & &\qquad \cup O^\pm(a\ a{-}c)     \quad &&(a>c).
\end{alignat*}

Similar products of $C_2$ orbits are of the form
\begin{alignat*}{2}
&C_2\,:\quad &O^\pm(a\ b)\otimes O(c\ 0) &= O^\pm(a{+}c\ b)\cup
           {-}O^\pm({-}a{-}2b{+}c\ b) \cup {-}O^\pm(c{-}a\ a{+}b) \\
        & & &\qquad \cup O^\pm(c{-}2b{-}a\ a{+}b)  \qquad (a{+}b{-}c{<}b),\\
          & &O^\pm(a\ b)\otimes O(c\ 0) &= O^\pm(a{+}c\ b)\cup
           O^\pm(a{+}2b{-}c\ c{-}a{-}b) \cup O^\pm(a{-}c\ b{+}c) \\
        & & &\qquad \cup {-}O^\pm(a{+}2b{-}c\ c{-}b)  \qquad (b{>}c{-}a{-}b,a{>}c),\\
        & &O^\pm(a\ b)\otimes O(c\ 0) &= O^\pm(a{+}c\ b)\cup
           O^\pm(a{+}2b{-}c\ c{-}a{-}b) \cup {-}O^\pm(c{-}a\ a{+}b) \\
        & & &\qquad \cup {-}O^\pm(a{+}2b{-}c\ c{-}b)  \qquad (b{>}c{-}a{-}b,c{>}a),\\
        & &O^\pm(a\ b)\otimes O(c\ 0) &= O^\pm(a{+}c\ b)\cup
           O^\pm(a{-}c\ b) \cup O^\pm(a{-}c\ b{+}c) \\
        & & &\qquad \cup {-}O^\pm(a{+}2b{-}c\ c{-}b)  \qquad (a{+}b{>}b{+}c),\\
        & &O^\pm(a\ b)\otimes O(c\ 0) &= O^\pm(a{+}c\ b)\cup
           -O^\pm(c{-}a\ a{+}b{-}c) \cup {-}O^\pm(c{-}a\ a{+}b) \\
        & & &\qquad \cup {-}O^\pm(a{+}2b{-}c\ c{-}b)  \qquad (b{+}c{>}a{+}b{>}c{-}b),\\
        & &O^\pm(a\ b)\otimes O(c\ 0) &= O^\pm(a{+}c\ b)\cup
           {-}O^\pm(c{-}a\ a{+}b{-}c) \cup {-}O^\pm(c{-}a\ a{+}b) \\
        & & &\qquad \cup O^\pm(c{-}a{-}2b\ a{+}b)  \qquad (a{+}b{<}c{-}b).
\end{alignat*}

The corresponding decompositions of products of antisymmetric and
symmetric orbit functions can also be easily written down.

\section[Decomposition of antisymmetric $W$-orbit functions into
antisymmetric $W'$-orbit functions]{Decomposition of antisymmetric
$\boldsymbol{W}$-orbit functions \\ into
antisymmetric $\boldsymbol{W'}$-orbit functions}\label{section8} 

As in Section~\ref{section7}, for these decompositions it is
enough to obtain the corresponding decompositions for signed
orbits. For this reason, we shall deal mainly with signed orbits.
Our reasoning here is very similar to that of Section~4 in
\cite{KP06}.

\subsection{Introduction}\label{section8.1}

Let $R$ be a root system with a Weyl group $W$, and let $R'$ be
another root system which is a~subset of the set $R$. Then a Weyl
group $W'$ for $R'$ can be considered as a subgroup of $W$.

Let $O^\pm_W(\lambda)$ be a signed $W$-orbit. The set of points of
the usual orbit $O_W(\lambda)$ is invariant with respect to $W'$.
This means that the signed orbit $O^\pm_W(\lambda)$ consists of
signed $W'$-orbits.  In this section we deal with representing
$O^\pm_W(\lambda)$ as a union of signed $W'$-orbits. Properties of
such a~representation depend on root systems $R$ and $R'$ (or on
Weyl groups $W$ and $W'$). We distinguish two cases:
\medskip

\noindent {\bf Case 1:} {\it Root systems $R$ and $R'$ span vector
spaces of the same dimension.} In this case Weyl chambers for~$W$
are smaller than Weyl chambers for~$W'$. Moreover, each Weyl
chamber for~$W'$ consists of $|W/W'|$ chambers for~$W$. Let $D_+$
be a dominant Weyl chamber for the root system~$R$. Then a
dominant Weyl chamber for $W'$ consists of $W$-chambers  $w_iD_+$,
$i=1,2,\dots ,k$, $k=|W/W'|$, where $w_i$, $i=1,2,\dots ,k$, are
representatives of cosets in $W/W'$. If $\lambda$ does not lie on
any wall of the dominant Weyl chamber $D_+$, then
  \begin{gather} \label{Decompo}
 O^\pm_W(\lambda)=\bigcup_{i=1}^k\; (\det w_i)\;
O^\pm_{W'}(w_i\lambda),
 \end{gather}
where $O^\pm_{W'}$ are signed $W'$-orbits. (Note that if ${\rm
det}\; w_i=-1$, then $(\det w_i)\, O^\pm_{W'}(w_i\lambda)$ means
the signed orbit $O^\pm_{W'}(w_i\lambda)$ in which each point is
taken with opposite sign.)

Representing $\lambda$ by coordinates in $\omega$-basis it is
necessary to take into account that coordinates of the same point
in $\omega$-bases related to the root systems $R$ and $R'$ are
dif\/ferent. There exist matrices connecting coordinates in these
dif\/ferent $\omega$-bases (see \cite{MPS77}).

For expanding an antisymmetric $W$-orbit function into
antisymmetric $W'$-orbit functions it is necessary to take into
account the formula \eqref{Decompo}. Namely,  the following
expansion
\[
\varphi^{W}_\lambda(x)=\sum_{i=1}^k (\det
w_i)\varphi_{w_i\lambda}^{W'}(x)
\]
corresponds to the decomposition \eqref{Decompo}.

\noindent {\bf Case 2:} {\it Root systems $R$ and $R'$ span vector
spaces of different dimensions.} This case is more complicated. In
order to represent $O^\pm_W(\lambda)$ as a union of signed
$W'$-orbits, it is necessary to project points $\mu$ of
$O^\pm_W(\lambda)$ to the vector subspace $E_{n'}$ spanned by $R'$
and to select in the set of these projected points dominant points
with respect to the root system $R'$. Note that under projection,
dif\/ferent points of $O^\pm_W(\lambda)$ can give the same point
in $E_{n'}$. This leads to appearing of coinciding signed
$W'$-orbits in a representation of $O^\pm_W(\lambda)$ as a union
of $W'$-orbits. Moreover, for some signed $W'$-orbits their points
must be taken with opposite signs.

As in the previous case, under expansion of an antisymmetric
$W$-orbit function $\varphi_\lambda(x)$ into antisymmetric
$W'$-orbit functions we have to consider $\varphi_\lambda(x)$ on
the subspace $E_{n'}\subset E_n$ and to take into account the
corresponding decomposition of the signed orbit $O_W^\pm
(\lambda)$. For this reason, below in this section we consider
decomposition of signed $W$-orbits into $W'$-orbits. They uniquely
determine the corresponding expansions for antisymmetric orbit
functions.

\subsection[Decomposition of signed $W_{A_n}$-orbits into
$W_{A_{n-1}}$-orbits]{Decomposition of signed
$\boldsymbol{W_{A_n}}$-orbits into
$\boldsymbol{W_{A_{n-1}}}$-orbits}\label{section8.2}

For such decomposition it is convenient to represent orbit
elements in orthogonal coordinates (see Section~\ref{section3}).
Let $O^\pm(\lambda)\equiv O^\pm(m_1,m_2,\dots,m_{n+1})$ be a
signed $W_{A_n}$-orbit with dominant element
$\lambda=(m_1,m_2,\dots, m_{n+1})$, where
\[
m_1>m_2>\cdots >m_n>m_{n+1}.
\]
The orthogonal coordinates $m_1,m_2,\dots,m_{n+1}$ satisfy the
conditions $m_1+m_2+\cdots +m_{n+1}=0$. However, we may add to all
coordinates $m_i$ the same real number, since under this procedure
the $\omega$-coordinates $\lambda_i=m_i-m_{i+1}$, $i=1,2,\dots, n$
do not change (see Section~\ref{section3}).

The signed orbit $O^\pm(\lambda)$ consists of all points
  \begin{gather} \label{A_n-1}
w(m_1,m_2,\dots
,m_{n+1})=(m_{i_1},m_{i_2},\dots,m_{i_{n+1}}),\qquad w\in W_{A_n},
 \end{gather}
where $(i_1,i_2,\dots ,i_{n+1})$ is a permutation of the numbers
$1,2,\dots,n+1$, determined by $w$. The sign of $(\det w)$ is
attached to such point. Points of $O^\pm(\lambda)$ belong to the
vector space $E_{n+1}$. We restrict these points to the vector
subspace $E_n$, spanned by the simple roots
$\alpha_1,\alpha_2,\dots, \alpha_{n-1}$ of $A_n$, which form a set
of simple roots of $A_{n-1}$. This restriction is reduced to
removing the last coordinate $m_{i_{n+1}}$ in points
$(m_{i_1},m_{i_2},\dots,m_{i_{n+1}})$ of the signed orbit
$O^\pm(\lambda)$ (see \eqref{A_n-1}). As a~result, we obtain a set
of points
  \begin{gather} \label{A_n-2}
(m_{i_1},m_{i_2},\dots,m_{i_{n}})
  \end{gather}
received from the points \eqref{A_n-1}. The point \eqref{A_n-2} is
dominant if and only if
\[
m_{i_1}\ge m_{i_2}\ge \cdots \ge m_{i_{n}}.
\]
It is easy to see that after restriction to $E_{n}$ (that is,
under removing the last coordinate) we obtain from the set of
points \eqref{A_n-1} the following set of dominant
elements:{\samepage
\[
(m_1,\dots,m_{i-1},\hat{m_i},m_{i+1},\dots, m_{n+1}),\qquad
i=1,2,\dots,n+1,
\]
where a hat over $m_i$ means that the coordinate $m_i$ must be
omitted.}

Thus, {\it the signed $W_{A_n}$-orbit $O^\pm
(m_1,m_2,\dots,m_{n+1})$ consists of the following signed
$W_{A_{n-1}}$-orbits:
\[
O^\pm (m_1,\dots,m_{i-1},\hat{m_i},m_{i+1},\dots, m_{n+1}),\qquad
i=1,2,\dots,n+1.
\]
Each of these signed orbits must be taken with a coefficient $+1$
or $-1$. Moreover, a coefficient at the orbit $O^\pm
(m_1,\dots,m_{i-1},\hat{m_i},m_{i+1},\dots, m_{n+1})$ is $1$, if
after $\hat{m_i}$ in the point
\[
(m_1,\dots,m_{i-1},\hat{m_i},m_{i+1},\dots, m_{n+1})
\]
a  number of coordinates is even, and $-1$ otherwise.} This
statement completely determines an expansion of the antisymmetric
orbit function $\varphi^{(W_{A_n})}_{(m_1,m_2,\dots,m_{n+1})}$
into antisymmetric $W_{A_{n-1}}$-orbit function:
\[
\varphi_{(m_1,m_2,\dots,m_{n+1})}(x)= \sum_{i=1}^{n+1} (\det
w(m_i)) \varphi_{(m_1,\dots,m_{i-1},\hat{m_i},m_{i+1},\dots,
m_{n+1})}(x),
\]
where $w(m_i)$ is the permutation which sends the coordinate $m_i$
to the end, not changing an order of other coordinates.

\subsection[Decomposition of signed $W_{A_{n-1}}$-orbits
into $W_{A_{p-1}}\times W_{A_{q-1}}$-orbits, $p+q=n$]{Decomposition of signed $\boldsymbol{W_{A_{n-1}}}$-orbits\\
into $\boldsymbol{W_{A_{p-1}}\times W_{A_{q-1}}}$-orbits,
$\boldsymbol{p+q=n}$}
\label{section8.3}

Again we use orthogonal coordinates for orbit elements. We take in
the system of simple roots $\alpha_1,\alpha_2,\dots, \alpha_{n-1}$
of $A_{n-1}$ two parts as $\alpha_1,\alpha_2,\dots,\alpha_{p-1}$
and
$\alpha_{p+1},\alpha_{p+2},\dots,\alpha_{p+q-1}\equiv\alpha_{n-1}$.
The f\/irst part determines $W_{A_{p-1}}$ and the second part
generates $W_{A_{q-1}}$. We consider a signed $W_{A_{n-1}}$-orbit
$O^\pm(\lambda)$, where $\lambda=(m_1,m_2,\dots,m_n)$, $
m_1>m_2>\cdots >m_n$. This orbit consists of points
  \begin{gather} \label{A_n-3}
w(m_1,m_2,\dots ,m_{n})=(m_{i_1},m_{i_2},\dots,m_{i_{n}}),\qquad
w\in W_{A_{n-1}},
 \end{gather}
where $(i_1,i_2,\dots ,i_{n})$ is a permutation of the numbers
$1,2,\dots,n$. We restrict points \eqref{A_n-3} to the vector
subspace $E_{p}\times E_{q}$ spanned by the simple roots
$\alpha_1,\alpha_2,\dots,\alpha_{p-1}$ and
$\alpha_{p+1},\alpha_{p+2},\dots,\alpha_{n-1}$, respectively.
Under restriction the point \eqref{A_n-3} turns into the point
\[
(m_{i_1},m_{i_2},\dots,m_{i_{p}})(m_{i_{p+1}},m_{i_{p+1}},\dots,m_{i_{n}}).
\]

In order to determine a set of signed $W_{A_{p-1}}\times
W_{A_{q-1}}$-orbits contained in the signed orbit~$O^\pm(\lambda)$
we have to choose from \eqref{A_n-3} all elements for which
\[
m_{i_1}>m_{i_2}>\cdots > m_{i_{p}}, \qquad
m_{i_{p+1}}>m_{i_{p+2}}>\cdots >m_{i_{n}}.
\]
To f\/ind this set of points we have to take all subsets
$m_{i_1},m_{i_2},\dots,m_{i_{p}}$ in the set $m_1,m_2,\dots,m_n$
such that $m_{i_1}>m_{i_2}>\cdots
>m_{i_{p}}$. Let $\Sigma$ denote the collection of such subsets.
Then $O^\pm(\lambda)$ {\it consists of signed $W_{A_{p-1}}\times
W_{A_{q-1}}$-orbits
 \begin{gather}\label{A_n -- A_p}
O^\pm((m_{i_1},m_{i_2},\dots,m_{i_{p}})
(m_{j_1},m_{j_2},\dots,m_{j_{q}})),\qquad
(m_{i_1},m_{i_2},\dots,m_{i_{p}})\in \Sigma,
  \end{gather}
where $(m_{j_1},m_{j_2},\dots,m_{j_{q}})$ is a supplement of the
subset $(m_{i_1},m_{i_2},\dots,m_{i_{p}})$ in the whole set
$(m_1,m_2,\dots,m_n)$, taken in such an order that}
$m_{j_1}>m_{j_2}>\cdots >m_{j_{q}}$. Each of these
$W_{A_{p-1}}\times W_{A_{q-1}}$-orbits is contained in $O^\pm
(\lambda)$ only once. Each such a signed orbit is contained in the
signed orbit $O^\pm(\lambda)$ with sign ``$+$'' if the set of
numbers
\[
(m_{i_1},m_{i_2},\dots,m_{i_{p}}, m_{j_1},m_{j_2},\dots,m_{j_{q}})
\]
from \eqref{A_n -- A_p} is obtained from the set $(1,2,\dots,n)$
by an even permutation and with sign ``$-$'' otherwise. The
corresponding expansions of antisymmetric orbit functions
$\varphi_\lambda^{(W_{A_{n-1}})}(x)$ into antisymmetric
$W_{A_{p-1}}\times W_{A_{q-1}}$-orbit functions is evident.

\subsection[Decomposition of signed $W_{B_{n}}$-orbits into
$W_{B_{n-1}}$-orbits and of signed $W_{C_{n}}$-orbits into
$W_{C_{n-1}}$-orbits]{Decomposition of signed
$\boldsymbol{W_{B_{n}}}$-orbits into
$\boldsymbol{W_{B_{n-1}}}$-orbits\\ and of signed
$\boldsymbol{W_{C_{n}}}$-orbits into
$\boldsymbol{W_{C_{n-1}}}$-orbits}\label{section8.4}

Decomposition of signed $W_{B_{n}}$-orbits and decomposition of
signed $W_{C_{n}}$-orbits are fulf\/illed in the same way. For
this reason, we give a proof only for the case of signed
$W_{C_{n}}$-orbits.

A set of simple roots of $C_n$ consists of roots
$\alpha_1,\alpha_2,\dots,\alpha_n$. The roots
$\alpha_2,\dots,\alpha_n$ constitute a set of simple roots of
$C_{n-1}$. They span the subspace $E_{n-1}$.

For determining elements $\lambda$ of $E_n$ we use orthogonal
coordinates $m_1,m_2,\dots,m_n$. Then $\lambda$ is strictly
dominant if and only if
\[
m_1>m_2>\cdots >m_n>0.
\]
Then the signed orbit $O^\pm(\lambda)$ consists of all points
  \begin{gather} \label{C_n-1}
w(m_1,m_2,\dots ,m_{n})=(\pm m_{i_1},\pm m_{i_2},\dots,\pm
m_{i_{n}}),\qquad w\in W_{C_n},
 \end{gather}
where $(i_1,i_2,\dots ,i_{n})$ is a permutation of the set
$1,2,\dots,n $, and all combinations of signs are possible.

Restriction of elements \eqref{C_n-1} to the vector subspace
$E_{n-1}$, def\/ined above, reduces to removing  the f\/irst
coordinate $\pm m_{i_1}$ in \eqref{C_n-1}. As a result, we obtain
from the set of points \eqref{C_n-1} the collection
\[
(\pm m_{i_2},\pm m_{i_3},\dots,\pm m_{i_{n}}),\qquad w\in W_{C_n}.
\]
Only the points $(m_{i_2}, m_{i_3},\dots,m_{i_{n-1}}, m_{i_{n}})$
with positive coordinates may be dominant. Moreover, such a point
is dominant if and only if
\[
 m_{i_2}> m_{i_3}>\cdots > m_{i_{n}}.
\]
Therefore, under restriction of points \eqref{C_n-1} to $E_{n-1}$
we obtain the following strictly $W_{C_{n-1}}$-dominant elements:
  \begin{gather} \label{C_n-2}
(m_1,m_2,\dots ,m_{i-1},\hat{m_i},m_{i+1},\dots ,m_{n}),\qquad
i=1,2,\dots,n,
  \end{gather}
where a hat over $m_i$ means that the coordinate $m_i$ must be
omitted. Moreover, the element \eqref{C_n-2} with f\/ixed $i$ can
be obtained from two elements in \eqref{C_n-1}, namely, from
$(m_1,m_2,\dots ,m_{i-1},\pm m_i$, $m_{i+1},\dots ,m_{n})$. In the
signed orbit $O^\pm (m_1,m_2,\dots,m_n)$ these two elements have
opposite signs.

Thus, the signed $W_{C_n}$-orbits $O^\pm(m_1,m_2,\dots,m_n)$
consists of the following signed $W_{C_{n-1}}$-orbits:
\begin{gather*}
O^\pm(m_1,m_2,\dots ,m_{i-1},\hat{m_i},m_{i+1},\dots
,m_{n}),\qquad i=1,2,\dots,n.
\end{gather*}
Each such signed $W_{C_{n-1}}$-orbit is contained in $O^\pm
(m_1,m_2,\dots,m_n)$ twice with opposite signs.

Therefore, {\it a restriction of the antisymmetric orbit function
$\varphi_{(m_1,m_2,\dots,m_n)}$ to the sub\-spa\-ce~$E_{n-1}$,
described above, vanishes.}

For $W_{B_n}$-orbits we have similar assertions. A signed
$W_{B_n}$-orbits $O^\pm (m_1,m_2,\dots,m_n)$, $m_1>m_2>\cdots
>m_n>0$, consists of $W_{B_{n-1}}$-orbits
\[
O^\pm (m_1,m_2,\dots ,m_{i-1},\hat{m_i},m_{i+1},\dots
,m_{n}),\qquad i=1,2,\dots,n,
\]
and each such orbit is contained in the decomposition two times
(with opposite signs), that is, {\it a restriction of the
antisymmetric orbit function $\varphi_{(m_1,m_2,\dots,m_n)}$ of
$B_n$ to the subspace $E_{n-1}$, described above, vanishes.}

\subsection[Decomposition of signed $W_{C_{n}}$-orbits into
$W_{A_{p-1}}\times W_{C_{q}}$-orbits, $p+q=n$]{Decomposition of
signed $\boldsymbol{W_{C_{n}}}$-orbits into
$\boldsymbol{W_{A_{p-1}}\times W_{C_{q}}}$-orbits,
$\boldsymbol{p+q=n}$}
\label{section8.5}

If $\alpha_1,\alpha_2,\dots,\alpha_n$ are simple roots for $C_n$,
then $\alpha_1,\alpha_2,\dots,\alpha_{p-1}$ are simple roots for
$A_{p-1}$ (they can be embedded into the linear subspace $E_{p}$)
and $\alpha_{p+1},\alpha_{p+2},\dots,\alpha_n$ are simple roots
for $C_q$ (they generate the linear subspace $E_{q}$).

We use orthogonal coordinates for elements of $E_n$ and consider a
signed $W_{C_{n}}$-orbit $O^\pm(\lambda)$, $m_1>m_2>\cdots
>m_n>0$. This orbit consists of all points
\eqref{C_n-1}. Restriction of these points to the vector subspace
$E_{p}\times E_q$ reduces to splitting of coordinates
\eqref{C_n-1} into two parts:
  \begin{gather} \label{C_n-3}
(\pm m_{i_1},\pm m_{i_2},\dots ,\pm m_{i_p})(\pm m_{i_{p+1}},\dots
,\pm m_{i_n}) .
  \end{gather}
Due to the condition $m_1>m_2>\cdots >m_n>0$, these elements do
not lie on walls of the $W_{A_{p-1}}\times W_{C_{q}}$-chambers. We
have to choose dominant elements (with respect to the Weyl group
$W_{A_{p-1}}\times W_{C_{q}}$) in the set of points \eqref{C_n-3}.
The conditions of dominantness for elements of $E_{p}$ and~$E_q$
show that only the elements
\[
( m_{i_1},\dots, m_{i_j},-m_{i_{j+1}} \cdots ,-m_{i_p})(
m_{i_{p+1}},\dots , m_{i_n}) , \qquad j=0,1,2,\dots, p,
\]
satisfying the conditions
\[
 m_{i_1}>m_{i_2}>\cdots > m_{i_j},\qquad m_{i_{j+1}}<m_{i_{j+2}}<
 \cdots <m_{i_p}, \qquad m_{i_{p+1}}> m_{i_{p+2}}>\cdots >m_{i_n},
 \]
are dominant. Moreover, each such point is contained in the signed
$W_{C_n}$-orbit $O^\pm(\lambda)$ only once. This assertion
completely determines a list of signed $W_{A_{p-1}}\times
W_{C_q}$-orbits in $O^\pm(\lambda)$. Each signed
$W_{A_{p-1}}\times W_{C_q}$-orbit is contained in $O^\pm
(\lambda)$ only once.

This assertion uniquely determines a list of antisymmetric
$W_{A_{p-1}}\times W_{C_q}$-orbit functions, contained in the
antisymmetric $W_{C_n}$-orbit function $\varphi_\lambda$. However,
it is necessary to determine signs of antisymmetric
$W_{A_{p-1}}\times W_{C_q}$-orbit functions in the decomposition.
It is easily made by using the description of the Weyl groups in
the Euclidean space $E_n$ with orthogonal coordinates.

\subsection[Decomposition of signed $W_{D_{n}}$-orbits into signed
$W_{D_{n-1}}$-orbits]{Decomposition of signed
$\boldsymbol{W_{D_{n}}}$-orbits into signed
$\boldsymbol{W_{D_{n-1}}}$-orbits}\label{section8.6}
Assume that $\alpha_1,\alpha_2,\dots,\alpha_n$ is the set of
simple roots of $D_n$, $n>4$. Then $\alpha_2,\dots,\alpha_n$ are
simple roots of $D_{n-1}$. The last roots span the subspace
$E_{n-1}$.

For elements $\lambda$ of $E_n$ we again use orthogonal
coordinates $m_1,m_2,\dots,m_n$. Then $\lambda$ is strictly
dominant if and only if $m_1> m_2> \cdots > m_{n-1}> |m_n|$. We
assume that $\lambda$ satisf\/ies the condition
\[
m_1>m_2>\cdots >m_n>0.
\]
Then the signed orbit $O^\pm(\lambda)$ consists of all points
  \begin{gather} \label{D_n-1}
w(m_1,m_2,\dots ,m_{n})=(\pm m_{i_1},\pm m_{i_2},\dots,\pm
m_{i_{n}}),\qquad w\in W_{D_n},
 \end{gather}
where $(i_1,i_2,\dots ,i_{n})$ is a permutation of the numbers
$1,2,\dots,n $ and there exists an even number of signs ``$-$''.
Restriction of elements \eqref{D_n-1} to the subspace $E_{n-1}$
reduces to removing the f\/irst coordinate $\pm m_{i_1}$ in
\eqref{D_n-1}. As a result, we obtain from the set of points
\eqref{D_n-1} the collection
\[
(\pm m_{i_2},\pm m_{i_3},\dots,\pm m_{i_{n}}),\qquad w\in W_{D_n},
\]
where a number of signs ``$-$'' may be either even or odd. Only
points of the form $(m_{i_2}, m_{i_3},\dots$, $m_{i_{n-1}},\pm
m_{i_{n}})$ may be dominant. Moreover, such a point is dominant if
and only if
\[
 m_{i_2}> m_{i_3}>\cdots >m_{i_{n-1}}> |m_{i_{n}}|.
\]
Therefore, under restriction of points \eqref{D_n-1} to $E_{n-1}$
we obtain the following $W_{D_{n-1}}$-dominant elements:
  \begin{gather} \label{D_n-2}
(m_1,m_2,\dots ,m_{i-1},\hat{m_i},m_{i+1},\dots ,m_{n-1},\pm
m_{n}),\qquad i=1,2,\dots,n,
  \end{gather}
where a hat over $m_i$ means that the coordinate $m_i$ must be
omitted. Moreover, the ele\-ment~\eqref{D_n-2} with f\/ixed $i$
can be obtained only from one element in \eqref{D_n-1}, namely,
from element $(m_1,m_2,\dots ,m_{i-1},\pm m_i,m_{i+1},\dots ,\pm
m_{n})$, where at $m_i$ and $m_n$ signs are coinciding.

Thus, {\it the signed $W_{D_n}$-orbit $O^\pm(m_1,m_2,\dots,m_n)$
with $m_1>m_2>\cdots >m_{n}>0$ consists of the following signed
$W_{D_{n-1}}$-orbits:
\[
O^\pm(m_1,m_2,\dots ,m_{i-1},\hat{m_i},m_{i+1},\dots ,\pm
m_{n}),\qquad i=1,2,\dots,n.
\]
Each such signed $W_{D_{n-1}}$-orbit is contained in
$O^\pm(m_1,m_2,\dots,m_n)$ only once $($with sign ``$+$'' or
``$-$''$)$.} A sign of such an orbit depends on a number $i$ and
on a sign at $m_n$. This sign is uniquely determined by the sign
$(\det w)$ of the corresponding element $w$ of $W_{D_n}$.

It is shown similarly that {\it the signed $W_{D_n}$-orbit
\[
O^\pm(m_1,\dots,m_{n-1},-m_n), \qquad m_1>m_2>\dots
>m_{n}>0
\]
consists of the same signed $W_{D_{n-1}}$-orbits as the
$W_{D_n}$-orbits $O^\pm(m_1,\dots,m_{n-1},m_n)$ with the same
numbers $m_1,\dots,m_{n-1},m_n$ does.}

The above assertions uniquely determine expansions for the
corresponding antisymmetric $W_{D_{n}}$-orbit functions.

\subsection[Decomposition of signed $W_{D_{n}}$-orbits into
$W_{A_{p-1}}\times W_{D_{q}}$-orbits, $p+q=n$, $q\ge
4$]{Decomposition of signed $\boldsymbol{W_{D_{n}}}$-orbits into
$\boldsymbol{W_{A_{p-1}}\times W_{D_{q}}}$-orbits,\\
$\boldsymbol{p+q=n}$, $\boldsymbol{q\ge 4}$}
\label{section8.7}
If $\alpha_1,\alpha_2,\dots,\alpha_n$ are simple roots for $D_n$,
then $\alpha_1,\alpha_2,\dots,\alpha_{p-1}$ are simple roots for
$A_{p-1}$ (they can be embedded into the Euclidean subspace
$E_{p}$) and $\alpha_{p+1},\alpha_{p+2},\dots,\alpha_n$ are simple
roots for~$D_q$ (they generate the Euclidean subspace $E_{q}$).

We use orthogonal coordinates in $E_n$ and consider signed
$W_{D_{n}}$-orbits $O^{\pm}(\lambda)$ with
$\lambda=(m_1,m_2,\dots,m_n)$ such that $m_1>m_2>\cdots >m_n>0$.
The orbit $O(\lambda)$ consists of all points~\eqref{D_n-1}.
Restriction of these points to the vector subspace $E_{p-1}\times
E_q$ reduces to splitting the set of coor\-dinates \eqref{D_n-1}
into two parts:
  \begin{gather} \label{D_n-3}
(\pm m_{i_1},\pm m_{i_2},\dots ,\pm m_{i_p})(\pm m_{i_{p+1}},\dots
,\pm m_{i_n}) .
  \end{gather}
Due to the condition $m_1>m_2>\cdots >m_n>0$, these elements do
not lie on walls of the  $W_{A_{p-1}}\times W_{D_{q}}$-chambers.
We have to choose dominant elements (with respect to the Weyl
group $W_{A_{p-1}}\times W_{D_{q}}$) in the set of points
\eqref{D_n-3}. The conditions of dominantness for elements of
$E_{p}$ and $E_q$ show that only the elements
\[
( m_{i_1},\dots, m_{i_j},-m_{i_{j+1}} \dots ,-m_{i_p})(
m_{i_{p+1}},\dots ,\pm m_{i_n}) , \qquad j=0,1,2,\dots, p,
\]
having even number of sign minus and satisfying the conditions
\[
 m_{i_1}>m_{i_2}>\cdots > m_{i_j},\qquad m_{i_{j+1}}<m_{i_{j+2}}<
 \cdots <m_{i_p}, \qquad m_{i_{p+1}}> m_{i_{p+2}}>\cdots >m_{i_n},
 \]
are dominant. Moreover, each such point is contained in the
$W_{D_n}$-orbit $O^\pm (\lambda)$ only once. These assertions
completely determine a list of signed $W_{A_{p-1}}\times
W_{D_{q}}$-orbits in the signed $W_{D_n}$-orbit $O^\pm(\lambda)$.
All signed $W_{A_{p-1}}\times W_{D_{q}}$-orbits are contained in
$O^\pm(\lambda)$ with multiplicity 1 and with sign ``$+$'' or
``$-$''. This determines uniquely expansions for the corresponding
antisymmetric $W_{D_{n}}$-orbit functions.

\section[Characters of representations and antisymmetric orbit functions]{Characters
of representations and antisymmetric\\ orbit
functions}\label{section9}

Antisymmetric orbit functions $\varphi_\lambda(x)$ with
$\lambda\in P^+_+$ are closely related to characters of
irreducible representations of the corresponding compact Lie group
$G$. This relation serves for derivations of some properties of
antisymmetric orbit functions.

\subsection{Connection of characters with orbit functions}\label{section9.1}
{\sloppy To each Coxeter--Dynkin diagram there corresponds a connected
compact semisimple Lie group~$G$. Let us f\/ix
a Coxeter--Dynkin diagram and, therefore, a connected compact Lie
group~$G$. A comp\-lex valued function $f(g)$ on $G$  satisfying
the condition
\[
f(g)=f(hgh^{-1}), \qquad h\in G,
\]
is called a {\it class function}. It is constant on classes of
conjugate elements.}

For simplicity, we assume that $G$ is realized by matrices such
that the set of its diagonal matrices constitute a Cartan
subgroup, which will be denoted by $H$. This subgroup can be
identif\/ied with the $n$-dimensional torus ${\sf T}$, where $n$
is a rank of the group $G$. The subgroup $H$ can be represented as
$H=\exp ({\rm i}{\mathfrak h})$, where ${\mathfrak h}$ is a real
form of an appropriate Cartan subalgebra of the complex semisimple
Lie algebra, determined by the Coxeter--Dynkin diagram.

It is well-known that each element $g$ of $G$ is conjugate to some
element of $H$, that is, class functions are uniquely determined
by their values on $H$.

There exists a one-to-one correspondence between irreducible
unitary representations of the group $G$ and integral highest
weights $\lambda\in P_+$, where $P_+$ is determined by the
Coxeter--Dynkin diagram. The irreducible representation,
corresponding to a highest weight $\lambda$, will be denoted by
$T_\lambda$. The representation $T_\lambda$ and its properties are
determined by its character $\chi_\lambda(g)$, which is def\/ined
as the trace of $T_\lambda(g)$:
\[
\chi_\lambda(g)={\rm Tr}\; T_\lambda(g),\qquad g\in G.
\]
Since ${\rm Tr}\; T_\lambda(hgh^{-1})={\rm Tr}\; T_\lambda(g)$,
$h\in G$, the character $\chi_\lambda$ is a class function, that
is, it is uniquely determined by its values on the subgroup $H$.

All the operators $T_\lambda(h)$, $h\in H$, are diagonal with
respect to an appropriate basis of the representation space (this
basis is called a weight basis) and their diagonal matrix elements
are of the form $e^{2\pi {\rm i}\langle \mu,x \rangle}$, where
$\mu\in P$ is a weight of the representation $T_\lambda$,
$x=(x_1,x_2,\dots ,x_n)$ are coordinates of an element $t$ of the
Cartan subalgebra ${\mathfrak h}$ in an appropriate coordinate
system (that is, coordinates on the torus ${\sf T}$) and $\langle
\cdot ,\cdot \rangle$ is an appropriate bilinear form, which can
be chosen coinciding with the scalar product on $E_n$, considered
above. Then the character $\chi_\lambda(h)$ is a linear
combination of the diagonal matrix elements:
 \begin{gather}\label{trace}
 \chi_\lambda(h)=\sum_{\mu\in D(\lambda)} c_\lambda^\mu
e^{2\pi {\rm i}\langle \mu,x \rangle} ,
 \end{gather}
where $D(\lambda)$ is the set of all weights of the irreducible
representation $T_\lambda$ and $c_\lambda^\mu$ is a multiplicity
of the weight $\mu\in D(\lambda)$ in the representation
$T_\lambda$. It is known from representation theory that the
weight system $D(\lambda)$ of $T_\lambda$ is invariant with
respect to the Weyl group $W$, corresponding to the
Coxeter--Dynkin diagram, and $c_\lambda^{w\mu} =c_\lambda^\mu$,
$w\in W$, for each $\mu\in D(\lambda)$. This means that the
character $\chi_\lambda(h)$ can be represented as
 \begin{gather}\label{char}
 \chi_\lambda(h)=\sum_{\mu\in D_+(\lambda)}
c_\lambda^\mu \phi_\mu(x) ,
 \end{gather}
where $D_+(\lambda)$ is the set of all dominant weights in
$D(\lambda)$ and $\phi_\mu(x)$ is a symmetric  orbit function,
corresponding to the weight $\mu\in D_+(\lambda)$. Representing
$\chi_\lambda(h)$ as $\chi_\lambda(x)$, where $x=(x_1,x_2,\dots
,x_n)$ are coordinates corresponding to the element $t\in
{\mathfrak h}$ such that $h=\exp 2\pi{\rm i}t$, we can make an
analytic continuation of both sides of \eqref{char} to the
$n$-dimensional Euclidean space $E_n$. Since the right hand side
of \eqref{char} is invariant under transformations from the
af\/f\/ine Weyl group $W^{\rm aff}$, corresponding to the Weyl
group $W$, the function $\chi_\lambda(x)$ is also invariant under
the af\/f\/ine Weyl group $W^{\rm aff}$. That is, it is enough to
def\/ine $\chi_\lambda(x)$ only on the fundamental domain $F$ of
the group $W^{\rm aff}$. To this fundamental domain $F$ there
corresponds a fundamental domain (we denote it by $\tilde
F$) in the subgroup $H$ (and on the torus ${\sf T}$).

Many properties of orbit functions follow from properties of
characters $\chi_\lambda$, and we consider them   as known from
representation theory.

The well-known Weyl formula for characters of irreducible
representations of the group $G$ states \cite{Z} that under
appropriate selection of coordinates $(x_1,x_1,\dots,x_n)$ of
points $h\in H$ we have
 \begin{gather}\label{char-W}
 \chi_\lambda(h)=\frac{\sum\limits_{w\in W} (\det w)e^{2\pi{\rm i}\langle
 \lambda+\rho,x \rangle}}{\sum\limits_{w\in W} (\det w)e^{2\pi{\rm i}\langle
 \rho,x \rangle}}=\frac{\varphi_{\lambda+\rho}(x)}{\varphi_{\rho}(x)} ,
 \end{gather}
where, as before, $\rho=\frac 12 \sum\limits_{\alpha>0}\alpha$.
This is why antisymmetric  orbit functions are so important for
representation theory. Note that for dominant $\lambda\in P_+$ the
element $\lambda+\rho$ is strictly dominant; for this reason, the
antisymmetric  orbit function $\varphi_{\lambda+\rho}(x)$ in
\eqref{char-W} does not vanish for $\lambda\in P_+$.

\subsection{Orthogonality of characters}\label{section9.2}
The relation \eqref{char-W} between characters of irreducible
representations of compact Lie groups and the corresponding
antisymmetric orbit functions leads to a simple proof of
orthogonality of irreducible characters $\chi_\lambda(h)\equiv
\chi_\lambda(x)$. Indeed, due to the orthogonality
\eqref{ortog-tor} for antisymmetric orbit functions and to the
relation \eqref{char-W} we have
\[
 \int_{F} \varphi_{\lambda+\rho}(x)
\overline{ \varphi_{\lambda'+\rho}(x)} dx = \int_{F}
\chi_\lambda(x) \chi_{\lambda'}(x) |\varphi_{\rho}(x)|^2 dx
=\delta_{\lambda\lambda'},
\]
that is, irreducible characters are orthogonal on $F$ with respect
to the measure $|\varphi_{\rho}(x)|^2 dx$. The expressions for the
function $\varphi_{\rho}(x)$ for the classical compact Lie groups
are given in Subsection~\ref{section5.9}.

\subsection{Relations for antisymmetric orbit functions}\label{section9.3}
The formulas \eqref{char} and \eqref{char-W} are a source of
relations for antisymmetric orbit functions.

Comparing the expressions \eqref{char} and \eqref{char-W} for
characters, one gets the relation
 \begin{gather}\label{rela-1}
\varphi_{\lambda+\rho}(x)=\sum_{\mu\in D_+(\lambda)} c_\lambda^\mu
\phi_{\mu}(x)  \varphi_{\rho}(x),
 \end{gather}
where, as before, $c_\lambda^\mu$ are multiplicities of weights
$\mu$ in the irreducible representation $T_\lambda$ of $G$.

Let $T_\lambda$ and $T_\mu$ be two irreducible representations of
$G$. Then their tensor product decomposes as a direct sum of
irreducible representations of $G$ as
 \begin{gather}\label{rela-2}
T_\lambda\otimes T_\mu=\sum_{\nu\in P_+} m^{\lambda\mu}_{\nu}
T_\nu,
 \end{gather}
where $m^{\lambda\mu}_{\nu}$ is a multiplicity of the irreducible
representation $T_\nu$ in the tensor product. Since $T_\lambda$
and $T_\mu$ are f\/inite dimensional representations, the sum on
the right hand side of \eqref{rela-2} is f\/inite. To the
decomposition \eqref{rela-2} there corresponds
 a relation for characters,
 \begin{gather}\label{rela-3}
\chi_\lambda(x) \chi_\mu(x)=\sum_{\nu\in P_+} m^{\lambda\mu}_{\nu}
\chi_\nu(x).
 \end{gather}
Due to \eqref{char-W} it can be written as
\[
\frac{\varphi_{\lambda+\rho}(x)}{\varphi_\rho(x)}
\frac{\varphi_{\mu+\rho}(x)}{\varphi_\rho(x)}= \sum_{\nu\in P_+}
m^\lambda_{\mu\nu} \frac{\varphi_{\nu+\rho}(x)}{\varphi_\rho(x)},
\]
where summation is such as in \eqref{rela-3}. Therefore, we have
the following expression for a product of two antisymmetric  orbit
functions:
 \begin{gather}\label{rela-4}
\varphi_{\lambda+\rho}(x)\varphi_{\mu+\rho}(x)=  \sum_{\nu\in P_+}
m^{\lambda\mu}_{\nu} \varphi_{\nu+\rho}(x)\varphi_\rho(x).
 \end{gather}
In particular, if $\mu=\omega_i$, where $\omega_i$ is $i$-th
fundamental wight of the group $G$, then this formula takes the
form
\[
\varphi_{\lambda+\rho}(x)\varphi_{\omega_i+\rho}(x)=  \sum_{\nu\in
P_+} m^{\lambda\omega_i}_\nu \varphi_{\nu+\rho}(x)\varphi_\rho(x).
\]
Since multiplicities of irreducible constituents in the tensor
product $T_\lambda\otimes T_{\omega_i}$ for many groups can be
found in a simple way (in many cases these multiplicities are
equal to 1; for example, if $i=1$, then for the groups $SU(n)$ and
$SO(n)$ they do not exceed 1), then this formula can be considered
as a recurrence relation for antisymmetric  orbit functions.

\section{Antisymmetric orbit function transforms}\label{section10}

As in the case of symmetric orbit functions, antisymmetric orbit
functions determine certain orbit function transforms which
generalize the sine transform (in the case of symmetric orbit
functions these transforms generalize the cosine transform)
\cite{P-SIG-05,MP06}.

As in the case of symmetric orbit functions, antisymmetric orbit
functions determine three types of orbit function transforms: the
f\/irst one is related to the antisymmetric orbit
functions~$\varphi_\lambda(x)$ with integral $\lambda$, the second
one is related to $\varphi_\lambda(x)$ with dominant $\lambda\in
E_n$, and the third one is the related discrete transforms.

\subsection[Decomposition in antisymmetric orbit functions on $F$]{Decomposition in antisymmetric orbit
functions on $\boldsymbol{F}$}\label{section10.1}
Let $f(g)$ be a continuous class function on $G$ (see
Subsection~\ref{section9.1}). It def\/ines a continuous function
on the commutative subgroup $H$. We assume that this function on
$H$ has continuous partial derivatives of all orders with respect
to analytic parameters on $H$. Such function $f$ can be decomposed
in characters of irreducible unitary representations of $G$:
 \begin{gather}\label{decomp}
f(h)=\sum_{\lambda\in P_+} c_\lambda \chi_\lambda(h).
 \end{gather}
We see from this decomposition that each class function is
symmetric with respect to the corresponding af\/f\/ine Weyl group
$W^{\rm aff}$ (since characters $\chi_\lambda$ admit this
symmetry) and, therefore, is uniquely determined by its values on
the fundamental domain $\tilde F$ in $H$.

Going to the coordinate description of the points $h\in H$ (see
Subsection~\ref{section9.1}), we obtain
\[
f(x)=\sum_{\lambda\in P_+} c_\lambda \chi_\lambda(x),\qquad x\in
E_n.
\]
Taking into account formula \eqref{char-W} for the characters we
receive
\begin{gather}\label{decomp-coord}
\varphi_\rho(x)f(x)=\sum_{\lambda\in P_+} c_\lambda
\varphi_{\lambda+\rho}(x).
 \end{gather}
Due to Proposition~\ref{prop7}, we may state that any
antisymmetric (with respect to the af\/f\/ine Weyl group $W^{\rm
aff}$) continuous function $f$ on $E_n$, which has continuous
derivatives and vanishes on the boundary $\partial F$ of the
fundamental domain $F$, can be represented in the form
$f(x)=\varphi_\rho(x)\tilde f(x)$, where $\tilde f(x)$ is a
symmetric (with respect to $W^{\rm aff}$) continuous function on
$E_n$ with continuous derivatives. Thus, due to
\eqref{decomp-coord} we may state that {\it any antisymmetric
(with respect to $W^{\rm aff}$) continuous function $f$ on $E_n$,
which has continuous derivatives and vanishes on the
boundary~$\partial F$, can be decomposed in antisymmetric orbit
functions} $\varphi_{\lambda}$, $\lambda\in P^+_+$,
 \begin{gather}\label{decom-f1}
f(x)=\sum_{\lambda\in P^+_+} c_\lambda \varphi_{\lambda}(x).
 \end{gather}

By the orthogonality relation \eqref{ortog-tor} for antisymmetric
orbit functions, the coef\/f\/icients $c_\lambda$ in this
decomposition are determined by the formula
 \begin{gather}\label{decom-2}
c_\lambda =\int_F f(x)\overline{\varphi_\lambda(x)}dx .
 \end{gather}
Moreover, the Plancherel formula{\samepage
 \begin{gather}\label{decom-3}
\sum_{\lambda\in P_+}  |c_\lambda|^2=  \int_F |f(x)|^2dx
 \end{gather}
holds, which means that the Hilbert spaces with the appropriate
scalar products are isometric.}

Formula \eqref{decom-2} is the antisymmetrized Fourier transform
of the function $f(x)$. Formula~\eqref{decom-f1} gives an inverse
transform. Formulas \eqref{decom-f1} and \eqref{decom-2} give the
{\it orbit function transforms} corresponding to antisymmetric
orbit functions $\varphi_\lambda$, $\lambda\in P^+_+$.

Let ${\mathcal L}_0^2(F)$ denote the Hilbert space of functions on
the fundamental domain $F$, which vanish on the boundary $\partial
F$ of the fundamental domain, with the scalar product
\[
\langle f_1,f_2\rangle = \int_F f_1(x)\overline{f_2(x)} dx.
\]
The set of continuous functions on $F$ (vanishing on the boundary
$\partial F$) with continuous derivatives is dense in ${\mathcal
L}_0^2(F)$. Therefore, the formulas
\eqref{decom-f1}--\eqref{decom-3} show that {\it the set of orbit
functions $\varphi_\lambda$, $\lambda\in P^+_+$, form an
orthogonal basis of ${\mathcal L}_0^2(F)$.}

\subsection{Symmetric and antisymmetric multivariate sine and cosine series}\label{section10.2}
Symmetric and antisymmetric orbit functions for the
Coxeter--Dynkin diagram $C_n$ can be expressed in terns of
symmetric and antisymmetric multivariate sine and cosine functions
(see formulas \eqref{det-C} and \eqref{symm-cos}). Their
application in the formulas for symmetric and antisymmetric orbit
function transforms gives antisymmetric and symmetric multivariate
sine and cosine series expansions.

The formulas \eqref{decom-f1} and \eqref{decom-2}, applied to the
case $C_n$, determine expansions of functions, given on the
fundamental domain
\[
 F=\{ 1/2>x_1>x_2>\cdots >x_n>0\}
\]
for the Coxeter--Dynkin diagram $C_n$, into antisymmetric
multivariate sine functions:
 \begin{gather}\label{decom-anti-s}
f(x)=\sum_{{\bf m}\in P^+_+} c_{\bf m} \det \left( \sin 2\pi
m_ix_j\right)_{i,j=1}^n ,
 \end{gather}
where ${\bf m}=(m_1,m_2,\dots,m_n)$ are {\it integer} $n$-tuples
such that $m_1>m_2>\cdots >m_n>0$, and the coef\/f\/icients
$c_{\bf m}$ are given by the formula
 \begin{gather}\label{decom-anti-s-}
 c_{\bf m}=2^{2n} \int_F f(x)
\det \left( \sin 2\pi m_ix_j\right)_{i,j=1}^n dx.
 \end{gather}
The Plancherel formula is of the form
\[
 \sum_{{\bf m}\in P^+_+} |c_{\bf m}|^2=2^{2n} \int_F f(x)
|f(x)|^2 dx.
\]

Similarly, using the symmetric orbit function transform on the
fundamental domain $F$ (see Subsection~8.2 in \cite{KP06}),
determined by symmetric orbit functions for the Coxeter--Dynkin
diag\-ram~$C_n$ given by the formulas \eqref{det-B-symm} and
\eqref{symm-cos}, one obtains symmetric multivariate cosine
expansion on $F$:
\begin{gather}\label{decom-anti-c}
f(x)=\sum_{{\bf m}\in P_+} c_{\bf m} {\det}^+(\cos 2\pi
m_ix_j)_{i,j=1}^n,
 \end{gather}
where ${\bf m}=(m_1,m_2,\dots,m_n)$ are {\it integer} $n$-tuples
such that $m_1\ge m_2\ge \cdots \ge m_n\ge 0$, and the
coef\/f\/icients $c_{\bf m}$ are given by the formula
 \begin{gather}\label{decom-anti-c-}
 c_{\bf m}=2^{2n} \int_F f(x) {\det}^+(\cos 2\pi m_ix_j)_{i,j=1}^n dx.
 \end{gather}
The Plancherel formula is of the form
\[
 \sum_{{\bf m}\in P_+} |c_{\bf m}|^2=2^{2n} \int_F f(x)
|f(x)|^2 dx.
\]

\subsection{Orbit function transform on the dominant Weyl chamber}\label{section10.3}
The expansion \eqref{decom-f1} of functions on the fundamental
domain $F$ is an expansion in the antisymmetric orbit functions
$\varphi_\lambda (x)$ with integral strictly dominant weights
$\lambda$.  The antisymmetric orbit functions $\varphi_\lambda
(x)$ with $\lambda$ lying in the dominant Weyl chamber (and not
obligatory integral) are not invariant with respect to the
corresponding af\/f\/ine Weyl group $W^{\rm aff}$. They are
invariant only with respect to the Weyl group $W$. A fundamental
domain of $W$ coincides with the dominant Weyl chamber $D_+$. For
this reason, the orbit functions $\varphi_\lambda (x)$,
$\lambda\in D^+_+$, determine another orbit function transform (a
transform on $D_+$).

We began with the usual Fourier transforms on ${\mathbb R}^n$:
 \begin{gather}\label{F-1}
\tilde f (\lambda)=\int_{{\mathbb R}^n} f(x) e^{2\pi {\rm
i}\langle \lambda,x \rangle} dx,
  \\
  \label{F-2}
 f (x)=\int_{{\mathbb R}^n} \tilde f(\lambda) e^{-2\pi {\rm i}\langle
\lambda,x \rangle} d\lambda.
  \end{gather}
Let the function $f(x)$ be anti-invariant with respect to the Weyl
group $W$, that is, $f(wx)=(\det w)f(x)$, $w\in W$. It is easy to
check that the function $\tilde f (\lambda)$ is also
anti-invariant with respect to the Weyl group $W$. Replace in
\eqref{F-1} $\lambda$ by $w\lambda$, $w\in W$, multiply both sides
by $\det w$, and sum these both side over $w\in W$. Then instead
of \eqref{F-1} we obtain
 \begin{gather}\label{F-3}
\tilde f (\lambda)= \int_{D_+} f(x) \varphi_\lambda(x) dx,\qquad
\lambda\in D^+_+,
  \end{gather}
where we have taken into account that $f(x)$ is anti-invariant
with respect to $W$.

Similarly, starting from \eqref{F-2}, we obtain the inverse
formula:
 \begin{gather}\label{F-4}
 f (x)= \int_{D_+} \tilde f(\lambda)
 \overline{\varphi_\lambda(x)} d\lambda .
  \end{gather}
For the transforms \eqref{F-3} and \eqref{F-4} the Plancherel
formula
 \[
 \int_{D_+} |f(x)|^2 dx=
\int_{D_+} |\tilde f(\lambda) |^2  d\lambda
 \]
holds.

\subsection[Symmetric and antisymmetric multivariate sine and
cosine integral transforms]{Symmetric and antisymmetric
multivariate sine\\ and cosine integral
transforms}\label{section10.4}
The orbit function transforms \eqref{F-3} and \eqref{F-4} in the
case of the Coxeter--Dynkin diag\-ram~$C_n$ lead to symmetric and
antisymmetric multivariate sine and cosine integral transforms.

Taking into account the expression \eqref{det-C} for antisymmetric
orbit functions for $C_n$ we obtain the transform
 \begin{gather}\label{anty-1}
\tilde f (\lambda)= \int_{D_+} f(x) \det \left( \sin
2\pi\lambda_ix_j\right)_{i,j=1}^n  dx,
  \end{gather}
where $\lambda=(\lambda_1,\lambda_2,\dots,\lambda_n)\in D^+_+$,
that is, $\lambda_1>\lambda_2>\cdots >\lambda_n>0$ (the function
$\tilde f (\lambda)$ vanishes on the boundary of $D_+$). The
inverse transform is of the form
 \begin{gather}\label{anty-2}
 f (x)=2^{2n} \int_{D_+} \tilde f(\lambda)
\det \left( \sin 2\pi\lambda_ix_j\right)_{i,j=1}^n  d\lambda .
  \end{gather}
For these transforms the Plancherel formula
 \[
 \int_{D_+} |f(x)|^2 dx= 2^{2n}
\int_{D_+} |\tilde f(\lambda) |^2  d\lambda
 \]
holds.

Similar transformations hold for symmetric multivariate cosine
function:
\begin{gather}\label{anty-3}
\tilde f (\lambda)= \int_{D_+} f(x) {\det}^+ \left( \cos
2\pi\lambda_ix_j\right)_{i,j=1}^n  dx,
  \\
  \label{anty-4}
 f (x)=2^{2n} \int_{D_+} \tilde f(\lambda)
{\det}^+ \left( \cos 2\pi\lambda_ix_j\right)_{i,j=1}^n  d\lambda ,
  \end{gather}
where $\lambda=(\lambda_1,\lambda_2,\dots,\lambda_n)\in D_+$ and
the function ${\det}^+ \left( \cos \lambda_ix_j\right)_{i,j=1}^n$
is given by formula \eqref{symm-cos}.

\section[Multivariate generalization of the finite Fourier transform
and of finite sine and cosine transforms]{Multivariate generalization of the f\/inite Fourier transform\\
and of f\/inite sine and cosine transforms}\label{section11}

Along with the integral Fourier transform there exists a discrete
Fourier transform. Similarly, it is possible to introduce a
f\/inite orbit function transform, based on antisymmetric orbit
functions. It is done in the same way as in the case of symmetric
orbit functions in \cite{KP06} by using the results of the paper
\cite{MP87} (see \cite{MP06}). We f\/irst consider the f\/inite
Fourier transform. Then we consider a general theory (appropriate
for any connected Coxeter--Dynkin diagram). In the last
subsections we give concrete antisymmetric and symmetric
generalizations of the f\/inite Fourier transform. In particular,
we consider antisymmetric and symmetric multivariate f\/inite
Fourier transforms, discrete sine and cosine transforms, and
antisymmetric and symmetric multivariate discrete sine and cosine
transforms.

\subsection{Finite Fourier transform}\label{section11.1}
Let us f\/ix a positive integer $N$ and consider the numbers
 \begin{gather}\label{f-F-1}
e_{mn}:=N^{-1/2}\exp (2\pi {\rm i}mn/N),\qquad m,n=1,2,\dots,N.
 \end{gather}
The matrix $(e_{mn})_{m,n=1}^N$ is unitary, that is,
 \begin{gather}\label{f-F-2}
\sum_k e_{mk}\overline{e_{nk}} =\delta_{mn},\qquad \sum_k
e_{km}\overline{e_{kn}} =\delta_{mn}.
 \end{gather}
Indeed, according to the formula for a sum of a geometric
progression we have
\begin{gather*}
t^a+t^{a+1}+\cdots +t^{a+r}=(1-t)^{-1}t^a(1-t^{r+1}),\qquad t\ne
1,
\\
t^a+t^{a+1}+\cdots +t^{a+r}=r+1,\qquad t=1.
\end{gather*}
Setting $t=\exp (2\pi{\rm i}(m-n)/N)$, $a=1$ and $r=N-1$, we prove
\eqref{f-F-2}.

Let $f(n)$ be a function of $n\in \{ 1,2,\dots ,N\}$. We may
consider the transform
 \begin{gather}\label{f-F-3}
\sum_{n=1}^N f(n)e_{mn}\equiv N^{-1/2} \sum_{n=1}^N f(n) \exp
(2\pi{\rm i}mn/N) =\tilde f (m).
 \end{gather}
Then due to unitarity of the matrix $(e_{mn})_{m,n=1}^N$, we
express $f(n)$ as a linear combination of conjugates of the
functions \eqref{f-F-1}:
 \begin{gather}\label{f-F-4}
f(n)= N^{-1/2} \sum_{m=1}^N {\tilde f}(m) \exp (-2\pi{\rm i}mn/N)
.
 \end{gather}
The function ${\tilde f}(m)$ is a {\it finite Fourier transform}
of $f(n)$. This transform is a linear map. The formula
\eqref{f-F-4} gives an inverse transform. The Plancherel formula
\[
\sum_{m=1}^N |\tilde f(m)|^2=\sum_{n=1}^N | f(n)|^2
\]
holds for transforms \eqref{f-F-3} and \eqref{f-F-4}. This means
that the f\/inite Fourier transform preserves the norm introduced
in the space of functions on $\{ 1,2,\dots,N\}$.

The f\/inite Fourier transform on the $r$-dimensional linear space
$E_r$ is def\/ined similarly. We again f\/ix a positive integer
$N$. Let ${\bf m}=(m_1,m_2,\dots,m_r)$ be an $r$-tuple of integers
such that each $m_i$ runs over the integers $1,2,\dots,N$. Then
the f\/inite Fourier transform on $E_r$ is given by the kernel
\[
e_{\bf mn}:=e_{m_1n_1}e_{m_2n_2}\cdots e_{m_rn_r}=N^{-r/2} \exp
(2\pi{\rm i}{\bf m\cdot n}/N),
\]
where ${\bf m\cdot n}=m_1n_1+m_2n_2+\cdots +m_rn_r$. If $F({\bf
m})$ is a function of $r$-tuples ${\bf m}$, $m_i\in\{
1,2,\dots,N\}$, then the f\/inite Fourier transform of $F$ is
given by
\[
{\tilde F}({\bf n})=N^{-r/2}\sum_{\bf m}F({\bf m}) \exp (2\pi{\rm
i}{\bf m\cdot n}/N).
\]
The inverse transform is
\[
F({\bf m})=N^{-r/2}\sum_{\bf n}{\tilde F}({\bf n}) \exp (-2\pi{\rm
i}{\bf m\cdot n}/N).
\]
The corresponding Plancherel formula is of the form
$\sum\limits_{\bf m} |F({\bf m})|^2=\sum\limits_{\bf n} |\tilde F
({\bf n})|^2$.

\subsection[$W$-invariant lattices]{$\boldsymbol{W}$-invariant lattices}\label{section11.2}
In order to determine an analogue of the f\/inite Fourier
transform, based on antisymmetric orbit functions, we need an
analogue of the set
\[
\{ {\bf m}=\{ m_1,m_2,\dots,m_n\}\ |\  m_i\in \{1,2,\dots,N\}\},
\]
used for multidimensional f\/inite Fourier transform. Such a set
has to be invariant with respect to the Weyl group $W$ (see
\cite{MP87}).

We know that $Q^\vee$ is a discrete $W$-invariant subset of $E_n$.
Clearly, the set $\frac 1m Q^\vee$ is also $W$-invariant, where
$m$ is a f\/ixed positive integer. Then the set
\[
T_m={\textstyle \frac 1m} Q^\vee / Q^\vee
\]
is f\/inite and $W$-invariant. If $\alpha_1,\alpha_2,\dots
,\alpha_l$ is the set of simple root for the Weyl group $W$, then
$T_m$ can be identif\/ied with the set of elements
 \begin{gather}\label{lat-1}
m^{-1}\sum_{i=1}^l d_i\alpha_i^\vee,\qquad d_i=0,1,2,\dots, m-1.
 \end{gather}

We select from $T_m$ the set of elements which belongs to the
fundamental domain $\overline{F}$. These elements lie in the
collection $\frac 1m Q^\vee \cap \overline{F}$.

Let $\mu\in \frac 1m Q^\vee \cap \overline{F}$ be an element
determining an element of $T_m$ and let $M$ be the least positive
integer such that $M\mu \in P^\vee$. Then there exists the least
positive integer $N$ such that $N\mu\in Q^\vee$. One has $M | N$
and $N | m$.

The collection of points of $T_m$ which belong to $F$ (we denote
the set of these points by $F_M$) coincides with the set of
elements
\begin{gather}\label{fin-orb-1}
s=\frac{s_1}M \omega^\vee_1+\cdots +\frac{s_l}M \omega^\vee_l,
\qquad
 \omega_i^\vee =\frac{2\omega_i}{\langle \alpha_i,\alpha_i \rangle},
 \end{gather}
where $s_1,s_2,\dots,s_l$ runs over values from $\{ 0,1,2,\dots
\}$ and satisfy the following condition: there exists a
non-negative integer $s_0$ such that
\begin{gather}\label{fin-orb-2}
s_0+\sum_{i=1}^l s_im_i=M,
 \end{gather}
where $m_1,m_2,\dots, m_l$ are positive integers from formula
\eqref{highestroot}. Values of $m_i$ for all simple Lie algebras
can be found in Subsection~\ref{section2.4}.

Indeed, the fundamental domain consists of all points $y$ from the
dominant Weyl chamber for which $\langle y,\xi \rangle \le 1$,
where $\xi$ is the highest (long) root, $\xi=\sum\limits_{i=1}^l
m_i\alpha_i$. Since for elements $s$ of~\eqref{fin-orb-1} one has
$s_i/M \ge 0$ and
\[
\langle s,\xi \rangle= \frac 1M \sum_{i=1}^l s_im_i=  \frac 1M
(M-s_0)\le 1,
\]
then $s\in \overline{F}$. The converse reasoning shows that points
of $\frac 1m Q^\vee \cap \overline{F}$ must be of the form
\eqref{fin-orb-1}.

To every positive integer $M$ there corresponds the  grid
$F_M$ of points \eqref{fin-orb-1} in $\overline{F}$ which
corresponds to some set $T_n$ such that $M | m$. The precise
relation between $M$ and $n$ can be def\/ined by the grid $F_M$
(see \cite{MP87}). Acting upon the grid $F_M$ by elements of the
Weyl group $W$ we obtain the whole set $T_m$.

Since antisymmetric orbit functions vanish on the boundary of a
fundamental domain $F$, it make sense to consider also a subgrid
$F^-_M$ consisting of all points of $F_M$ which do not lie on
a~wall of $F$.
 \medskip

\noindent {\bf Example.} {\it Grids $F_M$ for $A_1$.}  We take
into account results of Example in Subsection~\ref{section4.1}.
For~$A_1$ we have $\omega^\vee =\omega= \alpha/2$, where $\alpha$
is the simple root. Elements of $P_+$ coincide with $m\omega$,
$m\in {\mathbb Z}_+$. Fixing $M\in {\mathbb Z}_+$ we have
 \[
  F_M=\left\{ s=\tfrac{s_1}M,\ \ {\rm where}\ \ s_0+s_1=M\ \ {\rm for}\ \
s_0,s_1\in {\mathbb Z}^{\ge 0} \right\} .
 \]
Therefore,
\[
 F_M=\left\{ 0, \tfrac1M, \tfrac2M, ,\dots ,\tfrac{M-1}M,1
\right\} .
\]

\subsection[Grids $F_M$ for $A_2$, $C_2$ and $G_2$]{Grids $\boldsymbol{F_M}$ for $\boldsymbol{A_2}$,
$\boldsymbol{C_2}$ and $\boldsymbol{G_2}$}\label{section11.3}
In this section we give some examples of grids $F_M$ for the rank
two cases (see \cite{Pat-Z-1} and \cite{Pat-Z-2}). Since the long
root $\xi$ of $A_2$ is representable in the form
$\xi=\alpha_1+\alpha_2$, where $\alpha_1$ and $\alpha_2$ are
simple roots, that is, $m_1=m_2=1$ (see formula
\eqref{fin-orb-2}), then
\[
 F_M(A_2)=\left\{ \tfrac{s_1}M \omega_1+\tfrac{s_2}M \omega_2;\ \
s_0+s_1+s_2=M,\ \ s_0,s_1,s_2\in {\mathbb Z}^{\ge 0} \right\}.
  \]
It is seen from here that the vertices $0,\omega_1,\omega_2$ of
the fundamental domain $F(A_2)$ belong to each grid $F_M(A_2)$. A
direct computation shows that in the $\omega$-coordinates we have
\begin{gather*}
F_2(A_2)=\left\{
(0,0),(1,0),(0,1),\big(\tfrac12,0\big),\big(0,\tfrac12\big),\big(\tfrac12,\tfrac12\big)
\right\},
\\
F_3(A_2)=\left\{
(0,0),(1,0),(0,1),\big(\tfrac13,0\big),\big(0,\tfrac13\big),\big(\tfrac23,0\big),
\big(0,\tfrac23\big),\big(\tfrac23,\tfrac13\big),\big(\tfrac13,\tfrac23\big),\big(\tfrac13,\tfrac13\big)\right\}.
\end{gather*}

In $F_2(A_2)$, only the point $\left(\frac12,\frac12\right)$ does
not belong to a wall of the fundamental domain $F(A_2)$. In
$F_3(A_2)$, three points $\left(\frac13,\frac23\right)$,
$\left(\frac23,\frac13\right)$, $\left(\frac13,\frac13\right)$ do
not belong to a wall of $F(A_2)$.

The set $F^-_5(A_2)$ consists of the points
\[
F^-_5(A_2)=\left\{
\big(\tfrac15,\tfrac35\big),\big(\tfrac25,\tfrac25\big),
\big(\tfrac35,\tfrac15\big),\big(\tfrac15,\tfrac25\big),\big(\tfrac25,\tfrac15\big),\big(\tfrac15,\tfrac15\big)\right\}.
 \]

Since the long root $\xi$ of $C_2$ is representable in the form
$\xi=2\alpha_1+\alpha_2$, where $\alpha_1$ and $\alpha_2$ are
simple roots, that is, $m_1=2,m_2=1$, then
\[
 F_M(C_2)=\left\{ \tfrac{s_1}M \omega^\vee_1+\tfrac{s_2}M \omega^\vee_2;
s_0+2s_1+s_2=M,\ \ s_0,s_1,s_2\in {\mathbb Z}^{\ge 0} \right\}.
 \]
A direct computation shows that in the $\omega^\vee$-coordinates
we have
\begin{gather*}
F_2(C_2)=\left\{
(0,0),(0,1),\big(\tfrac12,0\big),\big(0,\tfrac12\big) \right\},
\\
F_3(C_2)=\left\{
(0,0),(0,1),\big(\tfrac13,0\big),\big(0,\tfrac13\big),
\big(0,\tfrac23\big),\big(\tfrac13,\tfrac13\big)\right\},
\\
F^-_7(C_2)=\left\{
\big(\tfrac17,\tfrac47\big),\big(\tfrac27,\tfrac27\big),\big(\tfrac17,\tfrac37\big),
\big(\tfrac27,\tfrac17\big),\big(\tfrac17,\tfrac27\big),\big(\tfrac17,\tfrac17\big)\right\}.
\end{gather*}

Since the long root $\xi$ of $G_2$ is representable in the form
$\xi=2\alpha_1+3\alpha_2$, where $\alpha_1$ and $\alpha_2$ are
simple roots, that is, $m_1=2$, $m_2=3$, then
\[
 F_M(G_2)=\left\{ \tfrac{s_1}M \omega^\vee_1+\tfrac{s_2}M \omega^\vee_2;
s_0+2s_1+3s_2=M,\ \ s_0,s_1,s_2\in {\mathbb Z}^{\ge 0} \right\}.
 \]
A computation shows that in the $\omega^\vee$-coordinates we have
\begin{gather*}
F_2(G_2)=\left\{ (0,0),(1,0)\right\},
\\
F_3(G_2)=\left\{
(0,0),\big(0,\tfrac13\big),\big(\tfrac13,0\big)\right\},
\\
F_4(G_2)=F_2(G_2)\cup\left\{
\big(\tfrac14,0\big),\big(0,\tfrac14\big)\right\},
\\
F_5(G_2)=\left\{
(0,0),\big(0,\tfrac15\big),\big(\tfrac15,0\big),\big(\tfrac15,\tfrac15\big),
\big(\tfrac25,0\big)\right\},
\\
F_8(G_2)=F_4(G_2)\cup\left\{\big(\tfrac18,0\big),\big(0,\tfrac18\big),
\big(\tfrac18,\tfrac18\big),\big(\tfrac14,\tfrac18\big)\right\},
\\
F^-_{14}(G_2)=\left\{
\big(\tfrac17,\tfrac3{14}\big),\big(\tfrac5{14},\tfrac1{14}\big),
\big(\tfrac3{14},\tfrac17\big),\big(\tfrac1{14},\tfrac3{14}\big),
\big(\tfrac27,\tfrac1{14}\big),\big(\tfrac17,\tfrac17\big), \right.   \\
\left. \phantom{F^-_{14}(G_2)=}{}
\big(\tfrac3{14},\tfrac1{14}\big),
\big(\tfrac1{14},\tfrac17\big),\big(\tfrac17,\tfrac1{14}\big),\big(\tfrac1{14},\tfrac1{14}\big)
\right\}.
\end{gather*}

\subsection[Expanding in antisymmetric orbit functions through finite sets]{ Expanding in antisymmetric orbit functions through f\/inite sets}\label{section11.4}

Let us give an analogue of the f\/inite Fourier transform when
instead of exponential functions we use antisymmetric orbit
functions. This analogue is not so simple as f\/inite Fourier
transform. For this reason, we consider some weak form of the
transform. In fact, we consider this weak form in order to be able
to recover (at least approximately) the decomposition
$f(x)=\sum\limits_\lambda a_\lambda \varphi_\lambda(x)$ by values
of $f(x)$ on a f\/inite set of points.

Considering the f\/inite Fourier transform in
Section~\ref{section11.1}, we have restricted the exponential
function to a discrete set. Similarly, in order to determine a
f\/inite transform, based on antisymmetric orbit functions, we
have to restrict orbit functions $\varphi_\lambda(x)$ to
appropriate f\/inite sets of values of $x$. Candidates for such
f\/inite sets are sets $T_m$. However, antisymmetric orbit
functions $\varphi_\lambda(x)$ with integral $\lambda$ are
invariant with respect to the af\/f\/ine Weyl group $W^{\rm aff}$.
For this reason, we consider orbit functions $\varphi_\lambda(x)$
on grids $F_M$.

On the other side, we also have  to choose a f\/inite number of
antisymmetric orbit functions, that is, a f\/inite number of
integral strictly dominant elements $\lambda$. The best choice is
when a~number of orbit functions coincides with $|F_M|$. These
antisymmetric orbit functions must be selected in such a way that
the matrix
\begin{gather}\label{determi}
\left( \varphi_{\lambda_i}(x_j)\right)_{\lambda_i\in \Omega,x_j\in
F_M}
\end{gather}
(where $\Omega$ is our f\/inite set of strictly dominant elements
$\lambda$) is not singular. In order to have non-singularity of
this matrix some conditions must be satisf\/ied. In general, they
are not known. For this reason, we consider some, more weak, form
of the transform (when $|\Omega|\ge |F_M|$) and then explain how
the set $|\Omega|$ of $\lambda\in P^+_+$ can be chosen in such a
way that $|\Omega|= |F_M|$.

Let $O(\lambda)$ and $O(\mu)$ be two dif\/ferent $W$-orbits for
integral strictly dominant elements~$\lambda$ and~$\mu$. We say
that the group $T_m$ {\it separates} $O(\lambda)$ and $O(\mu)$ if
for any two elements $\lambda_1\in O(\lambda)$ and $\mu_1\in
O(\mu)$ there exists an element $x\in T_m$ such that $\exp
(2\pi{\rm i}\langle \lambda_1,x \rangle) \ne \exp (2\pi{\rm
i}\langle \mu_1,x \rangle)$ (we use here orbits, not signed
orbits, since signs of points are not important for this
reasoning). Note that $\lambda$ may coincide with $\mu$.

Let $f_1$ and $f_2$ be two functions on $E_n$ which are f\/inite
linear combinations of orbit functions. We introduce a
$T_m$-scalar product for them by the formula
\[
\langle f_1,f_2 \rangle_{T_m}=\sum_{x\in T_m}
f_1(x)\overline{f_2(x)} .
\]
Then the following proposition is true (see \cite{MP87} and
\cite{MP06}):

\begin{proposition}\label{prop17}
If $T_m$ separates the orbits $O(\lambda)$ and $O(\mu)$,
$\lambda,\mu\in P^+_+$, then
\begin{gather}\label{Fuir}
\langle \varphi_\lambda,\varphi_\mu \rangle_{T_m}=m^n
|W|\delta_{\lambda\mu}.
 \end{gather}
 \end{proposition}

\begin{proof} We have
\begin{gather*}
\langle \varphi_\lambda,\varphi_\mu \rangle_{T_m}= \sum_{x\in T_m}
\sum_{w\in W} \sum_{w'\in W} (\det w)(\det w') \exp (2\pi{\rm
i}\langle w\lambda-w'\mu,x\rangle)
\notag\\
\phantom{\langle \varphi_\lambda,\varphi_\mu \rangle_{T_m}}{}  =
\sum_{w\in W} \sum_{w'\in W} (\det ww') \left( \sum_{x\in T_m}
\exp (2\pi{\rm i}\langle w\lambda-w'\mu,x\rangle)\right). \notag
\end{gather*}
Since $T_m$ separates $O(\lambda)$ and $O(\mu)$, then none of the
dif\/ferences $w\lambda-w'\mu$ in the last sum vanishes on $T_m$.
Since $T_m$ is a group, one has
\[
\sum_{x\in T_m} \exp (2\pi{\rm i}\langle
w\lambda-w'\mu,x\rangle)=m^n\delta_{w\lambda,w'\mu}.
\]
Therefore, $\langle \varphi_\lambda,\varphi_\mu \rangle_{T_m}=m^n
|W|\delta_{\lambda\mu}$. Proposition is proved.
\end{proof}

Let $f$ be an anti-invariant (with respect to $W^{\rm aff}$)
function on $E_n$ which is a f\/inite linear combination of
antisymmetric orbit functions:
\begin{gather}\label{Fuir-2}
 f(x)=\sum_{\lambda_j\in
P^+_+} a_{\lambda_j}\varphi_{\lambda_j}(x).
 \end{gather}
Our aim is to determine $f(x)$ by its values on a f\/inite subset
of $E_n$, namely, on $T_m$.

We suppose that $T_m$ separate orbits $O(\lambda_j)$ with
$\lambda_j$ from the right hand side of \eqref{Fuir-2}. Then
taking the $T_m$-scalar product of both sides of \eqref{Fuir-2}
with $\varphi_{\lambda_j}$ and using the relation \eqref{Fuir} we
obtain
\[
a_{\lambda_j}=\left( m^n |W|\right)^{-1}\langle f,
\varphi_{\lambda_j} \rangle_{T_m}.
\]
Let now $s^{(1)},s^{(2)},\dots,s^{(h)}$ be all elements of
$\overline{F}\cap \frac 1m Q^\vee$, which do not lie on some wall
of Weyl chambers. Then
 \begin{gather}\label{Fuir-3}
a_{\lambda_j}=m^{-n}|W|^{-1}\sum_{x\in T_m} f(x)
\overline{\varphi_{\lambda_j}(x)}= m^n \sum_{i=1}^h  f(s^{(i)})
\overline{\varphi_{\lambda_j}(s^{(i)})} .
 \end{gather}

Thus, {\it the finite number of values $f(s^{(i)})$,
$i=1,2,\dots,h$, of the function $f(x)$ determines the
coefficients $a_{\lambda_j}$ and, therefore, determines the
function $f(x)$ on the whole space $E_l$.}

This means that we can reconstruct a $W^{\rm aff}$-anti-invariant
function $f(x)$ on the whole space~$E_n$ by its values on the
f\/inite set $F_M$ under an appropriate value of $M$. Namely, we
have to expand this function, taken on $F_M$, into the series
\eqref{Fuir-2} by means of the coef\/f\/icients $a_{\lambda_j}$,
determined by formula \eqref{Fuir-3}, and then to continue
analytically the expansion \eqref{Fuir-2} to the whole fundamental
domain $F$ (and, therefore, to the whole space $E_n$), that is, to
consider the decomposition \eqref{Fuir-2} for all $x\in E_n$.

We have assumed that the function $f(x)$ is a f\/inite linear
combination of orbit functions. If $f(x)$ expands into inf\/inite
sum of orbit functions, then for applying the above procedure
we have to approximate the function $f(x)$ by taking a f\/inite
number of terms in this inf\/inite sum and then apply the
procedure. That is, in this case we obtain an approximate
expression of the function $f(x)$ by using a f\/inite number of
its values.

At last, we explain how to choose a set $\Omega$ in formula
\eqref{determi}. The set $F_M$ consists of the
points~\eqref{fin-orb-1}. This set determines the set of points
 \[
\lambda=s_1\omega_1+s_2\omega_2+\cdots +s_n\omega_n,
 \]
where $s_1,s_2,\dots,s_n$ run over the same values as for the set
$F_M$. The subset of this set, consisting of strictly dominant
elements, can be taken as the set $\Omega$ (see \cite{MP06}).

\subsection{Antisymmetric and symmetric multivariate discrete
Fourier transforms}\label{section11.5}
The discrete Fourier transform of Subsection~\ref{section11.1} can
be generalized to the $n$-dimensional case in a symmetric or
antisymmetric form without using the results of
Subsection~\ref{section11.4}.

We take the discrete exponential function \eqref{f-F-1},
\[
e_m(s):=N^{-1/2}\exp (2\pi{\rm i}ms),\qquad s\in F_N\equiv\left\{
\tfrac1N, \tfrac2N,\dots,\tfrac{N-1}N ,1\right\},\qquad m\in
{\mathbb Z}^{\ge 0},
 \]
and make a multivariate discrete exponential function  taking a
product of $n$ copies of functions~\eqref{f-F-1}:
\begin{gather}\label{mult-F-1}
{\rm EXP}_{\bf m}({\bf s}):= e_{m_1}(s_1) e_{m_2}(s_2)\cdots
e_{m_n}(s_n)\\
\phantom{{\rm EXP}_{\bf m}({\bf s}):}{}\!\!= N^{-n/2}\exp
(2\pi{\rm i}m_1s_1)\exp (2\pi{\rm i}m_2s_2) \cdots \exp (2\pi{\rm
i}m_ns_n),\quad
 s_j\in F_N,\quad m_i\in {\mathbb Z}^{\ge 0},\notag
\end{gather}
where ${\bf s}=(s_1,s_2,\dots,s_n)$ and ${\bf
m}=(m_1,m_2,\dots,m_n)$. Now we take these multivariate functions
for integers $m_i$ such that $m_1>m_2>\cdots>m_n\ge 0$ and make an
antisymmetrization. As a~result, we obtain a f\/inite version of
the antisymmetric orbit function \eqref{det-A}:
\begin{gather}\label{mult-F-2}
e_{\bf m}({\bf s}):=|S_n|^{-1/2} \det (e_{m_i}(s_j))_{i,j=1}^n,
 \end{gather}
where $|S_n|$ is the order of the symmetric group $S_n$.

The $n$-tuples ${\bf s}$ in \eqref{mult-F-2} runs over $F_N^n
\equiv F_N\times \cdots \times F_N$ ($n$ times). We denote by
$\hat F_N^n$ the subset of $F_N^n$ consisting of ${\bf s}\in
F_N^n$ such that
\[
 s_1>s_2>\cdots >s_n.
\]
Note that acting by the permutations $w\in S_n$ upon $\hat F_N^n$
we obtain the whole set $F_N^n$ without those points which are
invariant under some nontrivial permutation $w\in S_n$. Clearly,
the function~\eqref{mult-F-2} vanishes on the last points.

Since the discrete exponential functions $e_m(s)$ satisfy the
equality $e_m(s)=e_{m+N}(s)$, we do not need to consider them for
all values $m\in {\mathbb Z}^{\ge 0}$. It is enough to consider
them for $m\in \{ 1,2,\dots,N\}$. By $D_N^+$ we denote the set of
integer $n$-tuples ${\bf m}=(m_1,m_2,\dots,m_n)$ such that
\[
 N\ge m_1>m_2>\cdots>m_n>0.
\]

We need a scalar product in the space of linear combinations of
the functions \eqref{mult-F-1}. It can be given by the
formula{\samepage
\begin{gather}\label{mult-F-3a}
\langle {\rm EXP}_{\bf m}({\bf s}),{\rm EXP}_{{\bf m}'}({\bf s})
\rangle \equiv \prod_{i=1}^n \langle e_{m_i}(s_i),
e_{m'_i}(s_i)\rangle :=\prod_{i=1}^n \sum_{s_i\in
F_N}e_{m_i}(s_i)\overline{e_{m'_i}(s_i)} =\delta_{{\bf m}{\bf
m}'},
\\
\phantom{\langle {\rm EXP}_{\bf m}({\bf s}),{\rm EXP}_{{\bf
m}'}({\bf s})   \rangle \equiv}{} m_i,m_i'\in \{ 1,2,\dots,
N\},\notag
\end{gather}
where we used the relation \eqref{f-F-2}.}

\begin{proposition}\label{prop18}
For ${\bf m},{\bf m}'\in D^+_N$ the discrete functions
\eqref{mult-F-2}
 satisfy the orthogonality relation
\begin{gather}\label{mult-F-3}
\langle e_{\bf m}({\bf s}),e_{{\bf m}'}({\bf s})\rangle =|S_n|
\sum_{{\bf s}\in \hat F_N^n} e_{\bf m}({\bf s})\overline{e_{{\bf
m}'}({\bf s})} =\delta_{{\bf m}{\bf m}'},
 \end{gather}
where the scalar product is determined by formula
\eqref{mult-F-3a}.
\end{proposition}

\begin{proof} Since $m_1>m_2>\cdots>m_n>0$, then due to
the def\/inition of the scalar product we have
\begin{gather}
\langle e_{\bf m}({\bf s}),e_{{\bf m}'}({\bf s})\rangle =
\sum_{{\bf s}\in F_N^n}e_{\bf m}({\bf s})
\overline{e_{{\bf m}'}({\bf s})}\nonumber\\
\phantom{\langle e_{\bf m}({\bf s}),e_{{\bf m}'}({\bf
s})\rangle}{} =|S_n|^{-1}\sum_{w\in S_n} \prod_{i=1}^n
\sum_{s_i\in F_N} e_{m_{w(i)}}(s_i)\overline{e_{m'_{w(i)}}(s_i)} =
\delta_{{\bf m}{\bf m}'},\label{mult-F-30}
\end{gather}
where $(m_{w(1)},m_{w(2)},\dots ,m_{w(n)})$ is obtained from
$(m_1,m_2,\dots, m_n)$ by action by the permutation $w\in S_n$.
Since functions $e_{\bf m}({\bf s})$ are antisymmetric with
respect to $S_n$, then
\[
 \sum_{{\bf s}\in F_N^n}e_{\bf m}({\bf s})
\overline{e_{{\bf m}'}({\bf s})} =|S_n| \sum_{{\bf s}\in \hat
F_N^n}e_{\bf m}({\bf s}) \overline{e_{{\bf m}'}({\bf s})} ,
\]
where we have taken into account that $e_{\bf m}({\bf s})$
vanishes on those ${\bf s}\in F_N^n$ for which there exists $w\in
S_n$, $w\ne 1$, such that $w {\bf s}={\bf s}$. This proves the
proposition.
\end{proof}

Let $f$ be a function on $\hat F^n_N$ (or an antisymmetric
function on $F_N^n$). Then it can be expanded in the functions
\eqref{mult-F-2} as
\begin{gather}\label{mult-F-4}
f({\bf s})=\sum_{{\bf m}\in D_N^+}a_{\bf m} e_{\bf m}({\bf s}).
 \end{gather}
The coef\/f\/icients $a_{\bf m}$ are determined by the formula
\begin{gather}\label{mult-F-5}
a_{\bf m}=|S_n|\sum_{{\bf m}\in \hat F_N^n} f({\bf s})
\overline{e_{\bf m}({\bf s})}.
 \end{gather}
The expansions \eqref{mult-F-4} and \eqref{mult-F-5} follow from
the facts that numbers of elements in $D_N^+$ and in $\hat F_N^n$
are the same, and that the matrix
\[
 \left(  e_{\bf m}({\bf s})\right)_{ {\bf m}\in D_N^+,{\bf s}\in \hat F_N^n}
\]
is unitary. We call expansions \eqref{mult-F-4} and
\eqref{mult-F-5}
 {\it antisymmetric multivariate discrete Fourier transforms}.

Let us also give a symmetric multivariate discrete Fourier
transforms. For this we take the multivariate exponential
functions \eqref{mult-F-1} for integers $m_i$ such that
\[
N\ge m_1\ge m_2\ge \cdots\ge m_n\ge 1
\]
and make a symmetrization. We obtain a f\/inite version of the
symmetric orbit function (6.11) in \cite{KP06} for the case $A_n$:
\begin{gather}\label{mult-F-10}
E_{\bf m}({\bf s}):=|S_n|^{-1/2} {\det}^+
(e_{m_i}(s_j))_{i,j=1}^n:= |S_n|^{-1/2} \sum _{w\in S_n}
\prod_{i=1}^n e_{m_{w(i)}}(s_i).
 \end{gather}

The $n$-tuples ${\bf s}$ in \eqref{mult-F-10} run over
$F_N^n\equiv F_N\times \cdots\times F_N$ ($n$ times). We denote by
$\breve F_N^n$ the subset of $F_N^n$ consisting of ${\bf s}\in
F_N^n$ such that
\[
 s_1\ge s_2\ge \cdots \ge s_n.
\]
Note that acting by the permutations $w\in S_n$ upon $\breve
F_N^n$ we obtain the whole set $F_N^n$, where each point, having
some coordinates $m_i$ coinciding, is repeated several times.
Namely, a point ${\bf s}$ is contained $|S_{\bf s}|$ times in $\{
w\breve F_N^n; w\in S_n\}$, where $S_{\bf s}$ is the subgroup of
$S_n$ consisting of elements $w\in S_n$ such that $w {\bf s}={\bf
s}$. The number $|S_{\bf s}|$ is called a {\it multiplicity} of
the point ${\bf s}$ in the set $\{ w\breve F_N^n; w\in S_n\}$.

By $\breve D_N^+$ we denote the set of integer $n$-tuples ${\bf
m}=(m_1,m_2,\dots,m_n)$ such that
\[
 N\ge m_1\ge m_2\ge \cdots\ge m_n\ge 1.
\]

\begin{proposition}\label{prop19}
For ${\bf m},{\bf m}'\in \breve D_N^+$ the discrete functions
\eqref{mult-F-10} satisfy the orthogonality relation
\begin{gather}\label{mult-F-11}
\langle E_{\bf m}({\bf s}),E_{{\bf m}'}({\bf s})\rangle =
|S_n|\sum_{{\bf s}\in \breve F_M^n} |S_{\bf s}|^{-1} E_{\bf
m}({\bf s}) \overline{E_{{\bf m}'}({\bf s})} =|S_{\bf m}|
\delta_{{\bf m}{\bf m}'}.
 \end{gather}
 \end{proposition}

\begin{proof} This proposition is proved in the same way as Proposition~\ref{prop18}, but
we have to take into account a dif\/ference between $\breve F_M^n$
and $\hat F_M^n$. Due to the def\/inition of the scalar product we
have
\begin{gather*}
\langle E_{\bf m}({\bf s}),E_{{\bf m}'}({\bf s})\rangle =
\sum_{{\bf s}\in F_N^n}E_{\bf m}({\bf s})
\overline{E_{{\bf m}'}({\bf s})} \\
\phantom{\langle E_{\bf m}({\bf s}),E_{{\bf m}'}({\bf
s})\rangle}{} = |S_n|^{-1}|S_{\bf m}| \sum_{w\in S_n}
\prod_{i=1}^n
\sum_{s_i\in F_N} e_{m_{w(i)}}(s_i)\overline{e_{m'_{w(i)}}(s_i)}\\
\phantom{\langle E_{\bf m}({\bf s}),E_{{\bf m}'}({\bf
s})\rangle}{} = |S_{\bf m}| \delta_{{\bf m}{\bf m}'},
\end{gather*}
where $(m_{w(1)},m_{w(2)},\dots ,m_{w(n)})$ is obtained from
$(m_1,m_2,\dots, m_n)$ by action by the permutation $w\in S_n$.
Here we have taken into account that additional summands appear
(with respect to \eqref{mult-F-30}) because some summands on the
right hand side of \eqref{mult-F-10} may coincide.

Since functions $E_{\bf m}({\bf s})$ are symmetric with respect to
$S_n$, then
\[
 \sum_{{\bf s}\in F_N^n} E_{\bf m}({\bf s})
\overline{E_{{\bf m}'}({\bf s})} =|S_n| \sum_{{\bf s}\in \breve
F_N^n} |S_{\bf s}|^{-1} E_{\bf m}({\bf s}) \overline{E_{{\bf
m}'}({\bf s})} ,
\]
where we have taken into account that under  action by $S_n$ upon
$\breve F_N^n$ a point ${\bf s}$ appears $|S_{\bf s}|$ times in
$F^n_N$. This proves the proposition.
\end{proof}

Let $f$ be a function on $\breve F^n_N$ (or a symmetric function
on $F_N^n$). Then it can be expanded in functions
\eqref{mult-F-10} as
\begin{gather}\label{mult-F-12}
f({\bf s})=\sum_{{\bf m}\in \breve D_N^+}a_{\bf m} E_{\bf m}({\bf
s}).
 \end{gather}
The coef\/f\/icients  $a_{\bf m}$ are determined by the formula
\begin{gather}\label{mult-F-13}
a_{\bf m}=|S_n| |S_{\bf m}|^{-1}\sum_{{\bf m}\in \breve F_N^n}
|S_{\bf s}|^{-1}f({\bf s}) \overline{E_{\bf m}({\bf s})}.
 \end{gather}

The expansions \eqref{mult-F-12} and \eqref{mult-F-13} follow from
the facts that numbers of elements in $\breve D_N^+$ and in
$\breve F_N^n$ are the same and from the orhogonality relation
\eqref{mult-F-11}. We call expansions \eqref{mult-F-12} and
\eqref{mult-F-13}
 {\it symmetric multivariate discrete Fourier transforms}.

\subsection{Discrete sine and cosine transforms}\label{section11.6}
The grid $F_M$ for $A_1$ is of the form
 \begin{gather}\label{grid-3}
F_M(A_1)=\left\{ 0,\tfrac1M, \tfrac2M,\dots,\tfrac{M-1}M,
1\right\} .
 \end{gather}
The points 0 and 1 belong to the boundary of the fundamental
domain $F(A_1)$ of $W^{\rm aff}(A_1)$. Therefore, antisymmetric
orbit functions of $A_1$ vanish on these points and
\[
 F^-_M(A_1)=\left\{ \tfrac1M, \tfrac2M,\dots,\tfrac{M-1}M\right\}
 \qquad (M-1\ {\rm points})
\]
(see Subsection~\ref{section11.2}). Since the antisymmetric orbit
functions for $A_1$ are of the form $\varphi_\lambda(x)=2{\rm
i}\sin (\pi m\theta)$ (see Example of
Subsection~\ref{section4.1}), these functions on the grid $F_M$
are given by
 \begin{gather}\label{grid-4}
\varphi_m(s)=2{\rm i}\sin (\pi ms) ,\qquad s\in F_M,\quad m\in
{\mathbb Z}^{\ge}.
 \end{gather}
Since $\varphi_m(s)=\varphi_{m+M}(s)$, we consider these discrete
functions only for $m\in D_M:=\{ 1,2,\dots$, $M-1\}$. The
orthogonality relation for these functions is of the form
\[
 \langle \varphi_m,\varphi_{m'}  \rangle=\sum_{s\in F_M^-}
\varphi_m(s)\overline{\varphi_{m'}(s)}=2M\delta_{mm'},\qquad
m,m'\in D_M
\]
(see \cite{PZ-06}). They determine the following expansion of
functions, given on the grid $F^-_M$:
 \begin{gather}\label{grid-5}
f(s)=\sum_{m=1}^{M-1} a_m \varphi_m(s) ,
 \end{gather}
where the coef\/f\/icients $a_m$ are given by{\samepage
 \begin{gather}\label{grid-6}
a_m=\frac1{2M} \sum_{s\in F_M^-} f(s)\overline{\varphi_m(s)}.
 \end{gather}
Formulas \eqref{grid-5} and \eqref{grid-6} determine the {\it
discrete sine transform}.}

The symmetric orbit functions for $A_1$ are of the form
$\phi_\lambda(x)=2\cos (\pi m\theta)$. Then these functions on the
grid $F_M(A_1)$ are
 \begin{gather}\label{grid-12}
 \phi_m(s)=2\cos (\pi ms),\qquad s\in F_M,\quad m\in \{ 0,1,2,\dots, M\} .
 \end{gather}
The scalar product of these functions is given by
 \begin{gather}\label{grid-13}
 \langle \phi_m,\phi_{m'}  \rangle=\sum_{s\in F_M} c_s
\phi_m(s)\overline{\phi_{m'}(s)}=r_m M\delta_{mm'},
 \end{gather}
where $r_m=4$ for $m=0,M$ and $r_m=2$ otherwise, $c_s=1/2$ for
$s=0, 1$ and $c_s=1$ otherwise.

The functions $\phi_m$, given by \eqref{grid-12}, determine an
expansion of functions on the grid $F_M$ as
 \begin{gather}\label{grid-7}
f(s)=\sum_{m=0}^{M} b_m \phi_m(s) ,\qquad s\in F_M,
 \end{gather}
where the coef\/f\/icients $b_m$ are given by
 \begin{gather}\label{grid-8}
b_m=r_m^{-1} \sum_{s\in F_M} c_s f(s)\phi_m(s).
 \end{gather}
Formulas \eqref{grid-7} and \eqref{grid-8} determine the {\it
discrete cosine transform}.

\subsection{Antisymmetric multivariate discrete sine transforms}\label{section11.7}
The discrete sine and cosine transforms of the previous subsection
can be generalized to the $n$-di\-men\-sional case in symmetric or
antisymmetric form. In fact, these generalizations are f\/inite
antisymmetric orbit function transforms \eqref{Fuir-2} and
\eqref{Fuir-3} for multivariate sine transforms and f\/inite
symmetric orbit function transforms (9.11) and (9.12)
in~\cite{KP06} for multivariate cosine transforms. However, we
give in this subsection a derivation of these multivariate sine
and cosine transforms, independent of the previous consideration.
We need only the 1-dimensional discrete sine and cosine transforms
of the previous subsection.

We take the discrete sine function \eqref{grid-4} and make a
multivariate discrete sine function by multiplying $n$ copies of
functions \eqref{grid-4}:
\begin{gather}\label{multi-1}
{\rm SIN}_{\bf m}({\bf s}):=(2{\rm i})^n \sin (\pi m_1s_1) \sin
(\pi m_2s_2)\cdots \sin (\pi m_ns_n),
\\
 s_j\in F_M\equiv F_M(A_1),\qquad m_i\in D_M\equiv \{ 1,2,\dots,N-1\},\nonumber
\end{gather}
where ${\bf s}=(s_1,s_2,\dots,s_n)$ and ${\bf
m}=(m_1,m_2,\dots,m_n)$. Now we take these multivariate functions
for integers $m_i$ such that $M>m_1>m_2>\cdots>m_n>0$ and make
antisymmetrization. As a result, we obtain a f\/inite version of
the orbit function \eqref{det-C}:
\begin{gather}\label{multi-2}
\varphi_{\bf m}({\bf s}):=(2{\rm i})^n |S_n|^{-1/2} \det (\sin \pi
m_is_j)_{i,j=1}^n,
 \end{gather}
where $|S_n|$ is the order of the symmetric group $|S_n|$. (We
have here expressions $\sin \pi m_is_j$, not $\sin 2\pi m_is_j$ as
in \eqref{det-C}. Note that in \eqref{det-B} $m_i$,
$i=1,2,\dots,n$, run over integers and half-integers, whereas in
\eqref{multi-2} $m_i$ run over integers. Thus, in fact we have
replaced $2m_i$ with half-integer values of $m_i$ by $m_i$ with
integer values of $m_i$.)

The $n$-tuple ${\bf s}$ in \eqref{multi-2} runs over $
F^-_M(A_1)^n \equiv F^-_M(A_1)\times \cdots \times F^-_M(A_1)$
($n$ times). We denote by $\hat F_M^n$ the subset of
$F^-_M(A_1)^n$ consisting of ${\bf s}\in F^-_M(A_1)^n$ such that
\[
 s_1>s_2>\cdots >s_n.
\]
Note that $s_i$ here may take the values $\frac1M,\frac2M,\dots,
\frac{M-1}M$. Acting by permutations $w\in S_n$ upon~$\hat F_M^n$
we obtain the whole set $F^-_M(A_1)^n$ without those points which
are invariant under some nontrivial permutation $w\in S_n$.
Clearly, the function \eqref{multi-2} vanishes on the last points.

We denote by $D_M^+$ the set of integer $n$-tuples ${\bf
m}=(m_1,m_2,\dots,m_n)$ such that
\[
 M>m_1>m_2>\cdots>m_n>0.
\]

We wish to have a scalar product of functions \eqref{multi-2}. For
this we def\/ine a scalar product of functions \eqref{multi-1} as
\[
\langle {\rm SIN}_{\bf m}({\bf s}),{\rm SIN}_{{\bf m}'}({\bf s})
\rangle =\prod_{i=1}^n \langle \varphi_{m_i}(s_i),
\varphi_{m'_i}(s_i)\rangle,
\]
where the scalar product $\langle \varphi_{m_i}(s_i),
\varphi_{m'_i}(s_i)\rangle$ is given in
Subsection~\ref{section11.6}. Since functions $\varphi_{\bf
m}({\bf s})$ are linear combinations of functions ${\rm SIN}_{{\bf
m}'}({\bf s})$, a scalar product for $\varphi_{\bf m}({\bf s})$ is
also def\/ined.

\begin{proposition}\label{prop20}
For ${\bf m},{\bf m}'\in D_M^+$, the discrete functions
\eqref{multi-2} satisfy the orthogonality relation
\begin{gather}\label{multi-3}
\langle \varphi_{\bf m}({\bf s}),\varphi_{{\bf m}'}({\bf
s})\rangle := \sum_{{\bf s}\in F^-_M(A_1)^n}\!\varphi_{\bf m}({\bf
s}) \overline{\varphi_{{\bf m}'}({\bf s})} = |S_n|\! \sum_{{\bf
s}\in \hat F_M^n} \varphi_{\bf m}({\bf s})\overline{\varphi_{{\bf
m}'}({\bf s})} =(2M)^n \delta_{{\bf m}{\bf m}'}.\!\!\!
 \end{gather}
 \end{proposition}

\begin{proof} Since $M>m_1>m_2>\cdots>m_n>0$, then due to the orthogonality relation for
the sine functions $2{\rm i}\sin (\pi ms)$ (see the previous
subsection) we have
\begin{gather*}
\sum_{{\bf s}\in F^-_M(A_1)^n}\varphi_{\bf m}({\bf s})
\overline{\varphi_{{\bf m}'}({\bf s})} =4^n |S_n|^{-1} \sum_{w\in
S_n} \prod_{i=1}^n \sum_{s_i=1}^{M-1} \sin (\pi m_{w(i)}s_i) \sin
(\pi m'_{w(i)}s_i) = (2M)^n \delta_{{\bf m}{\bf m}'},\!
\end{gather*}
where $(m_{w(1)},m_{w(2)},\dots ,m_{w(n)})$ is obtained from
$(m_1,m_2,\dots, m_n)$ by action by the permutation $w\in S_n$.
Since functions $\varphi_{\bf m}({\bf s})$ are antisymmetric with
respect to $S_n$, we have
\[
 \sum_{{\bf s}\in F^-_M(A_1)^n}\varphi_{\bf m}({\bf s})
\overline{\varphi_{{\bf m}'}({\bf s})} =|S_n| \sum_{{\bf s}\in
\hat F_M^n}\varphi_{\bf m}({\bf s}) \overline{\varphi_{{\bf
m}'}({\bf s})} .
\]
This proves the proposition.
\end{proof}

Let $f$ be a function on $\hat F^n_M$ (or an antisymmetric
function on $F^-_M(A_1)^n$). Then it can be expanded in functions
\eqref{multi-2} as
\begin{gather}\label{multi-4}
f({\bf s})=\sum_{{\bf m}\in D_M^+}a_{\bf m} \varphi_{\bf m}({\bf
s}),
 \end{gather}
where the coef\/f\/icients $a_{\bf m}$ are determined by the
formula
\begin{gather}\label{multi-5}
a_{\bf m}=(2M)^{-n}|S_n| \sum_{{\bf m}\in \hat F_M^n} f({\bf s})
\overline{\varphi_{\bf m}({\bf s})}.
 \end{gather}

A validity of the expansions \eqref{multi-4} and \eqref{multi-5}
follows from the facts that numbers of elements in $D_M^+$ and in
$\hat F_M^n$ are the same and from the orthogonality relation
\eqref{multi-3}.

\subsection{Symmetric multivariate discrete cosine transforms}\label{section11.8}
We take the discrete cosine functions \eqref{grid-12} and make
multivariate discrete cosine functions by multiplying $n$ copies
of these functions:
\begin{gather}
{\rm COS}_{\bf m}({\bf s}):=\phi_{m_1}(s_1)\phi_{m_2}(s_2)\cdots
\phi_{m_n}(s_n)=2^n \cos (\pi m_1s_1) \cos (\pi m_2s_2)\cdots \cos
(\pi m_ns_n),\notag
\\
 s_j\in F_M\equiv F_M(A_1),\qquad m_i\in \{ 0,1,2,\dots,M\} .\label{multiv-1}
\end{gather}
We take these functions for integers $m_i$ such that $M\ge m_1\ge
m_2\ge \cdots\ge m_n\ge 0$ and make a~symmetrization. As a result,
we obtain a f\/inite version of the orbit function
\eqref{det-B-symm}:
\begin{gather}\label{multiv-2}
\phi_{\bf m}({\bf s}):=2^n |S_n|^{-1/2} \sum_{w\in S_n} \cos \pi
m_{w(1)}s_1 \cdot \cos \pi m_{w(2)}s_2\cdots \cos \pi m_{w(n)}s_n.
 \end{gather}
(We have here expressions $\cos \pi m_is_j$, not $\cos 2\pi
m_is_j$ as in \eqref{det-B-symm}.)

The $n$-tuple ${\bf s}$ in \eqref{multiv-2} runs over $F_M^n\equiv
F_M(A_1)^n$. We denote by $\breve F_M^n$ the subset of $F_M^n$
consisting of ${\bf s}\in F_M^n$ such that
\[
 s_1\ge s_2\ge \cdots \ge s_n .
\]
Note that $s_i$ here may take the values $0,\frac1M,\frac2M,\dots,
\frac{M-1}M,1$. Acting by permutations $w\in S_n$ upon $\breve
F_M^n$ we obtain the whole set $F_M^n$, where points, invariant
under some nontrivial permutation $w\in S_n$, are repeated several
times. It is easy to see that a point ${\bf s}_0\in F_M^n$ is
repeated $|S_{{\bf s}_0}|$ times in the set $\{ w\breve F_M^n$;
$w\in S_n\}$, where $|S_{{\bf s}_0}|$ is an order of the subgroup
$S_{{\bf s}_0}\subset S_n$, whose elements leaves ${\bf s}_0$
invariant.

We denote by $\breve D_M^+$ the set of integer $n$-tuples ${\bf
m}=(m_1,m_2,\dots,m_n)$ such that
\[
 M\ge m_1\ge m_2\ge \cdots\ge m_n\ge 0.
\]

A scalar product of functions \eqref{multiv-1} is determined by
\[
\langle {\rm COS}_{\bf m}({\bf s}),{\rm COS}_{{\bf m}'}({\bf s})
\rangle =\prod_{i=1}^n \langle \phi_{m_i}(s_i),
\phi_{m'_i}(s_i)\rangle,
\]
where the scalar product $\langle \phi_{m_i}(s_i),
\phi_{m'_i}(s_i)\rangle$ is given by \eqref{grid-13}. Since
functions $\phi_{\bf m}({\bf s})$ are linear combinations of
functions ${\rm COS}_{{\bf m}'}({\bf s})$, then a scalar product
for $\phi_{\bf m}({\bf s})$ is also def\/ined.

\begin{proposition}\label{prop21}
For ${\bf m},{\bf m}'\in \breve D_M^+$, the discrete functions
\eqref{multiv-2} satisfy the orthogonality relation
\begin{gather}
\langle \phi_{\bf m}({\bf s}),\phi_{{\bf m}'}({\bf s})\rangle =
\sum_{{\bf s}\in  F_M^n} c_{\bf s}\phi_{\bf m}({\bf s})
\overline{\phi_{{\bf m}'}({\bf s})} = |S_n|\sum_{{\bf s}\in \breve
F_M^n} |S_{\bf s}|^{-1} c_{\bf s}
\phi_{\bf m}({\bf s})\overline{\phi_{{\bf m}'}({\bf s})}  \notag\\
\phantom{\langle \phi_{\bf m}({\bf s}),\phi_{{\bf m}'}({\bf
s})\rangle}{}= M^n  r_{\bf m} |S_{\bf m}|  \delta_{{\bf m}{\bf
m}'},\label{multiv-3}
\end{gather}
where $c_{\bf s}=c_{s_1}c_{s_2}\cdots c_{s_n}$, $r_{\bf
s}=r_{m_1}r_{m_2}\cdots r_{m_n}$, and $c_{s_i}$ and  $r_{m_i}$ are
such as in \eqref{grid-13}.
\end{proposition}

\begin{proof} Due to the orthogonality relation for
the cosine functions $\phi_m(s)=2\cos (\pi ms)$ (see formula
\eqref{grid-13}) we have
\begin{gather}
\sum_{{\bf s}\in F_M^n} c_{\bf s}\phi_{\bf m}({\bf s})
\overline{\phi_{{\bf m}'}({\bf s})} =
 4^n |S_n|^{-1} |S_{\bf m}|\sum_{w\in S_n} \prod_{i=1}^n
\sum_{s_i=0}^{M}c_{s_i} \cos (\pi m_{w(i)}s_i) \cos (\pi m'_{w(i)}s_i)\notag \\
\phantom{\sum_{{\bf s}\in F_M^n} c_{\bf s}\phi_{\bf m}({\bf s})
\overline{\phi_{{\bf m}'}({\bf s})}}{} = |S_{\bf m}| M^n r_{\bf m}
\delta_{{\bf m}{\bf m}'},\label{multiv-12}
\end{gather}
where $(m_{w(1)},m_{w(2)},\dots ,m_{w(n)})$ is obtained from
$(m_1,m_2,\dots, m_n)$ by action by the permutation $w\in S_n$.
Since functions $\phi_{\bf m}({\bf s})$ are symmetric with respect
to $S_n$, we have
\[
 \sum_{{\bf s}\in F_M(A_1)^n} c_{\bf s}\phi_{\bf m}({\bf s})
\overline{\phi_{{\bf m}'}({\bf s})} =|S_n| \sum_{{\bf s}\in \breve
F_M^n} |S_{\bf s}|^{-1} c_{\bf s}\phi_{\bf m}({\bf s})
\overline{\phi_{{\bf m}'}({\bf s})} .
\]
This proves the proposition.
\end{proof}

Let $f$ be a function on $\breve F^n_M$ (or an antisymmetric
function on $F_M^n$). Then it can be expanded in functions
\eqref{multiv-2} as
\begin{gather}\label{multiv-4}
f({\bf s})=\sum_{{\bf m}\in \breve D_M^+}a_{\bf m} \phi_{\bf
m}({\bf s}),
 \end{gather}
where the coef\/f\/icients $a_{\bf m}$ are determined by the
formula
\begin{gather}
a_{\bf m}= M^{-n}|S_{\bf m}|^{-1} r_{\bf m}^{-1}
\langle f({\bf s}),\phi_{\bf m}({\bf s})  \rangle  \notag\\
\phantom{a_{\bf m}}{}= M^{-n} |S_{\bf m}|^{-1} r_{\bf m}^{-1}
|S_n|\sum_{{\bf s}\in \breve F_M^n} |S_{\bf s}|^{-1} c_{\bf
s}f({\bf s}) \overline{\phi_{\bf m}({\bf s})}.\label{multiv-5}
 \end{gather}

A validity of the expansions \eqref{multiv-4} and \eqref{multiv-5}
follows from the fact that numbers of elements in $\breve  D_M^+$
and $\breve  F_M^n$ are the same and from the orthogonality
relation \eqref{multiv-3}.

\subsection{Other discrete cosine transforms}\label{section11.9}
Along with the discrete cosine transform of
Subsection~\ref{section11.6} there are other discrete transforms
with the discrete cosine function as a kernel. In \cite{Strang}
the discrete cosine transforms are called as DCT-1, DCT-2, DCT-3,
DCT-4. The transform DCT-1 is in fact the transform, considered in
Subsection~\ref{section11.6}. Let us describe all these transforms
(including the transform DCT-1). They are determined by a positive
integer $N$.
\medskip

\noindent {\bf DCT-1.} This transform is given by the kernel
 \begin{gather}\label{DCT-1-1}
\mu_r(k)=\cos \frac{\pi rk}N
 \end{gather}
(we preserve the notation used in the literature on signal
processing), where
 \begin{gather}\label{DCT-1-2}
k,r\in \{ 0,1,2,\dots, N\} .
 \end{gather}

The orthogonality relation for these discrete functions is given
by
 \begin{gather}\label{DCT-1-3}
\sum_{k=0}^N c_k  \cos \frac{\pi rk}N  \cos \frac{\pi r'k}N =h_r
\frac N2 \delta_{rr'},
 \end{gather}
where $c_k=1/2$ for $k=0,N$ and $c_k=1$ otherwise, $h_r=2$ for
$r=0,N$ and $h_r=1$ otherwise.

Thus, these functions give the expansion
 \begin{gather}\label{DCT-1-4}
f(k)=\sum_{r=0}^N a_r  \cos \frac{\pi rk}N ,\qquad {\rm
where}\qquad
 a_r= \frac2{h_rN} \sum_{k=0}^N c_kf(k) \cos \frac{\pi rk}N .
 \end{gather}

\noindent {\bf DCT-2.} This transform is given by the kernel
 \begin{gather}\label{DCT-2-1}
\omega_r(k)=\cos \frac{\pi (r+\frac12)k}N,
 \end{gather}
where
\[
 k,r\in \{ 0,1,2,\dots,N-1\}.
\]
The orthogonality relation for these discrete functions is given
by
 \begin{gather}\label{DCT-2-3}
\sum_{k=0}^{N-1} c_k  \cos \frac{\pi (r+\frac12)k}N \cos \frac{\pi
(r'+\frac12)k}N= \frac N2 \delta_{rr'},
 \end{gather}
where $c_k=1/2$ for $k=0$ and $c_k=1$ otherwise.

These functions determine the expansion
 \begin{gather}\label{DCT-2-4}
f(k)=\sum_{r=0}^{N-1} a_r  \omega_r(k), \qquad
 {\rm where} \qquad
 a_r= \frac2{N} \sum_{k=0}^{N-1} c_kf(k) \omega_r(k).
 \end{gather}

\noindent {\bf DCT-3.} This transform is determined by the kernel
 \begin{gather}\label{DCT-3-1}
\sigma_r(k)=\cos \frac{\pi r(k+\frac12)}N,
 \end{gather}
where $k$ and $r$ run over the values $\{ 0,1,2,\dots,N-1\}$. The
orthogonality relation for these discrete functions is given by
the formula{\samepage
 \begin{gather}\label{DCT-3-3}
\sum_{k=0}^{N-1}  \cos \frac{\pi r(k+\frac12)}N \cos \frac{\pi
r'(k+\frac12)}N=h_r \frac N2 \delta_{rr'},
 \end{gather}
where $h_k=2$ for $k=0$ and $h_k=1$ otherwise.}

These functions give the expansion
 \begin{gather}\label{DCT-3-4}
f(k)=\sum_{r=0}^{N-1} a_r  \cos \frac{\pi r(k+\frac12)}N, \qquad
 {\rm where} \qquad
 a_r= \frac2{h_rN} \sum_{k=0}^{N-1} f(k) \cos \frac{\pi r(k+\frac12)}N.
 \end{gather}

\noindent {\bf DCT-4.} This transform is given by the kernel
 \begin{gather}\label{DCT-4-1}
\tau_r(k)=\cos \frac{\pi (r+\frac12)(k+\frac12)}N,
 \end{gather}
where $k$ and $r$ run over the values $\{ 0,1,2,\dots,N-1\}$. The
orthogonality relation for these discrete functions is given by
 \begin{gather}\label{DCT-4-3}
\sum_{k=0}^{N-1}  \cos \frac{\pi (r+\frac12)(k+\frac12)}N \cos
\frac{\pi (r'+\frac12)(k+\frac12)}N= \frac N2 \delta_{rr'}.
 \end{gather}
These functions determine the expansion
 \begin{gather}\label{DCT-4-4}
f(k)=\sum_{r=0}^{N-1} a_r  \cos \frac{\pi
(r+\frac12)(k+\frac12)}N,
\end{gather}
where
\[
 a_r= \frac2{N} \sum_{k=0}^{N-1} f(k) \cos \frac{\pi (r+\frac12)(k+\frac12)}N.
\]

Note that there also exist  four discrete sine transforms,
corresponding to the above discrete cosine transforms. They are
obtained from the cosine transforms by replacing in
\eqref{DCT-1-4}, \eqref{DCT-2-4}, \eqref{DCT-3-4} and
\eqref{DCT-4-4} cosines discrete functions by sine discrete
functions (see \cite{R} and \cite{Sanch}).

\subsection{Other antisymmetric multivariate discrete cosine transforms}\label{section11.10}
Each of the discrete cosine transforms DCT-1, DCT-2, DCT-3, DCT-4
generates the corres\-pon\-ding antisymmetric multivariate
discrete cosine transforms. We call these transforms as AMDCT-1,
AMDCT-2, AMDCT-3 and AMDCT-4. Let us give these transforms without
proof. Their proofs are the same as in the case of antisymmetric
multivariate discrete cosine transforms of
Subsection~\ref{section11.7}. Below we use the notation
$D^{n,-}_N$ for the subset of the set $D_N^n\equiv D_N\times
D_N\times \dots \times D_N$ ($n$ times) with $D_N=\{
0,1,2,\dots,N\}$ consisting of points ${\bf
r}=(r_1,r_2,\dots,r_n)$, $r_i\in D_N$, such that
\[
 N\ge r_1>r_2>\cdots >r_n\ge 0.
\]

{\bf AMDCT-1.}  This transform is given by the kernel
 \begin{gather}\label{AMDCT-1-1}
M_{\bf r}({\bf k})=|S_n|^{-1/2} \det \left(
\mu_{r_i}(k_j)\right)_{i,j=1}^n,
 \end{gather}
where $\mu_{r}(k)=\cos \frac{\pi rk}N$ and ${\bf
k}=(k_1,k_2,\dots,k_n)$, $k_i\in \{ 0,1,2,\dots,N\}$. The
orthogonality relation for these kernels is
 \begin{gather}\label{AMDCT-1-2}
\langle M_{\bf r}({\bf k}),M_{{\bf r}'}({\bf k}) \rangle =
|S_n|\sum_{{\bf k}\in D^{n,-}_N}c_{\bf k} M_{\bf r}({\bf k})
M_{{\bf r}'}({\bf k}) =h_{\bf r}  \left( \frac N2\right)^n
\delta_{{\bf r}{\bf r}'},
 \end{gather}
where
\[
 c_{\bf k}=c_1c_2\cdots c_n,\qquad  h_{\bf k}=h_1h_2\cdots h_n,
\]
and $c_i$ and $h_j$ are such as in formula \eqref{DCT-1-3}.

This transform is given by the formula
 \begin{gather}\label{AMDCT-1-3}
f( {\bf k})=\sum _{{\bf r}\in D^{n,-}_N} a_{\bf r}
 M_{\bf r}({\bf k}) , \qquad {\rm where} \qquad
a_{\bf r}= h_{\bf r}^{-1}|S_n|\left( \frac2N\right)^n \sum_{{\bf
k}\in D^{n,-}_N} c_{\bf k} f({\bf k}) M_{\bf r}({\bf k}).
 \end{gather}
The Plancherel formula for this transform is
\[
 |S_n| \sum_{{\bf k}\in D^{n,-}_N} c_{\bf k} |f({\bf k})|^2=
\left( \frac N2\right)^n \sum _{{\bf r}\in D^{n,-}_N}  h_{\bf r}
|a_{\bf r}|^2.
\]


{\bf AMDCT-2.}  We use the subset $D_{N-1}^{n,-}$ of the set
$D^n_{N-1}$ with $D_{N-1}=\{ 0,1,2,\dots,N-1\}$ consisting of
points ${\bf r}=(r_1,r_2,\dots,r_n)$, $r_i\in D_{N-1}$, such that
\[
 N-1\ge r_1>r_2>\cdots >r_n\ge 0.
\]
This transform is given by the kernel
 \begin{gather}\label{AMDCT-2-1}
\Omega_{\bf r}({\bf k})=|S_n|^{-1/2}\det \left(
\omega_{r_i}(k_j)\right)_{i,j=1}^n,
 \end{gather}
where $\omega_{r}(k)=\cos \frac{\pi (r+\frac12)k}N$ and ${\bf k}=
(k_1,k_2,\dots,k_n)$, $k_i\in \{ 0,1,2,\dots,N-1\}$. The
orthogonality relation for these kernels is
 \begin{gather}\label{AMDCT-2-2}
\langle \Omega_{\bf r}({\bf k}),\Omega_{{\bf r}'}({\bf k}) \rangle
= |S_n|\sum_{{\bf k}\in D_{N-1}^{n,-}}c_{\bf k} \Omega_{\bf
r}({\bf k}) \Omega_{{\bf r}'}({\bf k}) = \left(\frac N2\right)^n
\delta_{{\bf r}{\bf r}'},
 \end{gather}
where $c_{\bf k}=c_1c_2\cdots c_n$ and $c_i$ are such as in
formula \eqref{DCT-2-3}.

This transform is given by the formula
 \begin{gather}\label{AMDCT-2-3}
f( {\bf k})=\sum _{{\bf r}\in D_{N-1}^{n,-}} a_{\bf r}
 \Omega_{\bf r}({\bf k}) ,\qquad {\rm where}\qquad
a_{\bf r}=|S_n| \left( \frac2N\right)^n \sum_{{\bf k}\in
D_{N-1}^{n,-}} c_{\bf k} f({\bf k}) \Omega_{\bf r}({\bf k}).
 \end{gather}
The Plancherel formula for this transform is of the form
\[
 |S_n| \sum_{{\bf k}\in D_{N-1}^{n,-}} c_{\bf k} |f({\bf k})|^2=
\left( \frac N2\right)^n \sum _{{\bf r}\in D_{N-1}^{n,-}} |a_{\bf
r}|^2.
\]

{\bf AMDCT-3.}  This transform is given by the kernel
 \begin{gather}\label{AMDCT-3-1}
\Sigma_{\bf r}({\bf k})=|S_n|^{-1/2} \det \left(
\sigma_{r_i}(k_j)\right)_{i,j=1}^n,\qquad {\bf r}\in
D_{N-1}^{n,-},\quad k_j\in D_{N-1},
 \end{gather}
where $\sigma_{r}(k)=\cos \frac{\pi r(k+\frac12)}N$. The
orthogonality relation for these kernels is
 \begin{gather}\label{AMDCT-3-2}
\langle \Sigma_{\bf r}({\bf k}),\Sigma_{{\bf r}'}({\bf k}) \rangle
= |S_n|\sum_{{\bf k}\in D_{N-1}^{n,-}} \Sigma_{\bf r}({\bf k})
\Sigma_{{\bf r}'}({\bf k}) =h_{\bf r} \left(\frac N2\right)^n
\delta_{{\bf r}{\bf r}'},
 \end{gather}
where $h_{\bf r}=h_1h_2\cdots h_n$ and $h_j$ are such as in
formula \eqref{DCT-3-3}.

This transform is given by the formula
 \begin{gather}\label{AMDCT-3-3}
f( {\bf k})=\sum _{{\bf r}\in D_{N-1}^{n,-}} a_{\bf r}
 \Sigma_{\bf r}({\bf k}) ,\qquad {\rm where}\qquad
a_{\bf r}=h_{\bf r}^{-1}|S_n| \left( \frac2N\right)^n \sum_{{\bf
k}\in D_{N-1}^{n,-}}  f({\bf k}) \Sigma_{\bf r}({\bf k}).
 \end{gather}
The Plancherel formula is of the form
\[
 |S_n| \sum_{{\bf k}\in D_{N-1}^{n,-}} |f({\bf k})|^2=
\left( \frac N2\right)^n \sum _{{\bf r}\in D_{N-1}^{n,-}} h_{\bf
r} |a_{\bf r}|^2.
\]

{\bf AMDCT-4.}  This transform is given by the kernel
 \begin{gather}\label{AMDCT-4-1}
T_{\bf r}({\bf k})=|S_n|^{-1/2} \det \left(
\tau_{r_i}(k_j)\right)_{i,j=1}^n, \qquad {\bf r}\in
D_{N-1}^{n,-},\quad k_j\in D_{N-1},
 \end{gather}
where $\tau_{r}(k)=\cos \frac{\pi (k+\frac12)(r+\frac12)}N$. The
orthogonality relation for these kernels is
 \begin{gather}\label{AMDCT-4-2}
\langle T_{\bf r}({\bf k}),T_{{\bf r}'}({\bf k}) \rangle =
|S_n|\sum_{{\bf k}\in D_{N-1}^{n,-}} T_{\bf r}({\bf k}) T_{{\bf
r}'}({\bf k}) = \left(\frac N2\right)^n \delta_{{\bf r}{\bf r}'}.
 \end{gather}

This transform is given by the formula
 \begin{gather}\label{AMDCT-4-3}
f( {\bf k})=\sum _{{\bf r}\in D_{N-1}^{n,-}} a_{\bf r}
 T_{\bf r}({\bf k}) ,\qquad {\rm where}\qquad
a_{\bf r}=|S_n| \left( \frac2N\right)^n \sum_{{\bf k}\in
D_{N-1}^{n,-}}  f({\bf k}) T_{\bf r}({\bf k}).
 \end{gather}
The Plancherel formula for this transform is
\[
 |S_n| \sum_{{\bf k}\in D_{N-1}^{n,-}} |f({\bf k})|^2=
\left( \frac N2\right)^n \sum _{{\bf r}\in D_{N-1}^{n,-}} |a_{\bf
r}|^2.
\]

\subsection{Other symmetric multivariate discrete cosine transforms}\label{section11.11}
To each of the discrete cosine transforms DCT-1, DCT-2, DCT-3,
DCT-4 there corresponds a~symmetric multivariate discrete
cosine transform. We denote the corresponding transforms as
SMDCT-1,  SMDCT-2, SMDCT-3, SMDCT-4. Below we give these
transforms without proof (proofs are the same as in the case of
symmetric multivariate discrete cosine transforms of
Subsection~\ref{section11.8}). We f\/ix a positive integer $N$ and
use the notation $D_{N}^{n,+}$ for the subset of the set
$D_N^n\equiv D_N\times D_N\times \dots \times D_N$ ($n$ times)
with $D_N=\{ 0,1,2,\dots,N\}$ consisting of points ${\bf
r}=(r_1,r_2,\dots,r_n)$, $r_i\in {\mathbb Z}^{\ge 0}$ such that
\[
 N\ge r_1\ge r_2\ge \cdots \ge r_n\ge 0.
\]
The set $D_{N}^{n,+}$ is an extension of the set $D_{N}^{n,-}$
from the previous subsection by adding points which are invariant
with respect of some elements of the permutation group $S_n$.

The set $D_N^n$ is obtained by action by elements of the group
$S_n$ upon $D_{N}^{n,+}$, that is, $D_N^n$~coincides with the set
$\{ wD_{N}^{n,+}; w\in S_n\}$. However, in $\{ wD_{N}^{n,+}; w\in
S_n\}$, some points occur  several times. Namely, a point  ${\bf
k}_0\in D_{N}^{n,+}$ occurs $|S_{{\bf k}_0}|$ times in the set $\{
w D_{N}^{n,+}; w\in S_n\}$, where $|S_{{\bf k}_0}|$ is an order of
the subgroup $S_{{\bf k}_0}\subset S_n$ consisting of elements
$w\in S_n$ leaving ${\bf k}_0$ invariant.
\medskip

{\bf SMDCT-1.}  This transform is given by the kernel
 \begin{gather}\label{SMDCT-1-1}
\hat M_{\bf r}({\bf k})=|S_n|^{-1/2}\sum_{w\in S_n}
\mu_{r_{w(1)}}(k_1) \mu_{r_{w(2)}}(k_2)\cdots \mu_{r_{w(n)}}(k_n),
 \end{gather}
where, as before, $\mu_{r}(k)=\cos \frac{\pi rk}N$ is the discrete
cosine function from Subsection~\ref{section11.9}, ${\bf
k}=(k_1,k_2,\dots,k_n)$, $k_i\in \{ 0,1,2,\dots,N\}$, and the set
$(w(1),w(2),\dots,w(n))$ is obtained from the set $(1,2,\dots,n)$
by applying the permutation $w\in S_n$. The orthogonality relation
for these kernels is
 \begin{gather}\label{SMDCT-1-2}
\langle \hat M_{\bf r}({\bf k}),\hat M_{{\bf r}'}({\bf k})
\rangle =|S_n|\sum_{{\bf k}\in D_{N}^{n,+}} |S_{\bf k}|^{-1}
c_{\bf k} \hat M_{\bf r}({\bf k}) \hat M_{{\bf r}'}({\bf k})
=h_{\bf r} \left(\frac N2\right)^n |S_{\bf r}|  \delta_{{\bf
r}{\bf r}'},
 \end{gather}
where $S_{\bf r}$ is the subgroup of $S_n$ consisting of elements
leaving ${\bf r}$ invariant,
\[
 c_{\bf k}=c_1c_2\cdots c_n,\qquad  h_{\bf k}=h_1h_2\cdots h_n,
\]
and $c_i$ and $h_j$ are such as in formula \eqref{DCT-1-3}.

This transform is given by the formula
 \begin{gather}\label{SMDCT-1-3}
f( {\bf k})=\sum _{{\bf r}\in D_{N}^{n,+}} a_{\bf r}
 \hat M_{\bf r}({\bf k}) ,
 \end{gather}
where
\[
a_{\bf r}= h_{\bf r}^{-1}|S_{\bf r}|^{-1}|S_n| \left(
\frac2N\right)^n \sum_{{\bf k}\in D_{N}^{n,+}} |S_{\bf k}|^{-1}
 c_{\bf k} f({\bf k}) \hat M_{\bf r}({\bf k}).
\]
The Plancherel formula for this transform is
\[
 |S_n|\sum_{{\bf k}\in D_{N}^{n,+}} |S_{\bf k}|^{-1} c_{\bf k} |f( {\bf k})|^2=
 \left(\frac N2\right)^n
  \sum _{{\bf r}\in D_{N}^{n,+}} h_{\bf r} |S_{\bf r}| |a_{\bf r}|^2.
\]

This transform is in fact a variation of the symmetric
multivariate discrete cosine transforms from
Subsection~\ref{section11.8}.
\medskip

{\bf SMDCT-2.}  This transform is given by the kernel
 \begin{gather}\label{SMDCT-2-1}
\hat\Omega_{\bf r}({\bf k})=|S_n|^{-1/2} \sum_{w\in S_n}
\omega_{r_{w(1)}}(k_1) \omega_{r_{w(2)}}(k_2)\cdots
\omega_{r_{w(n)}}(k_n),
\\
 {\bf r}\in D_{N-1}^{n,+},\qquad r_j\in D_{N-1}\equiv \{ 0,1,2,\dots, N-1\},\nonumber
\end{gather}
where $\omega_{r}(k)=\cos \frac{\pi (r+\frac12)k}N$ and ${\bf
k}=(k_1,k_2,\dots, k_n)$, $k_i\in \{ 0,1,2,\dots, N-1\}$. The
orthogonality relation for these kernels is
 \begin{gather}\label{SMDCT-2-2}
\langle \hat\Omega_{\bf r}({\bf k}),\hat\Omega_{{\bf r}'}({\bf k})
\rangle =|S_n|\sum_{{\bf k}\in D_{N-1}^{n,+}} |S_{\bf
k}|^{-1}c_{\bf k} \hat\Omega_{\bf r}({\bf k}) \hat\Omega_{{\bf
r}'}({\bf k}) = \left(\frac N2\right)^n |S_{\bf r}| \delta_{{\bf
r}{\bf r}'},
 \end{gather}
where $D_{N-1}^{n,+}$ is the set $D_{N}^{n,+}$ with $N$ replaced
by $N-1$, $c_{\bf k}=c_1c_2\cdots c_n$ and $c_j$ are such as
in~\eqref{DCT-2-3}.

This transform is given by the formula
 \begin{gather}\label{SMDCT-2-3}
f( {\bf k})=\sum _{{\bf r}\in D_{N-1}^{n,+}} a_{\bf r}
 \hat\Omega_{\bf r}({\bf k}) ,
 \end{gather}
  where
\[
a_{\bf r}=|S_n| |S_{\bf r}|^{-1} \left( \frac2N\right)^n
\sum_{{\bf k}\in D_{N-1}^{n,+}} |S_{\bf k}|^{-1} c_{\bf k} f({\bf
k}) \hat\Omega_{\bf r}({\bf k}).
\]
The Plancherel formula for this transform is of the form
\[
 |S_n|\sum_{{\bf k}\in D_{N-1}^{n,+}} |S_{\bf k}|^{-1} c_{\bf k} |f( {\bf k})|^2=
 \left(\frac N2\right)^n
  \sum _{{\bf r}\in D_{N-1}^{n,+}}|S_{\bf r}| |a_{\bf r}|^2.
\]

{\bf SMDCT-3.}  This transform is given by the kernel
 \begin{gather}\label{SMDCT-3-1}
\hat\Sigma_{\bf r}({\bf k})=|S_n|^{-1/2} \sum_{w\in S_n}
\sigma_{r_{w(1)}}(k_1) \sigma_{r_{w(2)}}(k_2)\cdots
\sigma_{r_{w(n)}}(k_n),
\\
 {\bf r}\in D_{N-1}^{n,+},\quad r_j\in D_{N-1},\nonumber
 \end{gather}
where $\sigma_{r}(k)=\cos \frac{\pi r(k+\frac12)}N$. The
orthogonality relation for these kernels is
 \begin{gather}\label{SMDCT-3-2}
\langle \hat\Sigma_{\bf r}({\bf k}),\hat\Sigma_{{\bf r}'}({\bf k})
\rangle =|S_n|\sum_{{\bf k}\in D_{N-1}^{n,+}} |S_{\bf k}|^{-1}
\hat\Sigma_{\bf r}({\bf k}) \hat\Sigma_{{\bf r}'}({\bf k}) =h_{\bf
r} \left(\frac N2\right)^n |S_{\bf r}| \delta_{{\bf r}{\bf r}'},
 \end{gather}
where $h_{\bf r}=h_1h_2\cdots h_n$ and $h_i$ are such as in
formula \eqref{DCT-3-3}.

This transform is given by the formula
 \begin{gather}\label{SMDCT-3-3}
f( {\bf k})=\sum _{{\bf r}\in D_{N-1}^{n,+}} a_{\bf r}
 \hat\Sigma_{\bf r}({\bf k}) ,
 \end{gather}
 where
 \[
a_{\bf r}=h_{\bf r}^{-1}|S_{\bf r}|^{-1}|S_n| \left(
\frac2N\right)^n \sum_{{\bf k}\in D_{N-1}^{n,+}} |S_{\bf k}|^{-1}
f({\bf k}) \hat\Sigma_{\bf r}({\bf k}).
\]
The Plancherel formula is of the form
\[
 |S_n|\sum_{{\bf k}\in D_{N-1}^{n,+}} |S_{\bf k}|^{-1} |f( {\bf k})|^2=
 \left(\frac N2\right)^n
  \sum _{{\bf r}\in D_{N-1}^{n,+}}h_{\bf r} |S_{\bf r}|  |a_{\bf r}|^2.
\]

{\bf SMDCT-4.}  This transform is given by the kernel
 \begin{gather}\label{SMDCT-4-1}
\hat T_{\bf r}({\bf k})=|S_n|^{-1/2} \sum_{w\in S_n}
\tau_{r_{w(1)}}(k_1) \tau_{r_{w(2)}}(k_2)\cdots
\tau_{r_{w(n)}}(k_n),
\\
 {\bf r}\in D_{N-1}^{n,+},\qquad r_j\in D_{N-1},\nonumber
\end{gather}
where $\tau_{r}(k)=\cos \frac{\pi (k+\frac12)(r+\frac12)}N$. The
orthogonality relation for these kernels is
 \begin{gather}\label{SMDCT-4-2}
\langle \hat T_{\bf r}({\bf k}),\hat T_{{\bf r}'}({\bf k})
\rangle =|S_n|\sum_{{\bf k}\in D_{N-1}^{n,+}} |S_{\bf k}|^{-1}
\hat T_{\bf r}({\bf k}) \hat T_{{\bf r}'}({\bf k}) = \left(\frac
N2\right)^n |S_{\bf r}| \delta_{{\bf r}{\bf r}'}.
 \end{gather}

This transform is given by the formula
 \begin{gather}\label{SMDCT-4-3}
f( {\bf k})=\sum _{{\bf r}\in D_{N-1}^{n,+}} a_{\bf r}
 \hat T_{\bf r}({\bf k}) ,
 \end{gather}
 where
\[
a_{\bf r}= \left( \frac2N\right)^n  |S_{\bf r}|^{-1}|S_n|
\sum_{{\bf k}\in D_{N-1}^{n,+}} |S_{\bf k}|^{-1} f({\bf k})\hat
T_{\bf r}({\bf k}).
\]
The Plancherel formula for this transform is
\[
 |S_n|\sum_{{\bf k}\in D_{N-1}^{n,+}} |S_{\bf k}|^{-1} |f( {\bf k})|^2=
 \left(\frac N2\right)^n
  \sum _{{\bf r}\in D_{N-1}^{n,+}}|S_{\bf r}| |a_{\bf r}|^2.
\]

\section[Solutions of the Laplace equation on
$n$-dimensional simplexes]{Solutions of the Laplace equation\\ on
$\boldsymbol{n}$-dimensional simplexes}\label{section12}

We have seen in \cite{KP06} that symmetric orbit functions are
solutions of the Neumann boundary value problem on $n$-dimensional
simplexes. That is, they are solutions of the Laplace equation
$\Delta f(x)=\lambda f(x)$ on the fundamental domain $F$ of the
corresponding af\/f\/ine Weyl group $W^{\rm aff}$ satisfying the
Neumann boundary condition
 \begin{gather}\label{Neum}
 \left.\frac{\partial \phi_\lambda(x)}{\partial
m}\right|_{\partial F}=0\,, \qquad \lambda\in P_+,
\end{gather}
where $\partial F$ is the $(n-1)$-dimensional boundary of $F$ and
$m$ is the normal to the boundary. In this section we show that
antisymmetric orbit functions are solutions of the Laplace
operator, which vanish on the boundary $\partial F$ of the
fundamental domain $F$.

\subsection[The case of $n$-dimensional simplexes related to
$A_n$, $B_n$, $C_n$ and $D_n$]{The case of
$\boldsymbol{n}$-dimensional simplexes related to
$\boldsymbol{A_n}$, $\boldsymbol{B_n}$, $\boldsymbol{C_n}$ and
$\boldsymbol{D_n}$} \label{section12.1}
Let $F$ be the fundamental domain of one of the af\/f\/ine Weyl
groups $W^{\rm aff}(A_n)$, $W^{\rm aff}(B_n)$, $W^{\rm aff}(C_n)$,
$W^{\rm aff}(D_n)$ (see Subsection~\ref{section5.9} for an
explicit form of these domains). We use orthogo\-nal coordinates
$x_1,x_2,\dots ,x_{n+1}$ on $F$ in the case of $W^{\rm aff}(A_n)$
and the orthogonal coordinates $x_1,x_2,\dots ,x_n$ in other cases
(see Section~\ref{section3}). Thus the fundamental domain $F$ for
$W^{\rm aff}(A_n)$ is placed in the hyperplane $x_1+x_2+\cdots
+x_{n+1}=0$.

The Laplace operator on $F$ in the orthogonal coordinates has the
form
\[
\Delta=\frac{\partial^2}{\partial
x^2_1}+\frac{\partial^2}{\partial x^2_2}+\cdots
+\frac{\partial^2}{\partial x^2_r} ,
\]
where $r=n+1$ for $A_n$ and $r=n$ for $B_n$, $C_n$ and $D_n$. Let
us consider the case $B_n$. We take a~summand from the expression
\eqref{orb-B} for the antisymmetric orbit function
$\varphi_\lambda(x)$ of $B_n$ and act upon it by the operator
$\Delta$. We get
\begin{gather*}
\Delta  e^{2\pi{\rm i}((w(\varepsilon \lambda))_1x_1+
 \cdots + (w(\varepsilon \lambda))_nx_{n})}
\\
\qquad  {}=-4\pi^2[(\varepsilon_1m_1)^2+
 \cdots +(\varepsilon_nm_{n})^2]e^{2\pi{\rm i}((w(\varepsilon \lambda))_1x_1+
 \cdots + (w(\varepsilon \lambda))_nx_{n})}
\\
\qquad  {}=-4\pi^2(m_1^2+\cdots +m_n^2)e^{2\pi{\rm
i}((w(\varepsilon \lambda))_1x_1+
 \cdots + (w(\varepsilon \lambda))_nx_{n})}
\\
\qquad  {}=-4\pi^2 \langle \lambda,\lambda \rangle\, e^{2\pi{\rm
i}((w(\varepsilon \lambda))_1x_1+
 \cdots + (w(\varepsilon \lambda))_nx_{n})},
\end{gather*}
where $\lambda=(m_1,m_2,\dots ,m_n)$ is the weight, determining
$\varphi_\lambda(x)$, in the orthogonal coordinates and $w\in
S_n$. Since this action of $\Delta$ does not depend on a summand
from \eqref{orb-B}, we have
\begin{gather}\label{Lap}
\Delta \varphi_\lambda(x)= -4\pi^2\langle \lambda,\lambda \rangle
\varphi_\lambda(x).
 \end{gather}
For $A_n$, $C_n$ and $D_n$ this formula also holds and the
corresponding proofs are the same. Remark that in the case $A_n$
the scalar product $\langle \lambda,\lambda \rangle$ is equal to
\[
\langle \lambda,\lambda \rangle =m_1^2+m_2^2+\cdots +m_{n+1}^2.
\]

The formula \eqref{Lap} can be generalized in the following way.
Let $\sigma_k(y_1,y_2,\dots,y_r)$ be the $k$-th elementary
symmetric polynomial of degree $k$ of the variables
$y_1,y_2,\dots,y_r$, that is,
\[
 \sigma_k(y_1,y_2,\dots,y_r)=\sum_{1\le k_1<k_2<\cdots <k_r\le n}
y_{k_1} y_{k_2}\cdots y_{k_r}.
\]
Then
\begin{gather}\label{Lap-gen}
\sigma_k\left( \tfrac{\partial^2}{\partial
x^2_1},\tfrac{\partial^2}{\partial
x^2_2},\dots,\tfrac{\partial^2}{\partial x^2_r} \right)
\varphi_\lambda(x)= (-4\pi^2)^k \sigma_k(m_1^2,m_2^2,\dots,m_r^2)
\varphi_\lambda(x),\quad k=1,2,\dots,r,
 \end{gather}
where, as before, $r=n+1$ for $A_n$ and $r=n$ for $B_n$, $C_n$ and
$D_n$.
 $r$ dif\/ferential equations~\eqref{Lap-gen} are algebraically
independent.

Thus, antisymmetric orbit functions are eigenfunctions of the
operator $\sigma_k\left( \frac{\partial^2}{\partial
x^2_1},\frac{\partial^2}{\partial
x^2_2},\dots,\frac{\partial^2}{\partial x^2_r} \right)$,
$k=1,2,\dots,n$, on the fundamental domain $F$ satisfying the
boundary condition
\begin{gather}\label{Neum1}
 \varphi_\lambda(x) =0, \qquad \lambda\in D_+,
\end{gather}
(see Subsection~\ref{section5.3}).

\subsection[The Laplace operator in $\omega$-basis]{The Laplace operator in $\boldsymbol{\omega}$-basis}
\label{section12.2}
We may parametrize elements of $F$ by coordinates in the
$\omega$-basis: $x=\theta_1\omega_1+\cdots+ \theta_2\omega_2$.
Denoting by $\p_k$ partial derivative with respect to $\theta_k$,
we have the Laplace operator $\Delta$ in the form
\begin{gather}\label{operator}
\Delta = \frac12\sum_{i,j=1}^n\l\alpha_j ,\alpha_j\r ^{-1}
       M_{ij}\p_i\p_j,
\end{gather}
where $(M_{ij})$ is the corresponding Cartan matrix. One can see
that it is indeed the Laplace operator as follows. The matrix
$(S_{ij})=(\l\alpha_j ,\alpha_j\r ^{-1}M_{ij})$ is symmetric with
respect to transposition and its determinant is positive. Hence it
can be diagonalized, so that $\Delta$ becomes a sum of second
derivatives with no mixed derivative terms.

\subsection{Rank two and three special cases}\label{section12.3}
We write down explicit form of the Laplace operators in
coordinates in the $\omega$-basis for ranks 2 and 3. For rank two
the operator $\Delta$ is of the form
\begin{alignat}{3}
A_2\,&:&&\ (\p_1^2-\p_1\p_2+\p_2^2)\varphi
     =-\tfrac{4\pi^2}3(a^2+ab+b^2)\varphi,\quad
     &&F=\{0,\omega_1,\omega_2\},
\\
C_2\,&:&&\ (2\p_1^2-2\p_1\p_2+\p_2^2)\varphi
     =-2\pi^2(a^2+4ab+4b^2)\varphi,\quad
     &&F=\{0,\omega_1,\omega_2\},
\\
G_2\,&:&&\ (\p_1^2-3\p_1\p_2+3\p_2^2)\varphi
     =-\tfrac{4\pi^2}3(3a^2+3ab+b^2)\varphi,\quad
     &&F=\{0,\tfrac{\omega_1}2,\omega_2\}.
\end{alignat}
Here,  to simplify notation, $\varphi$ stands for
$\varphi_\lambda(x)$, $\lambda=(a\;b)$ and
$x=(\theta_1\;\theta_2)$. Although the same symbols are used for
analogous objects in the three cases, their geometric meaning is
very dif\/ferent. It is given by the appropriate Cartan matrix $M$
from \eqref{Mmatrix}. In particular, the vertices of $F$ form an
equilateral triangle in the case of $A_2$, for $C_2$ the triangle
is half of a square, and it is a half of an equilateral triangle
for $G_2$.

In the semisimple case $A_1\times A_1$ one has
$M=2\left(\begin{smallmatrix}1&0\\0&1\end{smallmatrix}\right)$,
therefore $\Delta=2\p_1^2+2\p_2^2$, and $\varphi_\lambda(x)$ is
the product of two antisymmetric orbit functions, one from each
$A_1$. The fundamental domain is a square.

There are three 3-dimensional cases to consider, namely $A_3$,
$B_3$, and $C_3$. In addition there are four cases involving
non-simple groups of the same rank. For $A_3$, $B_3$, and $C_3$
the result can be represented by the formulas
 \begin{alignat}{2} A_3\
&:&\quad\Delta&=\p_1^2+\p_2^2+\p_3^2-\p_1\p_2-\p_2\p_3,
\notag\\
B_3\ &:&\quad\Delta&=\p_1^2+\p_2^2+2\p_3^2-\p_1\p_2-2\p_2\p_3,
\notag
\\
C_3\ &:&\quad\Delta&=2 \p_1^2+2 \p_2^2+2\p_3^2-2
\p_1\p_2-2\p_2\p_3.
\end{alignat}

\subsection{Antisymmetric orbit functions as eigenfunctions of
other operators}\label{section12.4}
Antisymmetric orbit functions are eigenfunctions of many other
operators. We consider examples of such operators.

 We associate with each $y\in E_n$ the shift operator $T_y$ which
acts on the exponential functions $e^{2\pi{\rm i}\langle \lambda,x
\rangle}$ as
\[
T_y e^{2\pi{\rm i}\langle \lambda,x \rangle}=e^{2\pi{\rm i}\langle
\lambda,x+y \rangle}=e^{2\pi{\rm i}\langle \lambda,y
\rangle}e^{2\pi{\rm i}\langle \lambda,x \rangle}.
\]
The action of elements of the Weyl group $W$ on functions, given
on $E_n$, is given as $wf(x)=f(wx)$. For each $y\in E_n$ we
def\/ine an operator acting on orbit functions by the formula
\[
D_y=\sum_{w\in W} (\det w) wT_y.
\]
Then
\begin{gather*}
D_y\varphi_\lambda(x)=D_y\sum _{w\in W}(\det w) e^{2\pi{\rm
i}\langle w\lambda,x \rangle}
\notag \\
\phantom{D_y\varphi_\lambda(x)}{} =\sum_{w'\in W}(\det w') \sum
_{w\in W} (\det w) e^{2\pi{\rm i}\langle w\lambda,y \rangle}
e^{2\pi{\rm i}\langle w\lambda,w'x \rangle}
\notag \\
\phantom{D_y\varphi_\lambda(x)}{}=\sum _{w\in W} (\det w)
e^{2\pi{\rm i}\langle w\lambda,y \rangle} \sum _{w'\in W} (\det
w') e^{2\pi{\rm i}\langle w\lambda,w'x \rangle}
\notag \\
\phantom{D_y\varphi_\lambda(x)}{}=\sum _{w\in W} (\det w)
e^{2\pi{\rm i}\langle w\lambda,y \rangle} \sum _{w'\in W} (\det
w') e^{2\pi{\rm i}\langle {w'}^{-1}w\lambda,x \rangle}
\notag \\
\phantom{D_y\varphi_\lambda(x)}{}=\sum _{w\in W}  e^{2\pi{\rm
i}\langle w\lambda,y \rangle} \sum _{w'\in W} (\det {w'}^{-1}w)
e^{2\pi{\rm i}\langle {w'}^{-1}w\lambda,x \rangle}
\notag \\
\phantom{D_y\varphi_\lambda(x)}{}= \sum _{w\in W} e^{2\pi{\rm
i}\langle w\lambda,y \rangle}\varphi_\lambda(x)
=\phi_\lambda(y)\varphi_\lambda(x), \notag
\end{gather*}
that is, $\varphi_\lambda(x)$ is an eigenfunction of the operator
$D_y$ with eigenvalue $\phi_\lambda(y)$.

Now we consider the operator
\[
\hat D_y=\sum_{w\in W}  wT_y.
\]
Then, conducting the same reasoning as above, we receive the
relation
\[
\hat D_y \varphi_\lambda(x)= \varphi_\lambda(y)\varphi_\lambda(x),
\]
that is, $\varphi_\lambda(x)$ is an eigenfunction of the operator
$\hat D_y$ with eigenvalue $\varphi_\lambda(y)$.

It is proved in the same way that
\[
 D_y \phi_\lambda(x)= \varphi_\lambda(y)\phi_\lambda(x),
\]
that is, the symmetric orbit function $\phi_\lambda(x)$ is an
eigenfunction of the operator $ D_y$ with eigenvalue
$\varphi_\lambda(y)$.

It is shown similarly that in the cases of $A_n$, $B_n$, $C_n$,
$D_n$ antisymmetric orbit functions $\varphi_\lambda(x)$ are
eigenfunctions of the operators
\[
\sum_{w\in W} w\frac{\partial^2}{\partial x_i^2}, \qquad
 i=1,2,\dots ,r,
\]
where $x_1,x_2,\dots,x_r$ are orthogonal coordinates of the point
$x$, $r=n+1$ for $A_n$ and $r=n$ for other cases. In fact, these
operators are multiple to the Laplace operator $\Delta$.

It is easy to show that in the cases of $A_n$, $B_n$, $C_n$, and
also of $D_n$ with even $n$, antisymmetric orbit functions
$\varphi_\lambda(x)$ are solutions of the equations
\[
\sum_{w\in W} w\frac{\partial}{\partial x_i} f=0, \qquad
 i=1,2,\dots ,r.
\]


\section{Symmetric and antisymmetric functions}\label{section13}

Symmetric (antisymmetric) orbit functions are symmetrized
(antisymmetrized) versions of the exponential function, when
symmetrization (antisymmetrization) is fulf\/illed by a Weyl
group. Instead of the exponential function we can take any other
set of functions, for example, a set of orthogonal polynomials or
a countable set of functions. Then we obtain a corresponding set
of orthogonal symmetric (antisymmetric) polynomials or a set of
symmetric (antisymmetric) functions. Such sets of polynomials and
functions are a subject of investigation in this section.


\subsection[Symmetrization and antisymmetrization by (anti)symmetric
orbit functions]{Symmetrization and antisymmetrization\\ by
(anti)symmetric orbit functions}\label{section13.1}
Symmetric and antisymmetric orbit functions can be used for
symmetrization and antisymmetrization of functions. Let $u_m(x)$,
$m=0,1,2,\dots$, be a set of continuous functions of one
variables. We create functions of $n$ variables
\[
u_{i_1,i_2,\dots,i_n}(x_1,x_2,\dots,x_n)\equiv u_{i_1}(x_1)
u_{i_2}(x_2)\cdots u_{i_n}(x_n),\qquad i_k=0,1,2,\ldots.
\]
Then the functions
\begin{gather}\label{func-s}
{\tilde
u}_{i_1,i_2,\dots,i_n}(\lambda_1,\lambda_2,\dots,\lambda_n)=
\int_F u_{i_1,i_2,\dots,i_n}(x_1,x_2,\dots,x_n)
\phi_{\lambda}(x_1,x_2,\dots,x_n) d{\bf x},
\end{gather}
where $\lambda\equiv (\lambda_1,\lambda_2,\dots,\lambda_n)$,
$\phi_{\lambda}(x)$ is a symmetric orbit function, and $d{\bf x}$
is the Euclidean measure on $E_n$ (that is, $d{\bf x}=dx_1\cdots
dx_n$), is symmetric with respect to the action of the Weyl group
$W$. Indeed, for $w\in W$ we have
\begin{gather*}
{\tilde u}_{i_1,i_2,\dots,i_n}(w\lambda)= \int_F
u_{i_1,i_2,\dots,i_n}(x_1,x_2,\dots,x_n)
\phi_{w\lambda}(x_1,x_2,\dots,x_n)d{\bf x}
\notag\\
\phantom{{\tilde u}_{i_1,i_2,\dots,i_n}(w\lambda)}{}=\int_F
u_{i_1,i_2,\dots,i_n}(x_1,x_2,\dots,x_n)
\phi_{\lambda}(x_1,x_2,\dots,x_n)d{\bf x}={\tilde
u}_{i_1,i_2,\dots,i_n}(\lambda). \notag
\end{gather*}
Similarly, the functions
\begin{gather}\label{func-a}
{\hat u}_{i_1,i_2,\dots,i_n}(\lambda_1,\lambda_2,\dots,\lambda_n)=
\int_F u_{i_1,i_2,\dots,i_n}(x_1,x_2,\dots,x_n)
\varphi_{\lambda}(x_1,x_2,\dots,x_n)d{\bf x},
\end{gather}
where $\varphi_{\lambda}(x)$ is an antisymmetric orbit function,
are antisymmetric with respect to the action of the Weyl group
$W$. In particular, the functions ${\hat
u}_{i_1,i_2,\dots,i_n}(\lambda)$ vanish on Weyl chambers.

Formulas \eqref{func-s} and \eqref{func-a} are used for obtaining
symmetric and antisymmetric functions or polynomials.


\subsection[Eigenfunctions of (anti)symmetric orbit function transform for $W(A_n)$]{Eigenfunctions
of (anti)symmetric orbit function transform for
$\boldsymbol{W(A_n)}$} \label{section13.2}
Let $H_n(x)$, $n=0,1,2,\dots$, be the well-known Hermite
polynomials. They are def\/ined by the formula
\[
H_n(x)=n!\sum_{m=0}^{[n/2]} \frac{(-1)^m (2x)^{n-2m}}{m!(n-2m)!}
\]
where $[n/2]$ is an integral part of the number $n/2$. These
polynomials obey the dif\/ference equation
\begin{gather}\label{Her}
 \left( \frac{d^2}{dx^2}
-2x \frac{d}{dx} +2n\right) H_n(x)=0.
 \end{gather}
They satisfy the relation
\[
\frac1{\sqrt{2\pi}} \int_{-\infty}^\infty e^{{\rm
i}px}e^{-p^2/2}H_m(p)dp={\rm i}^{-m}e^{-x^2/2}H_m(x)
\]
(see, for example, Subsection 12.2.4 in \cite{KVII}), which can be
written in the form
 \begin{gather}\label{Herm-1}
 \int_{-\infty}^\infty e^{2\pi {\rm i}px}e^{-\pi p^2}
 H_m(\sqrt{2\pi}p)dp={\rm
i}^{-m}e^{-\pi x^2}H_m(\sqrt{2\pi}x).
 \end{gather}

Using the Hermite polynomials we create polynomials of many
variables
\begin{gather}\label{Her-m}
H_{\bf m}({\bf  x})\equiv H_{m_1,m_2,\dots,m_n}(x_1,x_2,\dots
,x_n):=H_{m_1}(x_1)H_{m_2}(x_2)\cdots H_{m_n}(x_n).
 \end{gather}
The functions
\begin{gather}\label{Her-2}
 e^{-|{\bf  x}|^2/2} H_{\bf m}({\bf  x}),\qquad
m_i=0,1,2,\dots,\quad i=1,2,\dots,n,
 \end{gather}
form an orthogonal basis of the Hilbert space $L^2(\mathbb{R}^n)$
with the scalar product{\samepage
\[
\langle f_1,f_2\rangle:=\int_{\mathbb{R}^n}
f_1(\mathbf{x})\overline{f_2(\mathbf{x})}d\mathbf{x} ,
\]
where $d\mathbf{x}=dx_1\,dx_2\cdots dx_n$.}

The polynomials $H_{\bf m}({\bf  x})$ satisfy the dif\/ferential
equation
\begin{gather}\label{Her-eq}
 \left( \Delta
-2\sum_{i=1}^n x_i \frac{\partial}{\partial x_i} +2|{\bf
m}|\right) H_{\bf m}({\bf  x})=0,
 \end{gather}
where $\Delta$ is the Laplace operator $\Delta=\sum\limits_{m=0}^n
\frac{\partial^2}{\partial x_m^2}$ and $|{\bf m}|=m_1+m_2+\cdots
+m_n$.

We make symmetrization and antisymmetrization of the functions
\[
{\mathcal H}_{\bf m}({\bf x}):=e^{-\pi |x|} H_{\bf m}(\sqrt{2\pi}
{\bf x})
\]
(obtained from \eqref{Her-2} by replacing ${\bf x}$ by
$\sqrt{2\pi}{\bf x}$) by means of orbit functions of $A_{n-1}$:
\begin{gather}\label{Her-3}
\int_{\mathbb{R}^n} \hat\phi_\lambda(\mathbf{x})
 e^{-\pi|{\bf  x}|^2} H_{\bf m}(\sqrt{2\pi}{\bf  x})=
 {\rm i}^{-|{\bf m}|} e^{-\pi|\lambda|^2}
 H^{\rm sym}_{\bf m}(\sqrt{2\pi}\lambda),
\\
\label{Her-4} \int_{\mathbb{R}^n} \varphi_\lambda(\mathbf{x})
 e^{-\pi|{\bf  x}|^2} H_{\bf m}(\sqrt{2\pi}{\bf  x})
 ={\rm i}^{-|{\bf m}|} e^{-\pi|\lambda|^2}
 H^{\rm anti}_{\bf m}(\sqrt{2\pi}\lambda),
\end{gather}
where $\hat\phi_\lambda(\mathbf{x})$ is a symmetric orbit function
of $A_{n-1}$, given by formula (6.2) in \cite{KP06},
$\varphi_\lambda(\mathbf{x})$ is an antisymmetric orbit function
of $A_{n-1}$, and
$\lambda=(\lambda_1,\lambda_2,\dots,\lambda_n)$.

The polynomials $H^{\rm sym}_{\bf m}$ and $H^{\rm anti}_{\bf m}$
are symmetric and antisymmetric, respectively, with respect to the
Weyl group $W\equiv S_n$ of $A_{n-1}$:
\[
H^{\rm sym}_{\bf m}(w\lambda)=H^{\rm sym}_{\bf m}(\lambda),\ \ \ \
H^{\rm anti}_{\bf m}(w\lambda)=(\det w)H^{\rm anti}_{\bf m}
(\lambda),\ \ \ \ w\in S_n.
\]
For this reason, $H^{\rm sym}_{\bf m}(\lambda)$ are uniquely
determined by their values of
$\lambda=(\lambda_1,\lambda_2,\dots,\lambda_n)$ such that
$\lambda_1\ge \lambda_2\ge \cdots \ge \lambda_n$ and $H^{\rm
anti}_{\bf m}(\lambda)$ by their values of $\lambda$ such that
$\lambda_1>\lambda_2 >\cdots >\lambda_n$. (Note that $H^{\rm
anti}_{\bf m}(\lambda)=0$ if $\lambda_i=\lambda_{i+1}$ for some
$i=1,2,\dots,n-1$.)

The polynomials $H^{\rm sym}_{\bf m}$ are of the form
 \begin{gather}\label{sym-H}
H^{\rm sym}_{\bf m}(\lambda)=\sum_{w\in S_n} H_{w\bf m}(\lambda),
 \end{gather}
where the polynomials $H_{w\bf m}(\lambda)$ are of the form
\eqref{Her-m}.  The polynomials $H^{\rm anti}_{\bf m}$ are of the
form
 \begin{gather}\label{anti-H}
H^{\rm anti}_{\bf m}(\lambda)=\sum_{w\in S_n}(\det w) H_{w\bf
m}(\lambda) ,
 \end{gather}
that is,
\[
H^{\rm anti}_{\bf m}(\lambda)= \det
\left(H_{m_i}(\lambda_j)\right)_{i,j=1}^n.
\]
Moreover, $H^{\rm anti}_{\bf m}(\lambda)=0$ if $m_i=m_{i+1}$ for
some $i=1,2,\dots,n-1$. For this reason, we may consider the
polynomials $H^{\rm sym}_{\bf m}(\lambda)$ for $n$-tuples ${\bf
m}$ such that $m_1\ge m_2\ge \cdots \ge m_n$ and the polynomials
$H^{\rm anti}_{\bf m}(\lambda)$ for $n$-tuples ${\bf m}$ such that
$m_1>m_2 >\cdots >m_n$.

{\sloppy Let us apply symmetric orbit function transform (8.10) of
\cite{KP06} to the symmetric func\-tion \eqref{sym-H}. Taking into
account formula~\eqref{Her-3} we obtain}
\begin{alignat*}{2}
\mathfrak{F}\left(  e^{-\pi|{\bf  x}|^2} H^{\rm sym}_{\bf m}
(\sqrt{2\pi}{\bf x})\right):=& \; \frac1{|S_n|}
\int_{\mathbb{R}^n} \hat\phi_\lambda({\bf x}) e^{-\pi|{\bf  x}|^2}
H^{\rm sym}_{\bf m} (\sqrt{2\pi}{\bf x}) d{\bf x}
 \notag\\
=& \; {\rm i}^{-|{\bf m}|} e^{-\pi|\lambda|^2}
 H^{\rm sym}_{\bf m}(\sqrt{2\pi}\lambda),
 \notag
\end{alignat*}
where $|S_n|$ is an order of the permutation group $S_n$, that is,
functions \eqref{sym-H} are eigenfunctions of the symmetric orbit
function transform $\mathfrak{F}$. Since the functions
\eqref{sym-H} for $m_i=0,1,2,\dots$, $i=1,2,\dots,n$, $m_1\ge
m_2\ge \cdots \ge m_n$, form an orthogonal basis of the Hilbert
space $L_{\rm sym}^2(\mathbb{R}^n)$ of functions from
$L^2(\mathbb{R}^n)$ symmetric with respect to $W$, then they
constitute a complete set of eigenfunctions of this transform.
Thus, this transform has only four eigenvalues ${\rm i}$, $-{\rm
i}$, $1$, $-1$ in $L_{\rm sym}^2(\mathbb{R}^n)$. This means that,
as in the case of the usual Fourier transform, we have
\[
\mathfrak{F}^4=1.
\]

Now we apply antisymmetric orbit function transform \eqref{F-3} to
the antisymmetric function~\eqref{anti-H}. Taking into account
formula~\eqref{Her-4} we obtain
\begin{alignat*}{2}
\tilde{\mathfrak{F}}\left(  e^{-\pi|{\bf  x}|^2} H^{\rm anti}_{\bf
m}(\sqrt{2\pi}{\bf x})\right):=& \; \frac1{|S_n|}
\int_{\mathbb{R}^n} \varphi_\lambda({\bf x}) e^{-\pi|{\bf  x}|^2}
H^{\rm anti}_{\bf m}(\sqrt{2\pi}{\bf x}) d{\bf x}
 \notag\\
=& \; {\rm i}^{-|{\bf m}|} e^{-\pi|\lambda|^2}
 H^{\rm anti}_{\bf m}(\sqrt{2\pi}\lambda),\ \ \ m_1>m_2>\cdots >m_n\ge 0,
 \end{alignat*}
that is, functions \eqref{anti-H} are eigenfunctions of the
symmetric orbit function transform $\tilde{\mathfrak{F}}$. Since
the functions \eqref{anti-H} for $m_i=0,1,2,\dots$;
$i=1,2,\dots,n$, $m_1>m_2>\cdots >m_n\ge 0$, form an orthogonal
basis of the Hilbert space $L_{\rm anti}^2(\mathbb{R}^n)$ of
functions from $L^2(\mathbb{R}^n)$ antisymmetric with respect to
$W$, then they constitute a complete set of eigenfunctions of this
transform. Thus, this transform has only four eigenvalues ${\rm
i}$, $-{\rm i}$, $1$, $-1$. This means that, as in the case of the
usual Fourier transform, we have
\[
\tilde{\mathfrak{F}}^4=1.
\]


\subsection{Symmetric and antisymmetric sets of polynomials}
\label{section13.3}
In the previous subsection we constructed symmetric and
antisymmetric sets of functions connected with Hermite
polynomials. Similarly other sets of orthogonal polynomials can be
constructed (see \cite{KMc62,Koor75} and \cite{BSX95}).

Let $p_m(x)$, $m=0,1,2,\dots$, be the set of orthogonal
polynomials in one variable such that
\[
\int_{\mathbb{R}} p_m(x)p_{m'}(x)d\sigma(x) =\delta_{mm'},
\]
where $d\sigma(x)$ is some orthogonality measure, which may be
f\/inite or discrete.

We create a set of symmetric polynomials of $n$ variables as
follows:
 \begin{gather}\label{sym-p}
p^{\rm sym}_{\bf m}({\bf x})= \sum_{w\in S_n/S_{\bf m}}
p_{m_{w(1)}}(x_1) p_{m_{w(2)}}(x_2)\cdots p_{m_{w(n)}}(x_n),
\\
 m_i=0,1,2,\dots,\qquad i=1,2,\dots,n,\nonumber
\end{gather}
where ${\bf m}=(m_1,m_2,\dots,m_n)$, $m_1\ge m_2\ge \cdots \ge
m_n\ge 0$, ${\bf x}=(x_1,x_2,\dots,x_n)$, and $w(1),w(2)$,
$\dots,w(n)$ is a set of numbers $1,2,\dots,n$ transformed by the
permutation $w\in S_n/S_{\bf m}$, where $S_{\bf m}$ is the
subgroup of $S_n$ consisting of elements leaving ${\bf m}$
invariant.

We also create the set of polynomials
\begin{gather}\label{sym-p'}
p^{\rm anti}_{\bf m}({\bf x})= \sum_{w\in S_n} (\det w)
p_{m_{w(1)}}(x_1) p_{m_{w(2)}}(x_2)\cdots p_{m_{w(n)}}(x_n)=\det
\left( p_{m_i}(x_j)\right)_{i,j=1}^n,
\\
m_i=0,1,2,\dots,\qquad i=1,2,\dots,n,\nonumber
\end{gather}
where notations are the same as in \eqref{sym-p} and the condition
$m_1>m_2>\cdots >m_n\ge 0$ is satisf\/ied.

It is easy to check that the polynomials $p^{\rm sym}_{\bf m}({\bf
x})$ are symmetric with respect to transformations of $S_n$:
\[
p^{\rm sym}_{\bf m}(w{\bf x}) = p^{\rm sym}_{\bf m}({\bf x}),\ \ \
\ w\in S_n.
\]
Similarly, the polynomials $p^{\rm anti}_{\bf m}({\bf x})$ are
antisymmetric with respect to transformations of $S_n$:
\[
p^{\rm anti}_{\bf m}(w{\bf x}) =(\det w) p^{\rm anti}_{\bf m}({\bf
x}),\ \ \ \ w\in S_n.
\]
Thus, we may consider the polynomials \eqref{sym-p} and
\eqref{sym-p'} on the closure of the fundamental domain of the
transformation group $W(A_{n-1})\equiv S_n$. This closure (we
denote it by $\overline{D_+}$) coincides with the set of points
${\bf x}=(x_1,x_2,\dots,x_n)$ for which
\[
x_1\ge x_2\ge \cdots \ge x_n\ge 0.
\]
The polynomials \eqref{sym-p'} vanish if for some $i$,
$i=1,2,\dots,n-1$, we have $x_i=x_{i+1}$.

The set of polynomials \eqref{sym-p}, as well as the set of the
polynomials \eqref{sym-p'}, is orthogonal with respect to the
product measure $d\sigma({\bf x})\equiv d\sigma(x_1)\,
d\sigma(x_2)\cdots d\sigma(x_n)$. Indeed, we have
\begin{gather*}
\int_{\overline{D_+}} p^{\rm sym}_{\bf m}({\bf x})
\overline{p^{\rm sym}_{{\bf m}'}({\bf x})}d\sigma({\bf
x})=\frac{|O({\bf m})|}{|S_n|} \delta_{{\bf mm}'}=\frac1{|S_{\bf
m}|} \delta_{{\bf mm}'},
\\
\int_{\overline{D_+}} p^{\rm anti}_{\bf m}({\bf x})
\overline{p^{\rm anti}_{{\bf m}'}({\bf x})}d\sigma({\bf x})=
\delta_{{\bf mm}'},
\end{gather*}
where $O({\bf m})$ is the $S_n$-orbit of the point ${\bf m}$.

Note that each polynomial $p^{\rm anti}_{\bf m}({\bf x})$ vanishes
at $x_i=x_j$ for any admitted $i$ and $j$. This means that $p^{\rm
anti}_{\bf m}({\bf x})$ can be divided by $x_i-x_j$. Therefore,
the functions
\[
P^{\rm anti}_{{\bf m}}({\bf x})=\frac{p^{\rm anti}_{{\bf m}}({\bf
x})}{\prod\limits_{1\le i<j\le n} (x_i-x_j)}
\]
are polynomials in $x_i$, $i=1,2,\dots,n$. These polynomials are
also orthogonal and the orthogonality relation is of the form
\[
\int_{\overline{D_+}} P^{\rm anti}_{\bf m}({\bf x})
\overline{P^{\rm anti}_{{\bf m}'}({\bf x})}\Xi({\bf
x})d\sigma({\bf x})= \delta_{{mm}'},
\]
where $\Xi({\bf x})={\prod\limits_{1\le i<j\le n} (x_i-x_j)}^2$.

\subsection*{Acknowledgements}

The f\/irst author (AK) acknowledges CRM of University of Montreal
for hospitality when this paper was under preparation. His
research was partially supported by Grant 10.01/015 of the State
Foundation of Fundamental Research of Ukraine. We are grateful for
partial support for this work to the National Research Council of
Canada, MITACS, the MIND Institute of Costa Mesa, California, and
Lockheed Martin, Canada.

\newpage

\pdfbookmark[1]{References}{ref}
\LastPageEnding
\end{document}